\documentclass[11pt]{article}%
\usepackage{amssymb}
\usepackage[singlespacing]{setspace}
\usepackage{amsmath}
\usepackage{amsfonts}
\usepackage{graphicx}
\usepackage{cite}%
\allowdisplaybreaks
\begin{document}

\author{\vspace{0.16in}Hartmut Wachter\thanks{e-mail:
Hartmut.Wachter@physik.uni-muenchen.de}\\Max-Planck-Institute\\for Mathematics in the Sciences\\Inselstr. 22, D-04103 Leipzig\\\hspace{0.4in}\\Arnold-Sommerfeld-Center\\Ludwig-Maximilians-Universit\"{a}t\\Theresienstr. 37, D-80333 M\"{u}nchen}
\title{Quantum kinematics on q-deformed quantum spaces I\\{\small Mathematical framework}}
\date{}
\maketitle

\begin{abstract}
\noindent The aim of these two papers (I and II) is to try to give
fundamental concepts of quantum kinematics to q-deformed quantum spaces. Paper
I introduces the relevant mathematical concepts. A short review of the basic
ideas of q-deformed analysis is given. These considerations are continued by
introducing q-deformed analogs of Fourier transformations and delta functions.
Their properties are discussed in detail. Furthermore, q-deformed versions of
sesquilinear forms are defined, their basic properties are derived, and
q-analogs of the Fourier-Plancherel identity are proved. In paper II these
reasonings are applied to wave functions on position and momentum
space.\strut\smallskip

\newpage

\end{abstract}
\tableofcontents

\section{Introduction}

In Refs. \cite{FLW96, Wes97, CW98} it was outlined that deformation of
classical\ spacetime symmetries can lead to discretizations of the spectra of
spacetime observables. This observation nourishes the hope for a new method to
regularize quantum field theories \cite{Schw, Heis, GKP96, MajReg, Oec99,
Blo03}. Let us recall that the concept of deforming spacetime symmetries is
heavily based on the Gelfand-Naimark\textsf{\ }theorem \cite{GeNe}, which
tells us that Lie groups can be naturally embedded in the category of
algebras. Realizing that spacetime symmetries are usually described by Lie
groups the utility of this interrelation lies in formulating the geometrical
structure of Lie groups in terms of a Hopf structure \cite{Hopf}. The point is
that during the last two decades generic methods have been discovered for
continuously deforming matrix groups and Lie algebras within the category of
Hopf algebras. This development finally led to the arrival of quantum groups
and quantum spaces \cite{Ku83, Wor87, Dri85, Jim85, Drin86, RFT90, Tak90}.

From a physical point of view the most realistic and interesting deformations
are given by q-deformed versions of Minkowski space and Euclidean spaces as
well as their corresponding symmetries, i.e. respectively Lorentz symmetry and
rotational symmetry \cite{CSSW90, PW90, SWZ91, Maj91, LWW97}. Further studies
even allowed to establish differential calculi on these q-deformed quantum
spaces \cite{WZ91, CSW91, Song92, OSWZ92} representing nothing other than
q-analogs of classical translational symmetry. In this sense we can say that
q-deformations of Euclidean and Poincar\'{e} symmetries are now available
\cite{Maj93-2}.

The aim of our previous work \cite{WW01, BW01, Wac02, Wac03, Wac04, Wac05,
MSW04, SW04} was to give a q-deformed version of analysis to quantum spaces of
physical interest, i.e. Manin plane, q-deformed Euclidean space in three or
four dimensions, and q-deformed Minkowski space. In this respect, attention
was focused on explicit formulae realizing the elements of q-analysis on
commutative coordinate algebras. In doing so, we obtained expressions for
calculating star products, operator representations, q-integrals,
q-exponentials, q-translations, and braided products on the quantum spaces
under consideration. In this manner we can say that our results established
multi-dimensional versions of the well-known q-calculus \cite{Kac00}.

In Ref. \cite{qAn}\textbf{ }we continued these considerations, but stress was
taken on a presentation that reveals the properties of the elements of
q-deformed analysis. It should be mentioned that in some sense our
examinations are based on the general ideas of Shahn Majid \cite{Maj91Kat,
Maj94Kat, Maj94-10, Maj93-Int}, but the considerations in Refs. \cite{CSW93}
and \cite{Schir94} go into the same direction.\ The key idea of this approach
is that all the quantum spaces to a given quantum symmetry form a braided
tensor category. Consequently, operations and objects concerning quantum
spaces must rely on this framework of a braided tensor category, in order to
guarantee their well-defined behavior under quantum group transformations.
This so-called principle of covariance can be seen as the essential guideline
for constructing a consistent theory.

In this article we collect the mathematical concepts to formulate quantum
kinematics within the framework of q-deformed quantum spaces. In
Sec.\ \ref{RevQAn}\ we first give a short review of the present development of
q-analysis. For the details we refer the reader to Ref. \cite{qAn}. In Sec.
\ref{FourTrans} we continue this work by introducing q-analogs of Fourier
transformations and delta functions, discuss different possibilities for their
definition, and exhibit their basic properties. In doing so, we adjust the
general ideas exposed in Ref.\ \cite{KM94} to our formalism and special needs.
Section \ref{SecPart} is devoted to sesquilinear forms on quantum spaces. We
first concern ourselves with their definition and derive the adjoint of
certain operators. Then we prove that the sesquilinear forms are invariant
under symmetry transformations and derive q-analogs of the Fourier-Plancherel
identity. Section \ref{SecCon} closes our considerations by a short
conclusion. For reference and for the purpose of introducing consistent and
convenient notation, we provide a review of key notation and results in App.
\ref{AppQuan}. Last but not least App. \ref{Proofs} contains some proofs that
are less interesting or more complicated.

\section{Review of q-deformed analysis\label{RevQAn}}

In this section, we collect definitions and basic constructions that will be
needed throughout the article. First of all, let us recall that q-analysis can
be regarded as a non-commutative analysis formulated within the framework of
quantum spaces \cite{Wess00}. Quantum spaces are defined as comodule algebras
of quantum groups and\ can be interpreted as deformations of ordinary
coordinate algebras. For our purposes, it is at first sufficient to consider a
quantum space as an algebra $\mathcal{A}_{q}$ of formal power series in
non-commuting coordinates $X^{1},X^{2},\ldots,X^{n},$ i.e.%

\begin{equation}
\mathcal{A}_{q}=\mathbb{C}\left[  \left[  X^{1},\ldots X^{n}\right]  \right]
/\mathcal{I},
\end{equation}
where $\mathcal{I}$ denotes the ideal generated by the relations of the
non-commuting coordinates.

The two-dimensional Manin plane is one of the simplest examples for a quantum
space \cite{Man88}. It consists of all the power series in two coordinates
$X^{1}$ and $X^{2}$ being subject to the commutation relations
\begin{equation}
X^{1}X^{2}=qX^{2}X^{1},\quad q>1.
\end{equation}
We can think of $q$ as a deformation parameter measuring the coupling among
different spatial degrees of freedom. In the classical case, i.e. if $q$
becomes $1$ we regain commutative coordinates.

Next, we would like to focus our attention on the question how to perform
calculations on an algebra of quantum space coordinates. This can be
accomplished by a kind of pullback transforming operations on non-commutative
coordinate algebras to those on commutative ones. For this to become more
clear, we have to realize that the non-commutative algebras\ we are dealing
with satisfy the \textit{Poincar\'{e}-Birkhoff-Witt property}. It tells us
that the dimension of a subspace of homogenous polynomials should be the same
as for commuting coordinates. This property is the deeper reason why monomials
of a given normal ordering constitute a basis of $\mathcal{A}_{q}$. Due to
this fact, we can establish a vector space isomorphism between $\mathcal{A}%
_{q}$ and a commutative algebra $\mathcal{A}$ generated by ordinary
coordinates $x^{1},x^{2},\ldots,x^{n}$:
\begin{align}
\mathcal{W}  &  :\mathcal{A}\longrightarrow\mathcal{A}_{q},\nonumber\\
\mathcal{W}((x^{1})^{i_{1}}\ldots(x^{n})^{i_{n}})  &  \equiv(X^{1})^{i_{1}%
}\ldots(X^{n})^{i_{n}}. \label{AlgIsoN}%
\end{align}

This vector space isomorphism can even be extended to an algebra isomorphism
by introducing a non-commutative product in $\mathcal{A}$, the so-called
\textit{star product} \cite{BFF78, Moy49, MSSW00}. This product is defined via
the relation
\begin{equation}
\mathcal{W}(f\circledast g)=\mathcal{W}(f)\cdot\mathcal{W}(g), \label{StarDef}%
\end{equation}
being tantamount to%
\begin{equation}
f\circledast g\equiv\mathcal{W}^{-1}\left(  \mathcal{W}\left(  f\right)
\cdot\mathcal{W}\left(  g\right)  \right)  ,
\end{equation}
where $f$ and $g$ are formal power series in $\mathcal{A}$. In the case of the
Manin plane, the star product is of the well-known form%
\begin{align}
&  f(x^{i})\circledast g(x^{j})=q^{-\hat{n}_{x^{2}}\hat{n}_{y^{1}}}\left.
\left[  f(x^{i})\,g(y^{j})\right]  \right\vert _{y\rightarrow x}\nonumber\\
&  =f(x^{i})\,g(x^{j})+O(h),\quad\text{with\textsf{\quad}}q=e^{h},
\label{Star2-dimN}%
\end{align}
where we have introduced the operators%
\begin{equation}
\hat{n}_{x^{i}}\equiv x^{i}\frac{\partial}{\partial x^{i}},\quad i=1,2.
\end{equation}
From the last equality in (\ref{Star2-dimN}) we see that star products on
quantum spaces lead to modifications of commutative products. Evidently, these
modifications vanish in the classical limit $q$ $\rightarrow$ $1$.

Next, we would like to deal with tensor products of quantum spaces. To this
end, we have to specify the commutation relations between generators of
distinct quantum spaces. These relations are determined by the requirement of
being invariant under the action of a Hopf algebra $\mathcal{H}$ describing
the symmetry of the quantum spaces. It is well-known that for the case of
$\mathcal{H}$ being quasitriangular the commutation relations between two
quantum space generators $X^{i}\in\mathcal{A}_{q}$ and $Y^{j}\in
\mathcal{A}_{q}^{\prime}$ have to be of the form%
\begin{align}
X^{i}Y^{j}  &  =(\mathcal{R}_{[2]}\triangleright Y^{j})\,(\mathcal{R}%
_{[1]}\triangleright X^{i})\nonumber\\
&  =(Y^{j}\triangleleft\mathcal{R}_{[2]})\,(X^{i}\triangleleft\mathcal{R}%
_{[1]})=k\,\hat{R}_{kl}^{ij}\,Y^{k}X^{l}, \label{VerRN}%
\end{align}
where $\mathcal{R=R}_{[1]}\otimes\mathcal{R}_{[2]}\in\mathcal{H}%
\otimes\mathcal{H}$ denotes the universal R-matrix of $\mathcal{H}$ and $k$
stands for a complex number. Alternatively, we can also take the transposed
inverse of $\mathcal{R}$, i.e. $\tau\circ\mathcal{R}^{-1}=\mathcal{R}%
_{[2]}^{-1}\otimes\mathcal{R}_{[1]}^{-1},$ giving us%
\begin{align}
X^{i}Y^{j}  &  =(\mathcal{R}_{[1]}^{-1}\triangleright Y^{j})\,(\mathcal{R}%
_{[2]}^{-1}\triangleright X^{i})\nonumber\\
&  =(Y^{j}\triangleleft\mathcal{R}_{[1]}^{-1})\,(X^{i}\triangleleft
\mathcal{R}_{[2]}^{-1})=k^{-1}(\hat{R}^{-1})_{kl}^{ij}\,Y^{k}X^{l}.
\label{VerRInN}%
\end{align}
It should be obvious that from relations (\ref{VerRN}) and (\ref{VerRInN}) it
also follows how arbitrary elements of distinct quantum spaces commute with
each other. By virtue of the algebra isomorphism $\mathcal{W}$ we are able to
realize this process of commutation on commutative coordinate algebras. To
this end we introduce the operations%
\begin{align}
f(x^{i})\,\overset{x|y}{\odot}_{\hspace{-0.01in}L}\,g(y^{j})  &
\equiv\mathcal{W}^{-1}\big (\mathcal{R}_{[1]}^{-1}\triangleright
\mathcal{W}(g)\big)\otimes\mathcal{W}^{-1}\big (\mathcal{R}_{[2]}%
^{-1}\triangleright\mathcal{W}(f)\big),\nonumber\\
f(x^{i})\,\overset{x|y}{\odot}_{\hspace{-0.01in}\bar{R}}\,g(y^{j})  &
\equiv\mathcal{W}^{-1}\big (\mathcal{W}(g)\triangleleft\mathcal{R}_{[1]}%
^{-1}\big)\otimes\mathcal{W}^{-1}\big (\mathcal{W}(f)\triangleleft
\mathcal{R}_{[2]}^{-1}\big), \label{BraidDef1}%
\end{align}
and
\begin{align}
f(x^{i})\,\overset{x|y}{\odot}_{\hspace{-0.01in}\bar{L}}\,g(y^{j})  &
\equiv\mathcal{W}^{-1}\big (\mathcal{R}_{[2]}\triangleright\mathcal{W}%
(g)\big)\otimes\mathcal{W}^{-1}\big (\mathcal{R}_{[1]}\triangleright
\mathcal{W}(f)\big),\nonumber\\
f(x^{i})\,\overset{x|y}{\odot}_{\hspace{-0.01in}R}\,g(y^{j})  &
\equiv\mathcal{W}^{-1}\big (\mathcal{W}(g)\triangleleft\mathcal{R}%
_{[2]}\big)\otimes\mathcal{W}^{-1}\big (\mathcal{W}(f)\triangleleft
\mathcal{R}_{[1]}\big), \label{BraidDef2N}%
\end{align}
where $f$ and $g$ denote formal power series in the commutative coordinate
algebras $\mathcal{A}_{x}$ and $\mathcal{A}_{y}$, respectively. The operations
in (\ref{BraidDef1}) and (\ref{BraidDef2N}) are referred to as \textit{braided
products.}

Using the explicit form of the identities in (\ref{VerRN}) and (\ref{VerRInN})
we derived in Ref. \cite{Wac05}\ commutation relations between normally
ordered monomials. With these results at hand we were able to write down
explicit formulae for computing braided products. As an example we give the
expression we obtained for the braided product of the Manin plane:%
\begin{align}
&  f(x^{1},x^{2})\,\overset{x|y}{\odot}_{\hspace{-0.01in}L}\,g(y^{1}%
,y^{2})\nonumber\\
&  =\sum_{i=0}^{\infty}q^{i^{2}}(-\lambda)^{i}\,\frac{(y^{2})^{i}\otimes
(x^{1})^{i}}{[[i]]_{q^{-2}}!}\,q^{-\hat{n}_{y^{1}}\otimes\,\hat{n}_{x^{2}%
}-2\hat{n}_{y^{2}}\otimes\,\hat{n}_{x^{2}}-2\hat{n}_{y^{1}}\otimes\,\hat
{n}_{x^{1}}-2\hat{n}_{y^{2}}\otimes\,\hat{n}_{x^{1}}}\nonumber\\
&  \qquad\times(D_{q^{-2}}^{1})^{i}g(q^{-i}y^{1},q^{-2i}y^{2})\otimes
(D_{q^{-2}}^{2})^{i}f(q^{-2i}x^{1},q^{-i}x^{2}),
\end{align}
where we introduced $\lambda\equiv q-q^{-1}$ and the
so-called\ \textit{Jackson derivatives} being defined by \cite{Jack08, Kac00}
\begin{equation}
D_{q^{a}}^{i}f(x^{i})\equiv\frac{f(x^{i})-f(q^{a}x^{i})}{(1-q^{a})x^{i}},\quad
a\in\mathbb{C}.
\end{equation}

Now, we are in a position to introduce the \textit{tensor product of quantum
spaces}. It is equipped with a multiplication being determined by%
\begin{equation}
(a\otimes a^{\prime})(b\otimes b^{\prime})=\big(a(\mathcal{R}_{[2]}%
\triangleright b)\big)\otimes\big((\mathcal{R}_{[1]}\triangleright a^{\prime
})b^{\prime}\big),
\end{equation}
or
\begin{equation}
(a\otimes a^{\prime})(b\otimes b^{\prime})=\big(a(\mathcal{R}_{[1]}%
^{-1}\triangleright b)\big)\otimes\big((\mathcal{R}_{[2]}^{-1}\triangleright
a^{\prime})b^{\prime}\big),
\end{equation}
where $a,b\in\mathcal{A}_{q}$ and $a^{\prime},b^{\prime}\in\mathcal{A}%
_{q}^{\prime}.$ We see that multiplication on a tensor product requires to
know the commutation relations between the elements of the two tensor factors.
Essentially for us is the fact that the algebra isomorphism $\mathcal{W}$
allows us to represent the tensor product of quantum spaces on a tensor
product of commutative coordinate algebras. To be more specific, this can be
achieved by extending the braided products in (\ref{BraidDef1}) and
(\ref{BraidDef2N}) as follows:%
\begin{align}
&  \big(f(x^{i})\otimes f^{\prime}(y^{j})\big)\,\overset{y|x}{\odot}%
_{\hspace{-0.01in}L}\,\big(g(x^{k})\otimes g^{\prime}(y^{l})\big)\nonumber\\
&  \equiv\big[f\overset{x}{\circledast}\mathcal{W}^{-1}\big (\mathcal{R}%
_{[1]}^{-1}\triangleright\mathcal{W}(g)\big)\big]\otimes\big[\mathcal{W}%
^{-1}\big (\mathcal{R}_{[2]}^{-1}\triangleright\mathcal{W}(f^{\prime
})\big)\overset{y}{\circledast}g^{\prime}\big], \label{BraTens1}%
\end{align}
and similarly%
\begin{align}
&  \big(f(x^{i})\otimes f^{\prime}(y^{j})\big)\,\overset{y|x}{\odot}%
_{\hspace{-0.01in}\bar{L}}\,\big(g(x^{k})\otimes g^{\prime}(y^{l}%
)\big)\nonumber\\
&  \equiv\big[f\overset{x}{\circledast}\mathcal{W}^{-1}\big (\mathcal{R}%
_{[2]}\triangleright\mathcal{W}(g)\big)\big]\otimes\big[\mathcal{W}%
^{-1}\big (\mathcal{R}_{[1]}\triangleright\mathcal{W}(f^{\prime}%
)\big)\overset{y}{\circledast}g^{\prime}\big], \label{BraTens2}%
\end{align}
where $f,g\in\mathcal{A}_{x}$ and $f^{\prime},g^{\prime}\in\mathcal{A}_{y}.$

Now, we come to q-deformed analogs of \textit{partial derivatives}, which act
upon the algebra of quantum space coordinates \cite{WZ91, CSW91, Song92}. In
analogy to (\ref{VerRN}) and (\ref{VerRInN}) there are several possibilities
for commutation relations between partial derivatives and quantum space
coordinates. However, the action of partial derivatives on quantum space
coordinates requires to modify the commutation relations in (\ref{VerRN}) and
(\ref{VerRInN}) in such a way that they take the form%
\begin{align}
\partial^{i}X^{j}  &  =g^{ij}+k(\hat{R}^{-1})_{kl}^{ij}\,X^{k}\partial
^{l},\nonumber\\
\hat{\partial}^{i}X^{j}  &  =\bar{g}^{ij}+k^{-1}\hat{R}_{kl}^{ij}\,X^{k}%
\hat{\partial}^{l},\label{LeibRuleAnfN}\\[0.1in]
X^{i}\partial^{j}  &  =-g^{ij}+k(\hat{R}^{-1})_{kl}^{ij}\,\partial^{k}%
X^{l},\nonumber\\
X^{i}\hat{\partial}^{j}  &  =-\bar{g}^{ij}+k^{-1}\hat{R}_{kl}^{ij}%
\,\hat{\partial}^{k}X^{l}, \label{LeibRuleEndN}%
\end{align}
where we introduced a conjugate quantum metric $\bar{g}^{ij}$. Notice that
$\partial^{i}$ and $\hat{\partial}^{i}$ differ from each other by a
normalization factor, only (see the discussion in Ref. \cite{qAn}).

From the q-deformed Leibniz rules in (\ref{LeibRuleAnfN}) and
(\ref{LeibRuleEndN}) we can derive left and right actions of partial
derivatives on quantum spaces, respectively. To this end, we repeatedly apply
the Leibniz rules in (\ref{LeibRuleAnfN}) to the product of a partial
derivative with a normally ordered monomial of quantum space coordinates,
until all partial derivatives stand to the right of all quantum space
coordinates. In the expression obtained this way we pick out the summands
without a partial derivative and bring them to normal ordering. This method
finally yields left actions of partial derivatives on normally ordered
monomials. Right actions of partial derivatives can be calculated in a similar
way. The only difference is that we start from a partial derivative standing
to the right of a normally ordered monomial and commute it to the left of all
quantum space coordinates.

The algebra isomorphism $\mathcal{W}$ allows us to introduce q-deformed
derivatives that act upon commutative functions. With the help of the
relations
\begin{align}
\mathcal{W}(\partial^{i}\triangleright f)  &  =\partial^{i}\triangleright
\mathcal{W}(f),\quad f\in\mathcal{A}\text{,}\nonumber\\
\mathcal{W}(f\triangleleft\partial^{i})  &  =\mathcal{W}(f)\triangleleft
\partial^{i},
\end{align}
or%
\begin{align}
\partial^{i}\triangleright f  &  \equiv\mathcal{W}^{-1}\left(  \partial
^{i}\triangleright\mathcal{W}(f)\right)  ,\nonumber\\
f\triangleleft\partial^{i}  &  \equiv\mathcal{W}^{-1}\left(  \mathcal{W}%
(f)\triangleleft\partial^{i}\right)  , \label{DefParAcN}%
\end{align}
the actions of partial derivatives on the quantum space algebra $\mathcal{A}%
_{q}$ carry over to the corresponding commutative algebra $\mathcal{A}$. It
should be obvious that each Leibniz rule in (\ref{LeibRuleAnfN}) and
(\ref{LeibRuleEndN}) leads to its own q-derivative:%
\begin{align}
\partial^{i}X^{j}  &  =g^{ij}+k(\hat{R}^{-1})_{kl}^{ij}\,X^{k}\partial^{l} &
&  \Rightarrow &  &  \partial^{i}\triangleright f,\nonumber\\
\hat{\partial}^{i}X^{j}  &  =\bar{g}^{ij}+k^{-1}\hat{R}_{kl}^{ij}\,X^{k}%
\hat{\partial}^{l} &  &  \Rightarrow &  &  \hat{\partial}^{i}\,\bar
{\triangleright}\,f,\label{FundWirkN}\\[0.16in]
X^{i}\partial^{j}  &  =-\bar{g}^{ij}+k(\hat{R}^{-1})_{kl}^{ij}\,\partial
^{k}X^{l} &  &  \Rightarrow &  &  f\,\bar{\triangleleft}\,\partial
^{i},\nonumber\\
X^{i}\hat{\partial}^{j}  &  =-g^{ij}+k^{-1}\hat{R}_{kl}^{ij}\,\hat{\partial
}^{k}X^{l} &  &  \Rightarrow &  &  f\triangleleft\hat{\partial}^{i}.
\label{FundWirk2N}%
\end{align}

In the work of Ref. \cite{BW01} we derived operator representations
for\ q-deformed partial derivatives by applying these ideas. Our results can
be viewed as multi-dimensional versions of the celebrated Jackson derivative
\cite{Jack08}. As an example we write down the left representations
for\ partial derivatives on the two-dimensional quantum plane:%
\begin{align}
\partial^{1}\triangleright f(x^{1},x^{2})  &  =-q^{-1/2}D_{q^{2}}^{2}%
f(qx^{1},x^{2}),\nonumber\\
\partial^{2}\triangleright f(x^{1},x^{2})  &  =q^{1/2}D_{q^{2}}^{1}%
f(x^{1},q^{2}x^{2}). \label{Par2dimNN}%
\end{align}

Next, we wish to introduce \textit{q-translations}. Before doing so, it is
useful to make contact with the notion of a \textit{cross product algebra
}\cite{Maj95, KS97,ChDe96}. It is well-known that we can combine a Hopf
algebra $\mathcal{H}$ with its representation space $\mathcal{A}_{q}$ to form
a left cross product algebra $\mathcal{A}_{q}\rtimes\mathcal{H}$ built on
$\mathcal{A}_{q}\otimes\mathcal{H}$ with product%
\begin{equation}
(a\otimes h)(b\otimes g)=a(h_{(1)}\triangleright b)\otimes h_{(2)}g,\quad
a,b\in\mathcal{A}_{q},\mathcal{\quad}h,g\in\mathcal{H}, \label{LefCrosPro}%
\end{equation}
where the coproduct of $h$ is written in the Sweedler notation, i.e.
$\Delta(h)=h_{(1)}\otimes h_{(2)}.$ There is also a right-handed version of
this notion called a right cross product algebra $\mathcal{H}\ltimes
\mathcal{A}$ and built on $\mathcal{H}\otimes\mathcal{A}$ with product%
\begin{equation}
(h\otimes a)(g\otimes b)=hg_{(1)}\otimes(a\triangleleft g_{(2)})b.
\label{RigCrosPro}%
\end{equation}

When $\mathcal{A}_{q}$ is a q-deformed quantum space the cross product
algebras have a Hopf structure. On quantum space generators the corresponding
coproduct, antipode, and counit take the form \cite{OSWZ92, Maj93-2}%
\begin{align}
\Delta_{\bar{L}}(X^{i})  &  =X^{i}\otimes1+(\mathcal{\bar{L}}_{x})_{j}%
^{i}\otimes X^{j},\nonumber\\
\Delta_{L}(X^{i})  &  =X^{i}\otimes1+(\mathcal{L}_{x})_{j}^{i}\otimes
X^{j},\label{HopfStrucN}\\[0.16in]
S_{\bar{L}}(X^{i})  &  =-S(\mathcal{\bar{L}}_{x})_{j}^{i}\,X^{j},\nonumber\\
S_{L}(X^{i})  &  =-S(\mathcal{L}_{x})_{j}^{i}\,X^{j},\label{SExplN}\\[0.16in]
\epsilon_{\bar{L}}(X^{i})  &  =\epsilon_{L}(X^{i})=0, \label{Coeins}%
\end{align}
where $\mathcal{L}_{x}$ and $\mathcal{\bar{L}}_{x}$ stand for the so-called
L-matrix and its conjugate. The entries of the L-matrices are elements of the
Hopf algebra $\mathcal{H}$, so $S\ $stands for the antipode of $\mathcal{H}$.
One should also notice that the above Hopf structures are related to opposite
Hopf structures via%
\begin{equation}
\Delta_{\bar{R}/R}=\tau\circ\Delta_{\bar{L}/L},\qquad S_{\bar{R}/R}=S_{\bar
{L}/L}^{-1},\qquad\epsilon_{\bar{R}/R}=\epsilon_{\bar{L}/L}, \label{RightHopf}%
\end{equation}
where $\tau$ shall\ denote the usual transposition of tensor factors.

An essential observation is that coproducts of coordinates imply their
translations \cite{Maj93-5, Maj93-2, Maj-93/3, Me95, Wac04, SW04}. This can be
seen as follows. The coproduct on coordinates is an algebra homomorphism. If
the coordinates constitute a module coalgebra then the behavior of the quantum
space coordinates $X^{i}$ under symmetry transformations carries over to their
coproducts $\Delta_{A}(X^{i}).$ More formally, we have
\begin{equation}
\Delta_{A}(X^{i}X^{j})=\Delta_{A}(X^{i})\Delta_{A}(X^{j})\quad\text{and\quad
}\Delta_{A}(hX^{i})=\Delta(h)\Delta_{A}(X^{i}), \label{Trans}%
\end{equation}
where $h\in\mathcal{H}$. Due to this fact we can think of (\ref{HopfStrucN})
as nothing other than an addition law for q-deformed vector components.

To proceed any further we have to realize that our algebra morphism
$\mathcal{W}^{-1}$ can be extended by
\begin{align}
\mathcal{W}_{L}^{-1}  &  :\mathcal{A}_{q}\rtimes\mathcal{H}\longrightarrow
\mathcal{A},\nonumber\\
\mathcal{W}_{L}^{-1}((X^{1})^{i_{1}}\ldots(X^{n})^{i_{n}}\otimes h)  &
\equiv\mathcal{W}^{-1}((X^{1})^{i_{1}}\ldots(X^{n})^{i_{n}})\,\epsilon(h),
\label{ExtAlgIsoL}%
\end{align}
or
\begin{align}
\mathcal{W}_{R}^{-1}  &  :\mathcal{H}\ltimes\mathcal{A}_{q}\longrightarrow
\mathcal{A},\nonumber\\
\mathcal{W}_{R}^{-1}(h\otimes(X^{1})^{i_{1}}\ldots(X^{n})^{i_{n}})  &
\equiv\epsilon(h)\,\mathcal{W}^{-1}((X^{1})^{i_{1}}\ldots(X^{n})^{i_{n}}),
\label{ExtAlgIsoR}%
\end{align}
with $\epsilon$ being the counit of the Hopf algebra $\mathcal{H}.$ With these
mappings at hand we are able to introduce q-deformed translations:%
\begin{align}
f(x^{i}\oplus_{L/\bar{L}}y^{j})  &  \equiv((\mathcal{W}_{L}^{-1}%
\otimes\mathcal{W}_{L}^{-1})\circ\Delta_{L/\bar{L}})(\mathcal{W}%
(f)),\nonumber\\
f(x^{i}\oplus_{R/\bar{R}}y^{j})  &  \equiv((\mathcal{W}_{R}^{-1}%
\otimes\mathcal{W}_{R}^{-1})\circ\Delta_{R/\bar{R}})(\mathcal{W}%
(f)),\label{DefCoProN}\\[0.16in]
f(\ominus_{L/\bar{L}}\,x^{i})  &  \equiv(\mathcal{W}_{R}^{-1}\circ
S_{L/\bar{L}})(\mathcal{W}(f)),\nonumber\\
f(\ominus_{R/\bar{R}}\,x^{i})  &  \equiv(\mathcal{W}_{L}^{-1}\circ
S_{R/\bar{R}})(\mathcal{W}(f)). \label{DefAntiN}%
\end{align}
Let us note that these operations make the algebra $\mathcal{A}$ equipped with
a star product into a \textit{braided Hopf algebra}. For the details we refer
the reader to Refs. \cite{Maj94Kat, Maj94-10, Maj93-Int, Maj95}. In the work
of Ref. \cite{Wac04} we found formulae that enable us to compute the results
of the above operations for arbitrary functions. The results for the Manin
plane, for example,\ reads as%
\begin{align}
f(x^{i}\oplus_{L}y^{j})  &  =\sum_{k_{1},k_{2}=0}^{\infty}\frac{(x^{1}%
)^{k_{1}}(x^{2})^{k_{2}}}{[[k_{1}]]_{q^{-2}}![[k_{2}]]_{q^{-2}}!}%
\,\big ((D_{q^{-2}}^{1})^{k_{1}}(D_{q^{-2}}^{2})^{k_{2}}f\big )(q^{-k_{2}%
}y^{1}),\label{CoForm2dim}\\[0.16in]
f(\ominus_{L}\,x^{i})  &  =q^{-(\hat{n}_{x^{1}})^{2}-(\hat{n}_{x^{2}}%
)^{2}-2\hat{n}_{x^{1}}\hat{n}_{x^{2}}}\,f(-qx^{1},-qx^{2}).
\end{align}
It should be mentioned that Eq. (\ref{CoForm2dim}) can be seen as q-deformed
version of the Taylor rule in two dimensions.

Important for us is the fact that q-deformed translations show a number of
properties that are very similar to those fulfilled by\ classical
translations:%
\begin{align}
f((x^{i}\oplus_{A}y^{j})\oplus_{A}z^{k})  &  =f(x^{i}\oplus_{A}(y^{j}%
\oplus_{A}z^{k})),\nonumber\\
f(0\oplus_{A}x^{i})  &  =f(x^{i}\oplus_{A}0)=f(x^{i}),\nonumber\\
f((\ominus_{A}\,x^{i})\oplus_{A}x^{j})  &  =f(x^{i}\oplus_{A}(\ominus
_{A}\,x^{j}))=f(0). \label{HopfAxiom}%
\end{align}
For a correct understanding of the relations in (\ref{HopfAxiom}) one has to
realize that
\begin{equation}
f(0)\equiv\epsilon(\mathcal{W}(f))=\left.  f(x^{i})\right\vert _{x^{i}=0}.
\end{equation}
Furthermore, we took the convention that%
\begin{equation}
f(x^{i}\oplus_{A}x^{j})=f_{(1)}\circledast f_{(2)},
\end{equation}
i.e. tensor factors which are addressed by the same coordinates have to be
multiplied via the star product. The relations in (\ref{HopfAxiom}) can be
interpreted as follows. The first relation is nothing else than the law of
associativity, whereas the second and third relation concern the existence of
the identity and that of an inverse, respectively.

Let us make contact with another important ingredient of q-analysis, i.e.
\textit{dual pairings} between the algebra of quantum space coordinates and
that of the corresponding partial derivatives. In Ref. \cite{Maj93-5} it was
shown that these pairings are given by
\begin{equation}
\left\langle .\,,.\right\rangle :\mathcal{A}_{q}^{\ast}\otimes\mathcal{A}%
_{q}\rightarrow\mathbb{C}\quad\text{with\quad}\big \langle f(\partial
^{i}),g(X^{j})\big \rangle\equiv\epsilon(f(\partial^{i})\triangleright
g(X^{j})), \label{DefParAb1}%
\end{equation}
or%
\begin{equation}
\left\langle .\,,.\right\rangle ^{\prime}:\mathcal{A}_{q}\otimes
\mathcal{A}_{q}^{\ast}\rightarrow\mathbb{C}\quad\text{with\quad}%
\big \langle f(X^{i}),g(\partial^{j})\big \rangle^{\prime}\equiv
\epsilon(f(X^{i})\triangleleft g(\partial^{j})). \label{DefParAb2}%
\end{equation}
As usual the above pairings carry over to commutative algebras by means of the
algebra isomorphism $\mathcal{W},$ i.e.%
\begin{equation}
\left\langle .\,,.\right\rangle :\mathcal{A}^{\ast}\otimes\mathcal{A}%
\rightarrow\mathbb{C}\quad\text{with\quad}\big \langle f(\partial^{i}%
),g(x^{j})\big \rangle\equiv\big \langle\mathcal{W}(f(\partial^{i}%
)),\mathcal{W}(g(x^{j}))\big \rangle,
\end{equation}
and likewise%
\begin{equation}
\left\langle .\,,.\right\rangle ^{\prime}:\mathcal{A}\otimes\mathcal{A}^{\ast
}\rightarrow\mathbb{C}\quad\text{with}\quad\big \langle f(x^{i}),g(\partial
^{j})\big \rangle^{\prime}\equiv\big \langle\mathcal{W}(f(x^{i})),\mathcal{W}%
(g(\partial^{j}))\big \rangle^{\prime}.
\end{equation}

Since partial derivatives can act on coordinates in four different ways, there
are four possibilities for defining a pairing between coordinates and
derivatives. More concretely, we have%
\begin{align}
\big \langle f(\partial^{i}),g(x^{j})\big \rangle_{L,\bar{R}}  &
\equiv(f(\partial^{i})\triangleright g(x^{j}))|_{x^{j}=\,0}=(f(\partial
^{i})\,\bar{\triangleleft}\,g(x^{j}))|_{\partial^{i}=\,0},\nonumber\\
\big \langle f(\hat{\partial}^{i}),g(x^{j})\big \rangle_{\bar{L},R}  &
\equiv(f(\hat{\partial}^{i})\,\bar{\triangleright}\,g(x^{j}))|_{x^{j}%
=\,0}=(f(\hat{\partial}^{i})\triangleleft g(x^{j}))|_{\hat{\partial}^{i}%
=\,0},\label{ExpDualAnfN}\\[0.16in]
\big \langle f(x^{i}),g(\partial^{j})\big \rangle_{L,\bar{R}}  &
\equiv(f(x^{i})\,\bar{\triangleleft}\,g(\partial^{j}))|_{x^{i}=\,0}%
=(f(x^{i})\triangleright g(\partial^{j}))|_{\partial^{j}=\,0},\nonumber\\
\big \langle f(x^{i}),g(\hat{\partial}^{j})\big \rangle_{\bar{L},R}  &
\equiv(f(x^{i})\triangleleft g(\hat{\partial}^{j}))|_{x^{i}=\,0}%
=(f(x^{i})\,\bar{\triangleright}\,g(\hat{\partial}^{j}))|_{\hat{\partial}%
^{j}=\,0}. \label{ExplDualEndN}%
\end{align}
Notice that in (\ref{ExpDualAnfN}) and (\ref{ExplDualEndN}) labels were
introduced to characterize the different pairings. Their meaning can be
understood in the following way. The left and right indices refer to the left
and right arguments of the dual pairings, respectively. If an argument of a
given dual paring acts on the other argument via the conjugate representation
the index for\ the acting argument is overlined. One should also notice that
each pairing can be calculated in two different ways. This observation is a
consequence of the fact that in our approach derivatives and coordinates play
symmetrical roles. As an example we give the explicit form of a pairing for
the Manin plane. In this case the first pairing in (\ref{ExpDualAnfN}) reads
on normally ordered monomials as follows:%
\begin{equation}
\big\langle(\partial_{1})^{n_{1}}(\partial_{2})^{n_{2}},(x^{2})^{m_{2}}%
(x^{1})^{m_{1}}\big\rangle_{L,\bar{R}}=\delta_{m_{1},n_{1}}\delta_{m_{2}%
,n_{2}}\,[[n_{1}]]_{q^{2}}!\,[[n_{2}]]_{q^{2}}!. \label{ExplDualn}%
\end{equation}

Now, we are ready to turn to a short discussion of
\textit{q-deformed\ exponentials}. From an abstract point of view an
exponential is nothing other than an object whose dualization is given by one
of the pairings in (\ref{ExpDualAnfN}) or (\ref{ExplDualEndN}). Thus, the
exponential can be introduced as the mapping%
\begin{equation}
\exp:\mathbb{C}\longrightarrow\mathcal{A}_{q}\otimes\mathcal{A}_{q}^{\ast
},\quad\text{with\quad}\exp(\alpha)=\alpha\sum_{a}e_{a}\otimes f^{a},
\label{AbsExp1}%
\end{equation}
or%
\begin{equation}
\exp^{\prime}:\mathbb{C}\longrightarrow\mathcal{A}_{q}^{\ast}\otimes
\mathcal{A}_{q},\quad\text{with\quad}\exp^{\prime}(\alpha)=\alpha\sum_{a}%
f^{a}\otimes e_{a}, \label{AbsExp2}%
\end{equation}
where $\{e_{a}\}$ is a basis in $\mathcal{A}_{q}$ and $\{f^{b}\}$ a dual basis
in $\mathcal{A}_{q}^{\ast}$, i.e. it holds%
\begin{equation}
\big \langle e_{a},f^{b}\big \rangle=\delta_{a}^{b},\quad\text{\quad
}\big \langle f^{b},e_{a}\big \rangle^{\prime}=\delta_{a}^{b}.
\end{equation}
It should be obvious that the algebra isomorphism $\mathcal{W}$ enables us to
introduce q-deformed exponentials that live on a tensor product of commutative
algebras. More concretely, this is achieved by the expressions%
\begin{equation}
\exp(x^{i}|\partial^{j})\equiv\sum_{a}\mathcal{W}(e_{a})\otimes\mathcal{W}%
(f^{a}), \label{ExpAll}%
\end{equation}
and%
\begin{equation}
\exp^{\prime}(\partial^{i}|x^{j})\equiv\sum_{a}\mathcal{W}(f^{a}%
)\otimes\mathcal{W}(e_{a}). \label{ExpAl2}%
\end{equation}

If we want to derive explicit formulae for q-deformed exponentials it is our
task to determine a basis of the coordinate algebra $\mathcal{A}_{q}$ being
dual to a given one of the derivative algebra $\mathcal{A}_{q}^{\ast}$.
Inserting the elements of the two bases into the formulae of (\ref{ExpAll})
and (\ref{ExpAl2}) will then provide us with expressions for q-deformed
exponentials. It should also be mentioned that in general the two bases being
dually paired depend on the choice of the pairing. Thus, each pairing in
(\ref{ExpDualAnfN}) and (\ref{ExplDualEndN}) leads to its own q-exponential:%
\begin{align}
&  \big\langle f(\partial^{i}),g(x^{j})\big\rangle_{L,\bar{R}} &  &
\Rightarrow &  &  \exp(x^{i}|\partial^{j})_{\bar{R},L},\nonumber\\
&  \big\langle f(\hat{\partial}^{i}),g(x^{j})\big\rangle_{\bar{L},R} &  &
\Rightarrow &  &  \exp(x^{i}|\hat{\partial}^{j})_{R,\bar{L}}%
,\label{DefExpAnfN}\\[0.16in]
&  \big\langle f(x^{i}),g(\partial^{j})\big\rangle_{L,\bar{R}} &  &
\Rightarrow &  &  \exp(\partial^{i}|x^{j})_{\bar{R},L},\nonumber\\
&  \big\langle f(x^{i}),g(\hat{\partial}^{j})\big\rangle_{\bar{L},R} &  &
\Rightarrow &  &  \exp(\hat{\partial}^{i}|x^{j})_{R,\bar{L}}.
\end{align}
In Ref. \cite{Wac03} we presented explicit formulae for q-deformed
exponentials. Once again, we would like to give a result from the
two-dimensional quantum plane. For the second exponential in (\ref{DefExpAnfN}%
) we found the expression%
\begin{equation}
\exp(x^{i}|\hat{\partial}^{j})_{R,\bar{L}}=\sum_{n_{1},n_{2}=\,0}^{\infty
}\frac{(x^{1})^{n_{1}}(x^{2})^{n_{2}}\otimes(\hat{\partial}_{2})^{n_{2}}%
(\hat{\partial}_{1})^{n_{1}}}{[[n_{1}]]_{q^{-2}}!\,[[n_{2}]]_{q^{-2}}!}.
\label{ExplExp}%
\end{equation}

There are two properties of q-exponentials worth recording here. First,
q-exponentials are subject to the addition laws%
\begin{align}
\exp(x^{i}\oplus_{L}y^{j}|\hat{\partial}^{k})_{R,\bar{L}}  &  =\exp(y^{j}%
|\hat{\partial}^{k})_{R,\bar{L}}\overset{y\partial|x}{\odot}_{\hspace
{-0.06in}\bar{R}}\exp(x^{i}|\hat{\partial}^{l})_{R,\bar{L}},\nonumber\\
\exp(x^{i}\oplus_{\bar{L}}y^{j}|\partial^{k})_{\bar{R},L}  &  =\exp
(y^{j}|\partial^{k})_{\bar{R},L}\overset{y\partial|x}{\odot}_{\hspace
{-0.06in}R}\exp(x^{i}|\partial^{l})_{\bar{R},L},\label{AddExp1N}\\[0.16in]
\exp(\hat{\partial}^{k}|x^{j}\oplus_{\bar{R}}y^{i})_{R,\bar{L}}  &  =\exp
(\hat{\partial}^{l}|y^{i})_{R,\bar{L}}\overset{y|\partial x}{\odot}%
_{\hspace{-0.06in}L}\exp(\hat{\partial}^{k}|x^{j})_{R,\bar{L}},\nonumber\\
\exp(\partial^{k}|x^{j}\oplus_{R}y^{i})_{\bar{R},L}  &  =\exp(\partial
^{k}|y^{i})_{\bar{R},L}\overset{y|\partial x}{\odot}_{\hspace{-0.06in}\bar{L}%
}\exp(\partial^{l}|x^{j})_{\bar{R},L}. \label{AddExp2N}%
\end{align}
Second, they are in some sense eigenfunctions of partial derivatives, since
they obey \cite{Wac03, Maj93-5, Schir94}%
\begin{align}
\partial^{i}\overset{x}{\triangleright}\exp(x^{k}|\partial^{l})_{\bar{R},L}
&  =\exp(x^{k}|\partial^{l})_{\bar{R},L}\overset{\partial}{\circledast
}\partial^{i},\nonumber\\
\hat{\partial}^{i}\,\overset{x}{\bar{\triangleright}}\,\exp(x^{k}%
|\hat{\partial}^{l})_{R,\bar{L}}  &  =\exp(x^{k}|\hat{\partial}^{l}%
)_{R,\bar{L}}\overset{\partial}{\circledast}\hat{\partial}^{i},\label{EigGl1N}%
\\[0.16in]
\exp(\partial^{l}|x^{k})_{\bar{R},L}\,\overset{x}{\bar{\triangleleft}%
}\,\partial^{i}  &  =\partial^{i}\overset{\partial}{\circledast}\exp
(\partial^{l}|x^{k})_{\bar{R},L},\nonumber\\
\exp(\hat{\partial}^{l}|x^{k})_{R,\bar{L}}\overset{x}{\triangleleft}%
\hat{\partial}^{i}  &  =\hat{\partial}^{i}\overset{\partial}{\circledast}%
\exp(\hat{\partial}^{l}|x^{k})_{R,\bar{L}}. \label{EigGl2}%
\end{align}

Lastly, we wish to introduce \textit{q-deformed integrals} as operations being
inverse to partial derivatives \cite{Wac02}. (For other concepts of
integration on quantum spaces see Ref. \cite{Fio93, KM94, Chry96, Sta96}.) To
this end, we first extend the algebra of partial derivatives by inverse
elements. In the case of the two-dimensional quantum plane the algebra of
partial derivatives is now characterized by the following relations:%
\begin{align}
\partial^{2}\partial^{1}  &  =q^{-1}\partial^{1}\partial^{2}%
,\label{VerExdParAl}\\[0.1in]
\partial^{i}(\partial^{i})^{-1}  &  =(\partial^{i})^{-1}\partial^{i}=1,\quad
i=1,2,\nonumber\\
\partial^{2}(\partial^{1})^{-1}  &  =q(\partial^{1})^{-1}\partial
^{2},\nonumber\\
\partial^{1}(\partial^{2})^{-1}  &  =q(\partial^{2})^{-1}\partial
^{1},\label{ZwVEPAN}\\[0.1in]
(\partial^{2})^{-1}(\partial^{1})^{-1}  &  =q^{-1}(\partial^{1})^{-1}%
(\partial^{2})^{-1}. \label{VerExdEnd}%
\end{align}

As next step we have to find representations for the inverse partial
derivatives. This can be achieved in the following way. In Ref. \cite{BW01} it
was shown that according to
\begin{equation}
\partial^{i}\triangleright F=\left(  \partial_{\text{cl}}^{i}+\partial
_{\text{cor}}^{i}\right)  F \label{SplitAbl}%
\end{equation}
representations of our partial derivatives can be divided up into a classical
part and corrections vanishing in the undeformed limit $q\rightarrow1$. For
this reason a solution to\ the difference equation
\begin{equation}
\partial^{i}\triangleright F=f
\end{equation}
for given $f$ can be written in the form%
\begin{align}
F  &  =(\partial^{i})^{-1}\triangleright f=\frac{1}{\partial_{\text{cl}}%
^{i}+\partial_{\text{cor}}^{i}}f\nonumber\\
&  =\sum_{k=\,0}^{\infty}\left(  -1\right)  ^{k}\left[  (\partial_{\text{cl}%
}^{i})^{-1}\partial_{\text{cor}}^{i}\right]  ^{k}(\partial_{\text{cl}}%
^{i})^{-1}f. \label{IntegralE3}%
\end{align}

To apply this formula, we need to identify the contributions $\partial
_{\text{cl}}^{i}$ and $\partial_{\text{cor}}^{i}$ in the representations of
partial derivatives. In the two-dimensional case, for example, we can read off
from the expressions in (\ref{Par2dimNN}) that%
\begin{gather}
(\partial_{\text{cl}}^{1})f=-q^{-1/2}D_{q^{2}}^{2}f(qx^{1}),\quad
(\partial_{\text{cor}}^{1})f=0,\nonumber\\
(\partial_{\text{cl}}^{2})f=q^{1/2}D_{q^{2}}^{1}f(q^{2}x^{2}),\quad
(\partial_{\text{cor}}^{2})f=0.
\end{gather}
Plugging this into formula (\ref{IntegralE3}) leaves us with
\begin{align}
(\partial^{1})^{-1}\big |_{-\infty}^{x^{2}}\triangleright f &  =(\partial
_{\text{cl}}^{1})^{-1}\big |_{-\infty}^{x^{2}}\triangleright f=-q^{1/2}%
(D_{q^{2}}^{2})^{-1}\big |_{-\infty}^{x^{2}}f(q^{-1}x^{1}),\nonumber\\
(\partial^{2})^{-1}\big |_{-\infty}^{x^{1}}\triangleright f &  =(\partial
_{\text{cl}}^{2})^{-1}\big |_{-\infty}^{x^{1}}\triangleright f=q^{-1/2}%
(D_{q^{2}}^{1})^{-1}\big |_{-\infty}^{x^{1}}f(q^{-2}x^{2}),\label{Int2dimN}%
\end{align}
where $(D_{q^{a}}^{i})^{-1}|_{-\infty}^{x^{i}}$ denotes the \textit{Jackson
integral} operator \cite{Jack27} over the interval $]-\infty,x^{i}]$.

What we have done so far applies to each of the representations in
(\ref{FundWirkN}) and (\ref{FundWirk2N}). In this manner, q-deformed integrals
can be placed into four categories:%
\begin{align}
&  \partial^{i}\triangleright f &  &  \Rightarrow &  &  \int_{-\infty
}^{x^{\overline{i}}}d_{L}y^{\overline{i}}\,f(y^{j})\equiv(\partial^{i}%
)^{-1}\big |_{-\infty}^{x^{\overline{i}}}\triangleright f, &  & \nonumber\\
&  \hat{\partial}^{i}\,\bar{\triangleright}\,f &  &  \Rightarrow &  &
\int_{-\infty}^{x^{\overline{i}}}d_{\bar{L}}y^{\overline{i}}\,f(y^{j}%
)\equiv(\hat{\partial}^{i})^{-1}\big |_{-\infty}^{x^{\overline{i}}}%
\,\bar{\triangleright}\,f, &  & \\[0.16in]
&  f\triangleleft\hat{\partial}^{i} &  &  \Rightarrow &  &  \int_{-\infty
}^{x^{\overline{i}}}d_{R}y^{\overline{i}}\,f(y^{j})\equiv f\triangleleft
(\hat{\partial}^{i})^{-1}\big |_{-\infty}^{x^{\overline{i}}}, &  & \nonumber\\
&  f\,\bar{\triangleleft}\,\partial^{i} &  &  \Rightarrow &  &  \int_{-\infty
}^{x^{\overline{i}}}d_{\bar{R}}y^{\overline{i}}\,f(y^{j})\equiv f\,\bar
{\triangleleft}\,(\partial^{i})^{-1}\big |_{-\infty}^{x^{\overline{i}}}. &  &
\end{align}
Notice that $\overline{i}$ denotes the so-called conjugate index of $i$. For
its definition see Ref. \cite{qAn}.

If we want to have integrals over the entire space we can apply
one-dimensional integrals in succession, i.e., for example,%
\begin{align}
\int\limits_{-\infty}^{+\infty}d_{L}^{n}x\,f  &  \equiv\int\limits_{-\infty
}^{+\infty}d_{L}x^{\bar{n}}\ldots\int\limits_{-\infty}^{+\infty}d_{L}%
x^{\bar{2}}\int\limits_{-\infty}^{+\infty}d_{L}x^{\bar{1}}\,f(x^{i}%
)\nonumber\\
&  =\lim_{x^{\bar{1}},\ldots,x^{\bar{n}}\rightarrow\infty}\big ((\partial
^{n})^{-1}\big |_{-\infty}^{x^{\bar{n}}}\ldots(\partial^{1})^{-1}%
\big |_{-\infty}^{x^{\bar{1}}}\triangleright f\big ),\\[0.1in]
\int\limits_{-\infty}^{+\infty}d_{R}^{n}x\,f  &  \equiv\int\limits_{-\infty
}^{+\infty}d_{R}x^{1}\int\limits_{-\infty}^{+\infty}d_{R}x^{2}\ldots
\int\limits_{-\infty}^{+\infty}d_{R}x^{n}\,f(x^{i})\nonumber\\
&  =\lim_{x^{1},\ldots,x^{n}\rightarrow\infty}\big (f\triangleleft
(\hat{\partial}^{\bar{1}})^{-1}\big |_{-\infty}^{x^{1}}\ldots(\hat{\partial
}^{\bar{n}})^{-1}\big |_{-\infty}^{x^{n}}\big ). \label{VolIntN}%
\end{align}
For the quantum spaces we are interested in these integrals take the form
\cite{qAn}

\begin{enumerate}
\item[(i)] (quantum plane)%
\begin{equation}
\int\limits_{-\infty}^{+\infty}d_{L}^{2}x\,f(x^{1},x^{2})=-\frac{q}%
{16}(D_{q^{1/2}}^{1})^{-1}\big |_{-\infty}^{\infty}(D_{q^{1/2}}^{2}%
)^{-1}\big |_{-\infty}^{\infty}f, \label{IntExp2}%
\end{equation}

\item[(ii)] (three-dimensional Euclidean space)%
\begin{gather}
\int\limits_{-\infty}^{+\infty}d_{L}^{3}x\,f(x^{+},x^{3},x^{-})=\nonumber\\
\frac{q^{-6}}{4}(D_{q^{2}}^{+})^{-1}\big |_{-\infty}^{\infty}(D_{q^{2}}%
^{3})^{-1}\big |_{-\infty}^{\infty}(D_{q^{2}}^{-})^{-1}\big |_{-\infty
}^{\infty}f,
\end{gather}

\item[(iii)] (four-dimensional Euclidean space)%
\begin{gather}
\int\limits_{-\infty}^{+\infty}d_{L}^{4}x\,f(x^{4},x^{3},x^{2},x^{1}%
)=\nonumber\\
\frac{1}{16}(D_{q}^{1})^{-1}\big |_{-\infty}^{\infty}(D_{q}^{2})^{-1}%
\big |_{-\infty}^{\infty}(D_{q}^{3})^{-1}\big |_{-\infty}^{\infty}(D_{q}%
^{4})^{-1}\big |_{-\infty}^{\infty}f,
\end{gather}

\item[(iv)] (q-deformed Minkowski space)%
\begin{gather}
\int\limits_{-\infty}^{+\infty}d_{L}^{4}x\,f(r^{2},x^{-},x^{3/0}%
,x^{+})=\nonumber\\
-\frac{1}{16}(q\lambda_{+})^{-3}(D_{q^{-1}}^{r^{2}})^{-1}\big |_{-\infty
}^{\infty}(D_{q^{-1}}^{+})^{-1}\big |_{-\infty}^{\infty}\nonumber\\
\times(D_{q^{-1}}^{3/0})^{-1}\big |_{-\infty}^{\infty}(D_{q^{-1}}^{-}%
)^{-1}\big |_{-\infty}^{\infty}f, \label{IntExpMin}%
\end{gather}

\end{enumerate}

\noindent where $(q>1,$ $a>0)$%

\begin{equation}
(D_{q^{\pm a}}^{i})^{-1}\big |_{-\infty}^{\infty}f\equiv\mp(1-q^{\pm a}%
)\sum_{k=-\infty}^{\infty}q^{ak}\big (f(q^{ak})+f(-q^{ak})\big ).
\label{JackDef}%
\end{equation}

The expressions in (\ref{IntExp2})-(\ref{IntExpMin}) behave like scalars,
which implies their trivial braiding with space and momentum coordinates, i.e.%
\begin{align}
f(x^{i})\overset{x|y}{\odot}_{\hspace{-0.01in}C}\int_{-\infty}^{+\infty}%
d_{A}^{n}y\,g(y^{j})  &  =\int_{-\infty}^{+\infty}d_{A}^{n}y\,f(\kappa
_{C}x^{i})\overset{x|y}{\odot}_{\hspace{-0.01in}C}g(y^{j})\nonumber\\
&  =\int_{-\infty}^{+\infty}d_{A}^{n}y\,g(y^{j})\otimes f(x^{i}),
\end{align}
and%
\begin{align}
f(p^{i})\overset{p|y}{\odot}_{\hspace{-0.01in}C}\int_{-\infty}^{+\infty}%
d_{A}^{n}y\,g(y^{j})  &  =\int_{-\infty}^{+\infty}d_{A}^{n}y\,f((\kappa
_{C})^{-1}p^{i})\overset{p|y}{\odot}_{\hspace{-0.01in}C}g(y^{j})\nonumber\\
&  =\int_{-\infty}^{+\infty}d_{A}^{n}y\,g(y^{j})\otimes f(p^{i}),
\end{align}
where $A,C\in\{L,\bar{L},R,\bar{R}\}.$ Notice that the scalings of position
and momentum coordinates,which in the above formulae are given by the
constants $\kappa_{C}$ and $(\kappa_{C})^{-1}$, respectively, result from the
braiding properties of the volume elements. With our conventions the values of
the constant $\kappa_{C}$ are determined as follows:

\begin{enumerate}
\item[(i)] (quantum plane)%
\begin{equation}
\kappa=\kappa_{\bar{L}}=\kappa_{R}=(\kappa_{L})^{-1}=(\kappa_{\bar{R}}%
)^{-1}=q^{3},
\end{equation}

\item[(ii)] (three-dimensional q-deformed Euclidean space)%
\begin{equation}
\kappa=\kappa_{\bar{L}}=\kappa_{R}=(\kappa_{L})^{-1}=(\kappa_{\bar{R}}%
)^{-1}=q^{6},
\end{equation}

\item[(iii)] (four-dimensional q-deformed Euclidean space)%
\begin{equation}
\kappa=\kappa_{\bar{L}}=\kappa_{R}=(\kappa_{L})^{-1}=(\kappa_{\bar{R}}%
)^{-1}=q^{4},
\end{equation}

\item[(iv)] (q-deformed Minkowski space)%
\begin{equation}
\kappa=\kappa_{\bar{L}}=\kappa_{R}=(\kappa_{L})^{-1}=(\kappa_{\bar{R}}%
)^{-1}=q^{-4}.
\end{equation}

\end{enumerate}

Important for us is the fact that q-deformed integrals over the whole space
are invariant under translations in the sense that%
\begin{align}
&  \int_{-\infty}^{\infty}d_{A}^{n}x(\partial^{i}\triangleright f)=\int
_{-\infty}^{\infty}d_{A}^{n}x(\hat{\partial}^{i}\,\bar{\triangleright
}\,f)=\nonumber\\
=  &  \int_{-\infty}^{\infty}d_{A}^{n}x(f\triangleleft\hat{\partial}^{i}%
)=\int_{-\infty}^{\infty}d_{A}^{n}x(f\,\bar{\triangleleft}\,\partial^{i})=0,
\label{TranProp}%
\end{align}
where $A\in\{L,\bar{L},R,\bar{R}\}.$ Finally, it should be mentioned that the
identities in (\ref{TranProp}) are equivalent to%
\begin{align}
\int\limits_{-\infty}^{+\infty}d_{A}x^{n}\,f(x^{i})  &  =\int\limits_{-\infty
}^{+\infty}d_{A}x^{n}\,f(y^{j}\oplus_{\bar{L}}x^{i})=\int\limits_{-\infty
}^{+\infty}d_{A}x^{n}\,f(y^{j}\oplus_{L}x^{i})\nonumber\\
&  =\int\limits_{-\infty}^{+\infty}d_{A}x^{n}\,f(x^{i}\oplus_{\bar{R}}%
y^{j})=\int\limits_{-\infty}^{+\infty}d_{A}x^{n}\,f(x^{i}\oplus_{R}y^{j}).
\label{TransGlob}%
\end{align}

\section{Fourier transformations on quantum spaces\label{FourTrans}}

Fourier transformations play a very important role in quantum physics, since
they allow decomposition of wave-functions into plane-waves. In this section
we present a detailed discussion of q-deformed Fourier transformations. In
doing so, we will follow the ideas exposed in Ref. \cite{KM94}, as it is our
aim to include them into our framework.

First, we use the objects of q-analysis to write down q-analogs of Fourier
transformations and delta functions. Then, it is shown that q-deformed Fourier
transformations intertwine actions of partial derivatives with star
multiplication by coordinates. In addition to this we give the proof that
q-deformed Fourier transformations are in some sense invertible. Towards this
end we introduce a second set of q-deformed Fourier transformations and
discuss their basic properties. For later purpose and to make our
considerations more concrete we calculate the Fourier transforms of
q-exponentials and q-deformed delta functions. For the sake of completeness a
subsection on convolution products has been added, but in this article we do
not make any further use of this notion. Lastly, we examine how Fourier
transformations behave under conjugation.

\subsection{Definition of Fourier transformations and related objects}

Since we have different versions of q-deformed integrals and q-deformed
exponentials, there are the following possibilities for defining q-deformed
Fourier transformations:%
\begin{align}
\mathcal{F}_{L}(f)(p^{k})  &  \equiv\int\nolimits_{-\infty}^{+\infty}d_{L}%
^{n}x\,f(x^{i})\overset{x}{\circledast}\exp(x^{j}|\text{i}^{-1}p^{k})_{\bar
{R},L},\nonumber\\
\mathcal{F}_{\bar{L}}(f)(p^{k})  &  \equiv\int\nolimits_{-\infty}^{+\infty
}d_{\bar{L}}^{n}x\,f(x^{i})\overset{x}{\circledast}\exp(x^{j}|\text{i}%
^{-1}p^{k})_{R,\bar{L}},\label{DefFourP1}\\[0.16in]
\mathcal{F}_{R}(f)(p^{k})  &  \equiv\int\nolimits_{-\infty}^{+\infty}d_{R}%
^{n}x\,\exp(\text{i}^{-1}p^{k}|x^{j})_{R,\bar{L}}\overset{x}{\circledast
}f(x^{i}),\nonumber\\
\mathcal{F}_{\bar{R}}(f)(p^{k})  &  \equiv\int\nolimits_{-\infty}^{+\infty
}d_{\bar{R}}^{n}x\,\exp(\text{i}^{-1}p^{k}|x^{j})_{\bar{R},L}\overset
{x}{\circledast}f(x^{i}), \label{DefFourP2}%
\end{align}
where we applied the substitutions $\partial^{k}=\,$i$^{-1}p^{k}$ and
$\hat{\partial}^{k}=\,$i$^{-1}p^{k}$ to our q-exponentials.

In complete analogy to the classical case we introduce q-deformed\ versions of
the\textit{ delta function} by%
\begin{align}
\delta_{L}^{n}(p^{k})  &  \equiv\mathcal{F}_{L}(1)(p^{k})=\int
\nolimits_{-\infty}^{+\infty}d_{L}^{n}x\exp(x^{j}|\text{i}^{-1}p^{k})_{\bar
{R},L},\nonumber\\
\delta_{\bar{L}}^{n}(p^{k})  &  \equiv\mathcal{F}_{\bar{L}}(1)(p^{k}%
)=\int\nolimits_{-\infty}^{+\infty}d_{\bar{L}}^{n}x\exp(x^{j}|\text{i}%
^{-1}p^{k})_{R,\bar{L}},\label{DefDelt1}\\[0.16in]
\delta_{R}^{n}(p^{k})  &  \equiv\mathcal{F}_{R}(1)(p^{k})=\int
\nolimits_{-\infty}^{+\infty}d_{R}^{n}x\,\exp(\text{i}^{-1}p^{k}%
|x^{j})_{R,\bar{L}},\nonumber\\
\delta_{\bar{R}}^{n}(p^{k})  &  \equiv\mathcal{F}_{\bar{R}}(1)(p^{k}%
)=\int\nolimits_{-\infty}^{+\infty}d_{\bar{R}}^{n}x\,\exp(\text{i}^{-1}%
p^{k}|x^{j})_{\bar{R},L}. \label{DefDelt2}%
\end{align}
As a consequence the q-deformed volume elements are then given by the
expressions%
\begin{align}
\text{vol}_{L}  &  \equiv\int\nolimits_{-\infty}^{+\infty}d_{\bar{R}}%
^{n}p\,\delta_{L}^{n}(p^{k})=\int\nolimits_{-\infty}^{+\infty}d_{L}^{n}%
x\int\nolimits_{-\infty}^{+\infty}d_{\bar{R}}^{n}p\exp(x^{j}|\text{i}%
^{-1}p^{k})_{\bar{R},L},\nonumber\\
\text{vol}_{\bar{L}}  &  \equiv\int\nolimits_{-\infty}^{+\infty}d_{R}%
^{n}p\,\delta_{\bar{L}}^{n}(p^{k})=\int\nolimits_{-\infty}^{+\infty}d_{\bar
{L}}^{n}x\int\nolimits_{-\infty}^{+\infty}d_{R}^{n}p\exp(x^{j}|\text{i}%
^{-1}p^{k})_{R,\bar{L}},\label{DefVol1}\\[0.16in]
\text{vol}_{R}  &  \equiv\int\nolimits_{-\infty}^{+\infty}d_{\bar{L}}%
^{n}p\,\delta_{R}^{n}(p^{k})=\int\nolimits_{-\infty}^{+\infty}d_{\bar{L}}%
^{n}p\int\nolimits_{-\infty}^{+\infty}d_{R}^{n}x\exp(\text{i}^{-1}p^{k}%
|x^{j})_{R,\bar{L}},\nonumber\\
\text{vol}_{\bar{R}}  &  \equiv\int\nolimits_{-\infty}^{+\infty}d_{L}%
^{n}p\,\delta_{\bar{R}}^{n}(p^{k})=\int\nolimits_{-\infty}^{+\infty}d_{L}%
^{n}p\int\nolimits_{-\infty}^{+\infty}d_{\bar{R}}^{n}x\exp(\text{i}^{-1}%
p^{k}|x^{j})_{\bar{R},L}. \label{DefVol2}%
\end{align}

As in the classical case \textit{q-deformed Fourier transformations} can be
viewed as mappings between position and momentum space. In our approach there
is a complete symmetry between position and momentum variables - apart from
occasional minus signs (see the discussion in Ref. \cite{qAn}). Thus, it is
always possible to interchange the roles of $x$ and $p$ in the above
definitions. Especially, this observation implies the identities%
\begin{equation}
\text{vol}_{L}=\text{vol}_{\bar{R}}\text{\quad and\quad vol}_{\bar{L}%
}=\text{vol}_{R}. \label{ConVol}%
\end{equation}

At this point, we should also make some comments on the convergence of the
integral expressions in (\ref{DefDelt1})-(\ref{DefVol2}). In general, we
cannot expect that a q-deformed integral of a q-exponential takes on a finite
value if the integral is taken over the whole space. The reason for this lies
in the fact that q-exponentials do not decrease rapidly enough at infinity. To
circumvent this problem we can try to modify q-exponentials in such a way that
they become functions with the necessary boundary conditions. In complete
analogy to the classical case this can be achieved by the requirement that the
values of momentum variables contain small imaginary parts. From a physical
point of view this means we deal with wave-packets instead of plane-waves and
the imaginary parts in the values of the momentum variables describe the
extension of the wave-packets in space and time.

With this modification our expressions in (\ref{DefDelt1})-(\ref{DefVol2})
should be well-defined. However, there is a price we have to pay, since our
results now depend on additional parameters. Thus, at the end of our
calculations we have to take the limit in which these parameters vanish. The
discrete structure of our formulae gives rise to the hope that such a method
can lead to finite expressions, but we do not want to investigate this problem here.

\subsection{Elementary properties of Fourier transformations}

From classical Fourier theory we know that Fourier transformations intertwine
the action of partial derivatives with multiplication by coordinates. For
q-deformed Fourier transformations a similar statement holds, since we have%
\begin{align}
\mathcal{F}_{L}(f\overset{x}{\triangleleft}\partial^{j})(p^{k})  &
=\mathcal{F}_{L}(f)(p^{k})\overset{p}{\circledast}(\text{i}^{-1}%
p^{j}),\nonumber\\
\mathcal{F}_{\bar{L}}(f\,\overset{x}{\bar{\triangleleft}}\,\hat{\partial}%
^{j})(p^{k})  &  =\mathcal{F}_{\bar{L}}(f)(p^{k})\overset{p}{\circledast
}(\text{i}^{-1}p^{j}),\label{FunProp1}\\[0.16in]
\mathcal{F}_{R}(\hat{\partial}^{j}\overset{x}{\triangleright}f)(p^{k})  &
=\text{i}^{-1}p^{j}\overset{p}{\circledast}\mathcal{F}_{R}(f)(p^{k}%
),\nonumber\\
\mathcal{F}_{\bar{R}}(\partial^{j}\,\overset{x}{\bar{\triangleright}%
}\,f)(p^{k})  &  =\text{i}^{-1}p^{j}\overset{p}{\circledast}\mathcal{F}%
_{\bar{R}}(f)(p^{k}), \label{FunProp1b}%
\end{align}
and%
\begin{align}
\mathcal{F}_{L}(f\overset{x}{\circledast}x^{j})(p^{k})  &  =\text{i}%
\mathcal{F}_{L}(f)(p^{k})\,\overset{p}{\bar{\triangleleft}}\,\partial
^{j},\nonumber\\
\mathcal{F}_{\bar{L}}(f\overset{x}{\circledast}x^{j})(p^{k})  &
=\text{i}\mathcal{F}_{\bar{L}}(f)(p^{k})\overset{p}{\triangleleft}%
\hat{\partial}^{j},\label{FunProp2a}\\[0.16in]
\mathcal{F}_{R}(x^{j}\overset{x}{\circledast}f)(p^{k})  &  =\text{i}%
\hat{\partial}^{j}\,\overset{p}{\bar{\triangleright}}\,\mathcal{F}%
_{R}(f)(p^{k}),\nonumber\\
\mathcal{F}_{\bar{R}}(x^{j}\overset{x}{\circledast}f)(p^{k})  &
=\text{i}\partial^{j}\overset{p}{\triangleright}\mathcal{F}_{\bar{R}}%
(f)(p^{k}). \label{FunProp2}%
\end{align}
Notice that the variable on top of the symbol for the action indicates upon
which space the partial derivatives shall act.

The above identities follow from the property of q-exponentials to be
eigenfunctions of q-deformed partial derivatives [cf. the identities in
(\ref{EigGl1N}) and (\ref{EigGl2})], as can be seen by the following
calculation:%
\begin{align}
\mathcal{F}_{L}(f\overset{x}{\triangleleft}\partial^{j})(p^{k})=  &
\,\int\nolimits_{-\infty}^{+\infty}d_{L}^{n}x\,(f\overset{x}{\triangleleft
}\partial^{j})\overset{x}{\circledast}\exp(x^{l}|\text{i}^{-1}p^{k})_{\bar
{R},L}\nonumber\\
=  &  \,\int\nolimits_{-\infty}^{+\infty}d_{L}^{n}x\,f\overset{x}{\circledast
}\big (\partial^{j}\overset{x}{\triangleright}\exp(x^{l}|\text{i}^{-1}%
p^{k})_{\bar{R},L}\big )\nonumber\\
=  &  \,\int\nolimits_{-\infty}^{+\infty}d_{L}^{n}x\,f\overset{x}{\circledast
}\exp(x^{l}|\text{i}^{-1}p^{k})_{\bar{R},L}\overset{p}{\circledast}%
(\text{i}^{-1}p^{j})\nonumber\\
=  &  \,\mathcal{F}_{L}(f)(p^{k})\overset{p}{\circledast}(\text{i}^{-1}p^{j}),
\label{HerFourId}%
\end{align}
and%
\begin{align}
\mathcal{F}_{L}(f\overset{x}{\circledast}x^{j})(p^{k})=  &  \,\int
\nolimits_{-\infty}^{+\infty}d_{L}^{n}x\,(f\overset{x}{\circledast}%
x^{j})\overset{x}{\circledast}\exp(x^{l}|\text{i}^{-1}p^{k})_{\bar{R}%
,L}\nonumber\\
=  &  \,\int\nolimits_{-\infty}^{+\infty}d_{L}^{n}x\,f\overset{x}{\circledast
}\big (x^{j}\overset{x}{\circledast}\exp(x^{l}|\text{i}^{-1}p^{k})_{\bar{R}%
,L}\big )\nonumber\\
=  &  \,\text{i}\int\nolimits_{-\infty}^{+\infty}d_{L}^{n}x\,f\overset
{x}{\circledast}\exp(x^{l}|\text{i}^{-1}p^{k})_{\bar{R},L}\,\overset{p}%
{\bar{\triangleleft}}\,\partial^{j}\nonumber\\
=  &  \,\text{i}\mathcal{F}_{L}(f)(p^{k})\,\overset{p}{\bar{\triangleleft}%
}\,\partial^{j}.
\end{align}
Notice that the second equality in (\ref{HerFourId}) can be recognized as
integration by parts. The other identities in (\ref{FunProp1})-(\ref{FunProp2}%
) can be proven in very much the same way.

\subsection{Invertibility of Fourier transformations}

In this section we would like to show that q-deformed Fourier transformations
are invertible in a certain sense. Towards this end, it is convenient to
introduce another type of q-deformed Fourier transformations:%
\begin{align}
\mathcal{F}_{L}^{\ast}(f)(x^{k})  &  \equiv\frac{1}{\text{vol}_{L}}%
\int\nolimits_{-\infty}^{+\infty}d_{L}^{n}p\exp(\text{i}^{-1}p^{l}%
|\!\ominus_{L}\!x^{k})_{\bar{R},L}\overset{x|p}{\odot}_{\hspace{-0.01in}%
\bar{L}}f(p^{j}),\nonumber\\
\mathcal{F}_{\bar{L}}^{\ast}(f)(x^{k})  &  \equiv\frac{1}{\text{vol}_{\bar{L}%
}}\int\nolimits_{-\infty}^{+\infty}d_{\bar{L}}^{n}p\exp(\text{i}^{-1}%
p^{l}|\!\ominus_{\bar{L}}\!x^{k})_{R,\bar{L}}\overset{x|p}{\odot}%
_{\hspace{-0.01in}L}f(p^{j}),\label{FTtype1}\\[0.16in]
\mathcal{F}_{R}^{\ast}(f)(x^{k})  &  \equiv\frac{1}{\text{vol}_{R}}%
\int\nolimits_{-\infty}^{+\infty}d_{R}^{n}p\,f(p^{j})\overset{p|x}{\odot
}_{\hspace{-0.01in}\bar{R}}\exp(\ominus_{R}\,x^{k}|\text{i}^{-1}p^{l}%
)_{R,\bar{L}},\nonumber\\
\mathcal{F}_{\bar{R}}^{\ast}(f)(x^{k})  &  \equiv\frac{1}{\text{vol}_{\bar{R}%
}}\int\nolimits_{-\infty}^{+\infty}d_{\bar{R}}^{n}p\,f(p^{j})\overset
{p|x}{\odot}_{\hspace{-0.01in}R}\exp(\ominus_{\bar{R}}\,x^{k}|\text{i}%
^{-1}p^{l})_{\bar{R},L}. \label{FTtype2}%
\end{align}
Essentially for us is the fact that this second set of Fourier transformations
enables us to invert the mappings in (\ref{DefFourP1})-(\ref{DefFourP2}).
Concretely, we have%
\begin{align}
(\mathcal{F}_{\bar{R}}^{\ast}\circ\mathcal{F}_{L})(f)(x^{k})  &  =f(\kappa
x^{k}),\nonumber\\
(\mathcal{F}_{R}^{\ast}\circ\mathcal{F}_{\bar{L}})(f)(x^{k})  &
=f(\kappa^{-1}x^{k}),\label{InvFourAnf2a}\\[0.1in]
(\mathcal{F}_{\bar{L}}^{\ast}\circ\mathcal{F}_{R})(f)(x^{k})  &
=f(\kappa^{-1}x^{k}),\nonumber\\
(\mathcal{F}_{L}^{\ast}\circ\mathcal{F}_{\bar{R}})(f)(x^{k})  &  =f(\kappa
x^{k}), \label{InvFourAnf2b}%
\end{align}
and%
\begin{align}
(\mathcal{F}_{L}\circ\mathcal{F}_{\bar{R}}^{\ast})(f)(x^{k})  &  =\kappa
^{-n}f(\kappa^{-1}x^{k}),\nonumber\\
(\mathcal{F}_{\bar{L}}\circ\mathcal{F}_{R}^{\ast})(f)(x^{k})  &  =\kappa
^{n}f(\kappa x^{k}),\label{InvFourAnf1a}\\[0.1in]
(\mathcal{F}_{R}\circ\mathcal{F}_{\bar{L}}^{\ast})(f)(x^{k})  &  =\kappa
^{n}f(\kappa x^{k}),\nonumber\\
(\mathcal{F}_{\bar{R}}\circ\mathcal{F}_{L}^{\ast})(f)(x^{k})  &  =\kappa
^{-n}f(\kappa^{-1}x^{k}). \label{InvFourAnf1b}%
\end{align}

Next, we would like to prove the relations in (\ref{InvFourAnf2a}%
)-(\ref{InvFourAnf1b}). To reach this goal, we need some useful identities
which we now collect. First of all, we have%
\begin{align}
&  \hspace{-0.16in}\int\nolimits_{-\infty}^{+\infty}d_{A}^{n}y\,f(y^{i}%
)\overset{y|x}{\odot}_{\hspace{-0.01in}B}g(x^{j}\oplus_{\bar{L}}%
y^{k})\nonumber\\
&  \qquad=\int\nolimits_{-\infty}^{+\infty}d_{A}^{n}y\,f(((\ominus_{B}%
\,x^{m})\oplus_{B}x^{l})\oplus_{B}y^{i})\overset{y|x}{\odot}_{\hspace
{-0.01in}B}g(x^{j}\oplus_{B}y^{k})\nonumber\\
&  \qquad=\int\nolimits_{-\infty}^{+\infty}d_{A}^{n}y\,f((\ominus_{B}%
\,x^{m})\oplus_{B}(x^{l}\oplus_{B}y^{i}))\overset{y|x}{\odot}_{\hspace
{-0.01in}B}g(x^{j}\oplus_{B}y^{k})\nonumber\\
&  \qquad=\int\nolimits_{-\infty}^{+\infty}d_{A}^{n}y\,f((\ominus_{B}%
\,x^{j})\oplus_{B}y^{i}))\overset{y}{\circledast}g(y^{k}),
\label{IntKoorTrans}%
\end{align}
where $A,B\in\{L,\bar{L},R,\bar{R}\}.$ The first and second equality are
applications of the axioms of q-translations [cf. the identities in
(\ref{HopfAxiom})]. The third equality uses the fact that
q-deformed\ integrals over the whole space are invariant under q-translations
[cf. the identities in (\ref{TransGlob})]. For the sake of completeness, it
should be noted that by a slight modification of these arguments we can also
verify\ that
\begin{equation}
\int\nolimits_{-\infty}^{+\infty}d_{A}^{n}y\,g(y^{i}\oplus_{B}x^{k}%
)\overset{x|y}{\odot}_{\hspace{-0.01in}B}f(y^{j})=\int\nolimits_{-\infty
}^{+\infty}d_{A}^{n}y\,g(y^{i})\overset{y}{\circledast}f(y^{j}\oplus
_{B}(\ominus_{B}\,x^{k})). \label{IntKoordTrans3}%
\end{equation}

However, the proof of the identities in (\ref{InvFourAnf2a}%
)-(\ref{InvFourAnf1b}) also requires the relations%
\begin{align}
\int_{-\infty}^{+\infty}d_{L}^{n}x\,g(x^{i})\overset{x}{\circledast}%
\delta_{\bar{R}}^{n}(x^{j})  &  =g(0)\,\text{vol}_{L},\nonumber\\
\int\nolimits_{-\infty}^{+\infty}d_{\bar{L}}^{n}x\,g(x^{i})\overset
{x}{\circledast}\delta_{R}^{n}(x^{j})  &  =g(0)\,\text{vol}_{\bar{L}}.
\label{CharPropDelN}%
\end{align}
Both relations follow from the same reasonings. Thus, it suffices to restrict
attention to the first identity in (\ref{CharPropDelN}), which can be derived
as follows:%
\begin{align}
&  \hspace{-0.16in}\int\nolimits_{-\infty}^{+\infty}d_{L}^{n}x\,g(x^{i}%
)\overset{x}{\circledast}\delta_{\bar{R}}^{n}(x^{j})\nonumber\\
&  \qquad=\,\int\nolimits_{-\infty}^{+\infty}d_{L}^{n}x\int\nolimits_{-\infty
}^{+\infty}d_{\bar{R}}^{n}p\,g(x^{i})\overset{x}{\circledast}\exp
(x^{j}|\text{i}^{-1}p^{k})_{\bar{R},L}\nonumber\\
&  \qquad=\,\int\nolimits_{-\infty}^{+\infty}d_{L}^{n}x\,g(x^{i})\overset
{x}{\circledast}e_{(x,\bar{R})}^{a}\otimes\int\nolimits_{-\infty}^{+\infty
}d_{\bar{R}}^{n}p\,e_{(p,L)}^{a}\nonumber\\
&  \qquad=\,\int\nolimits_{-\infty}^{+\infty}d_{L}^{n}%
x\,\big \langle e_{(\tilde{p},L)}^{b},g(x^{i})\big \rangle_{L,\bar{R}%
}\,e_{(x,\bar{R})}^{b}\overset{x}{\circledast}e_{(x,\bar{R})}^{a}\nonumber\\
&  \qquad\qquad\otimes\int\nolimits_{-\infty}^{+\infty}d_{\bar{R}}%
^{n}p\,e_{(p,L)}^{a}\nonumber\\
&  \qquad=\,\int\nolimits_{-\infty}^{+\infty}d_{L}^{n}%
x\,\big \langle(e_{(p,L)}^{a})_{(R,1)},g(x^{i})\big \rangle_{L,\bar{R}%
}\,e_{(x,\bar{R})}^{a}\nonumber\\
&  \qquad\qquad\otimes\int\nolimits_{-\infty}^{+\infty}d_{\bar{R}}%
^{n}p\,(e_{(p,L)}^{a})_{(R,2)}\nonumber\\
&  \qquad=\,\int\nolimits_{-\infty}^{+\infty}d_{L}^{n}x\,\big \langle1,g(x^{i}%
)\big \rangle_{L,\bar{R}}\,e_{(x,\bar{R})}^{a}\otimes\int\nolimits_{-\infty
}^{+\infty}d_{\bar{R}}^{n}p\,e_{(p,L)}^{a}\nonumber\\
&  \qquad=\,g(0)\int\nolimits_{-\infty}^{+\infty}d_{L}^{n}x\int
\nolimits_{-\infty}^{+\infty}d_{\bar{R}}^{n}p\,\exp(x^{j}|\text{i}^{-1}%
p^{k})_{\bar{R},L}=g(0)\,\text{vol}_{L}. \label{CharPropDel0}%
\end{align}
The first equality is the definition of the delta function in position space.
For the second equality the q-deformed exponential was rewritten as%
\begin{equation}
\exp(x^{j}|\text{i}^{-1}p^{k})_{\bar{R},L}=\sum_{a}e_{(x,\bar{R})}^{a}\otimes
e_{(p,L)}^{a}. \label{ExpTens}%
\end{equation}
Notice that for notational convenience the summation symbols were skipped in
the expressions of (\ref{CharPropDel0}). For the third equality we applied the
completeness relation
\begin{equation}
\text{id}=\big (\left\langle .\,,.\right\rangle _{L,\bar{R}}\otimes
\text{id}\big )\circ\big (\text{id}\otimes\exp(x^{j}|\text{i}^{-1}p^{k}%
)_{\bar{R},L}\big ). \label{CompRel}%
\end{equation}
The fourth equality follows from\ the addition law for q-exponentials written
in the form%
\begin{gather}
\sum_{a,b}e_{(x,\bar{R})}^{a}\overset{x}{\circledast}e_{(x,\bar{R})}%
^{b}\otimes e_{(p,L)}^{b}\otimes e_{(p,L)}^{a}=\nonumber\\
=\sum_{a}e_{(x,\bar{R})}^{a}\otimes(e_{(p,L)}^{a})_{(R,1)}\otimes
(e_{(p,L)}^{a})_{(R,2)},
\end{gather}
For the fifth equality we used translation invariance of the integral over the
whole momentum space and the sixth equality is a consequence of the
identities
\begin{equation}
g(0)\equiv\left.  g(x^{i})\right\vert _{x^{i}=\,0}=\epsilon(\mathcal{W}%
(g))=\big \langle1,g(x^{i})\big \rangle_{L,\bar{R}}.
\end{equation}

The relations in (\ref{CharPropDelN}) are formulated with left integrals and
the delta functions are multiplied from the right. However, we can write down
a variant of each relation in (\ref{CharPropDelN}) using right integrals and
multiplying delta functions from the left:%
\begin{align}
\int\nolimits_{-\infty}^{+\infty}d_{R}^{n}x\,\delta_{\bar{L}}^{n}%
(x^{j})\overset{x}{\circledast}g(x^{i})  &  =g(0)\,\text{vol}_{R},\nonumber\\
\int\nolimits_{-\infty}^{+\infty}d_{\bar{R}}^{n}x\,\delta_{L}^{n}%
(x^{j})\overset{x}{\circledast}g(x^{i})  &  =g(0)\,\text{vol}_{\bar{R}}.
\label{CharPropDel2N}%
\end{align}
These identities can be verified in a similar fashion as those in
(\ref{CharPropDelN}). We wish to illustrate this by the following calculation:%
\begin{align}
&  \hspace{-0.16in}\int\nolimits_{-\infty}^{+\infty}d_{R}^{n}x\,\delta
_{\bar{L}}^{n}(x^{j})\overset{x}{\circledast}g(x^{i})\nonumber\\
&  \qquad=\,\int\nolimits_{-\infty}^{+\infty}d_{R}^{n}x\int\nolimits_{-\infty
}^{+\infty}d_{\bar{L}}^{n}p\,\exp(\text{i}^{-1}p^{k}|x^{j})_{R,\bar{L}%
}\overset{x}{\circledast}g(x^{i})\nonumber\\
&  \qquad=\,\int\nolimits_{-\infty}^{+\infty}d_{\bar{L}}^{n}p\,e_{(p,R)}%
^{a}\otimes\int\nolimits_{-\infty}^{+\infty}d_{R}^{n}x\,e_{(x,\bar{L})}%
^{a}\overset{x}{\circledast}g(x^{i})\nonumber\\
&  \qquad=\,\int\nolimits_{-\infty}^{+\infty}d_{\bar{L}}^{n}p\,e_{(p,R)}%
^{a}\otimes\int\nolimits_{-\infty}^{+\infty}d_{R}^{n}x\,e_{(x,\bar{L})}%
^{a}\overset{x}{\circledast}e_{(x,\bar{L})}^{b}\big \langle g(x^{i}%
),e_{(\tilde{p},R)}^{b}\big \rangle_{\bar{L},R}\nonumber\\
&  \qquad=\,\int\nolimits_{-\infty}^{+\infty}d_{\bar{L}}^{n}p\,(e_{(p,R)}%
^{a})_{(L,1)}\otimes\int\nolimits_{-\infty}^{+\infty}d_{R}^{n}x\,e_{(x,\bar
{L})}^{a}\big \langle g(x^{i}),(e_{(p,R)}^{a})_{(L,2)}\big \rangle_{\bar{L}%
,R}\nonumber\\
&  \qquad=\,\int\nolimits_{-\infty}^{+\infty}d_{\bar{L}}^{n}p\,e_{(p,R)}%
^{a}\otimes\int\nolimits_{-\infty}^{+\infty}d_{R}^{n}x\,\big \langle g(x^{i}%
),1\big \rangle_{\bar{L},R}\,e_{(x,\bar{L})}^{a}\nonumber\\
&  \qquad=\,g(0)\int\nolimits_{-\infty}^{+\infty}d_{\bar{L}}^{n}%
p\int\nolimits_{-\infty}^{+\infty}d_{R}^{n}x\,\exp(\text{i}^{-1}p^{k}%
|x^{j})_{R,\bar{L}}=g(0)\,\text{vol}_{R}.
\end{align}

Remarkably, the arguments leading to the identities in (\ref{CharPropDelN})
and (\ref{CharPropDel2N}) do not really depend on the realizations of
q-deformed delta function and q-integral over the whole space. For this
reason, one can even show that we have%
\begin{align}
\int\nolimits_{-\infty}^{+\infty}d_{A}^{n}x\,g(x^{i})\overset{x}{\circledast
}\delta_{B}^{n}(x^{j})  &  =g(0)\,\text{vol}_{A,B},\nonumber\\
\int\nolimits_{-\infty}^{+\infty}d_{A}^{n}x\,\delta_{B}^{n}(x^{j})\overset
{x}{\circledast}g(x^{i})  &  =g(0)\,\text{vol}_{A,B}, \label{VeralDelt0}%
\end{align}
if we introduce as q-deformed volume elements%
\begin{equation}
\text{vol}_{A,B}\equiv\int\nolimits_{-\infty}^{+\infty}d_{A}^{n}x\,\delta
_{B}^{n}(x^{j}),\quad A,B\in\{L,\bar{L},R,\bar{R}\}. \label{VolAlg}%
\end{equation}

From what we have done so far we see that the integral of the product between
a given function and a q-deformed delta function is proportional to the value
of the given function at zero. This observation is in complete analogy to the
classical case and can be extended in the following way:%
\begin{align}
\int\nolimits_{-\infty}^{+\infty}d_{L}^{n}y\,f(y^{i})\overset{y|x}{\odot
}_{\hspace{-0.01in}\bar{L}}\delta_{\bar{R}}^{n}((\ominus_{\bar{R}}%
\,x^{j})\oplus_{\bar{L}}y^{k})  &  =\text{vol}_{L}\,f(x^{j}),\nonumber\\
\int\nolimits_{-\infty}^{+\infty}d_{\bar{L}}^{n}y\,f(y^{i})\overset{y|x}%
{\odot}_{\hspace{-0.01in}L}\delta_{R}^{n}((\ominus_{R}\,x^{j})\oplus_{L}%
y^{k})  &  =\text{vol}_{\bar{L}}\,f(x^{j}),\label{VeralDelt1}\\[0.16in]
\int\nolimits_{-\infty}^{+\infty}d_{R}^{n}y\,\delta_{\bar{L}}^{n}(y^{k}%
\oplus_{\bar{R}}(\ominus_{\bar{L}}\,x^{j}))\overset{x|y}{\odot\hspace
{-0.01in}}_{\bar{R}}f(y^{i})  &  =\text{vol}_{R}\,f(x^{j}),\nonumber\\
\int\nolimits_{-\infty}^{+\infty}d_{\bar{R}}^{n}y\,\delta_{L}^{n}(y^{k}%
\oplus_{R}(\ominus_{L}\,x^{j}))\overset{x|y}{\odot}_{\hspace{-0.01in}R}%
f(y^{i})  &  =\text{vol}_{\bar{R}}\,f(x^{j}). \label{VeralDelt2}%
\end{align}
The above identities are verified by rather simple calculations. Using
(\ref{IntKoorTrans}) together with the axioms for q-translations we have, for
example,%
\begin{align}
&  \hspace{-0.16in}\int\nolimits_{-\infty}^{+\infty}d_{L}^{n}y\,f(y^{i}%
)\overset{y|x}{\odot\hspace{-0.01in}}_{\bar{L}}\delta_{\bar{R}}^{n}%
((\ominus_{\bar{R}}\,x^{j})\oplus_{\bar{L}}y^{k})\nonumber\\
&  =\,\int\nolimits_{-\infty}^{+\infty}d_{L}^{n}y\,f((\ominus_{\bar{L}%
}(\ominus_{\bar{R}}\,x^{j}))\oplus_{\bar{L}}y^{i})\overset{y}{\circledast
}\delta_{\bar{R}}^{n}(y^{k})\nonumber\\
&  =\,\text{vol}_{L}\,f((\ominus_{\bar{L}}(\ominus_{\bar{R}}\,x^{j}%
))\oplus_{\bar{L}}0)=\text{vol}_{L}\,f(x^{j}).
\end{align}
Finally, let us note that there exist generalizations of the relations in
(\ref{VeralDelt1}) and (\ref{VeralDelt2}), which take the form%
\begin{align}
\int\nolimits_{-\infty}^{+\infty}d_{A}^{n}y\,f(y^{i})\overset{y|x}{\odot
}_{\hspace{-0.01in}B}\delta_{C}^{n}(x^{j}\oplus_{B}y^{k})  &  =\text{vol}%
_{A,C}\,f(\ominus_{B}\,x^{j}),\nonumber\\
\int\nolimits_{-\infty}^{+\infty}d_{A}^{n}y\,\delta_{C}^{n}(y^{k}\oplus
_{B}x^{j})\overset{x|y}{\odot}_{\hspace{-0.01in}B}f(y^{i})  &  =\text{vol}%
_{A,C}\,f(\ominus_{B}\,x^{j}).
\end{align}

For the sake of completeness we would like to give the identities%
\begin{align}
&  \int\nolimits_{-\infty}^{+\infty}d_{A}^{n}y\,f(y^{i})\overset
{y}{\circledast}\delta_{B}^{n}(y^{j}\oplus_{C}(\ominus_{C}\,x^{k}))\nonumber\\
&  \qquad=\,\int\nolimits_{-\infty}^{+\infty}d_{A}^{n}y\,f(y^{i}\oplus
_{C}x^{k})\overset{x|y}{\odot}_{\hspace{-0.01in}C}\delta_{B}^{n}%
(y^{j})\nonumber\\
&  \qquad=\,\text{vol}_{A,B}\,f(y^{i}\oplus_{C}((\kappa_{C})^{-1}%
x^{k}))|_{y^{i}=0}=\text{vol}_{A,B}\,f((\kappa_{C})^{-1}x^{k}),
\label{DeltProAlg0}%
\end{align}
and%
\begin{align}
&  \int\nolimits_{-\infty}^{+\infty}d_{A}^{n}y\,\delta_{B}^{n}((\ominus
_{C}\,x^{k})\oplus_{C}y^{j})\overset{y}{\circledast}f(y^{i})\nonumber\\
&  \qquad=\,\int\nolimits_{-\infty}^{+\infty}d_{A}^{n}y\,\delta_{B}^{n}%
(y^{j})\overset{y|x}{\odot}_{\hspace{-0.01in}C}f(x^{k}\oplus_{C}%
y^{i})\nonumber\\
&  \qquad=\,\text{vol}_{A,B}\,f(((\kappa_{C})^{-1}x^{k})\oplus_{C}%
y^{i})|_{y^{i}=0}=\text{vol}_{A,B}\,f((\kappa_{C})^{-1}x^{k}).
\label{DeltProAlg}%
\end{align}
Notice that for the second equality in (\ref{DeltProAlg0}) and
(\ref{DeltProAlg}) we exploited the identities (\ref{IntKoorTrans}) and
(\ref{IntKoordTrans3}), respectively. The third equality in (\ref{DeltProAlg0}%
) as well as in (\ref{DeltProAlg}) results from (\ref{VeralDelt0}) and the
braiding of q-deformed\ delta\ functions which is characterized by the
relations%
\begin{align}
f(x^{i})\overset{x|y}{\odot}_{\hspace{-0.01in}A}\delta_{B}(y^{j})  &
=\delta_{B}(y^{j})\otimes f((\kappa_{A})^{-1}x^{i}),\nonumber\\
\delta_{B}(y^{j})\overset{y|x}{\odot}_{\hspace{-0.01in}A}f(x^{i})  &
=f((\kappa_{A})^{-1}x^{i})\otimes\delta_{B}(y^{j}). \label{BraPropDelt}%
\end{align}
A short glance at the definitions in (\ref{DefDelt1}) and (\ref{DefDelt2})
tells us that q-deformed delta functions are made up out of q-integrals over
the entire space and q-deformed exponentials. The relations in
(\ref{BraPropDelt}) follow rather easily from the braiding properties of these objects.

Finally, we would like to mention that the functions $\delta_{A}^{n}%
(\ominus_{B}\,x^{i})$ give equally good delta functions, as they satisfy
\begin{align}
\int\nolimits_{-\infty}^{+\infty}d_{A}^{n}x\,g(x^{i})\overset{x}{\circledast
}\delta_{B}^{n}(\ominus_{C}\,x^{j})  &  =g(0)\,\text{vol}_{A,B}\times%
\begin{cases}
(\kappa_{C})^{n} & \text{if }B\in\{R,\bar{R}\},\\
(\kappa_{C})^{-n} & \text{if }B\in\{L,\bar{L}\},
\end{cases}
\nonumber\\[0.1in]
\int\nolimits_{-\infty}^{+\infty}d_{A}^{n}x\,\delta_{B}^{n}(\ominus_{C}%
\,x^{j})\overset{x}{\circledast}g(x^{i})  &  =g(0)\,\text{vol}_{A,B}\times%
\begin{cases}
(\kappa_{C})^{n} & \text{if }B\in\{L,\bar{L}\},\\
(\kappa_{C})^{-n} & \text{if }B\in\{R,\bar{R}\}.
\end{cases}
\label{DeltAnt}%
\end{align}
The above relations are needed to check the identities in (\ref{InvFourAnf1a})
and (\ref{InvFourAnf1b}). For a proof of the relations in (\ref{DeltAnt}) we
refer the reader to Appendix \ref{Proofs}.

Now, we have everything together to prove the identities in
(\ref{InvFourAnf2a})-(\ref{InvFourAnf1b}). In this section we give a proof of
the relations in (\ref{InvFourAnf2a}) and (\ref{InvFourAnf2b}), only.\ To
check the relations in (\ref{InvFourAnf1a}) and (\ref{InvFourAnf1b}) is a
little bit harder. Thus, the corresponding proof is presented in Appendix
\ref{Proofs}. Again, it suffices to restrict attention to one type of Fourier
transformations, since our considerations carry over to the other Fourier
transformations without any difficulties. Concretely, we have%
\begin{align}
&  \mathcal{F}_{\bar{R}}^{\ast}(\mathcal{F}_{L}(g)(p^{i}))(x^{j})\nonumber\\
&  \qquad=\,\frac{1}{\text{vol}_{\bar{R}}}\int\nolimits_{-\infty}^{+\infty
}d_{\bar{R}}^{n}p\,\mathcal{F}_{L}(g)(p^{i})\overset{p|x}{\odot}%
_{\hspace{-0.01in}R}\exp(\ominus_{\bar{R}}\,x^{j}|\text{i}^{-1}p^{k})_{\bar
{R},L}\\
&  \qquad=\,\frac{1}{\text{vol}_{\bar{R}}}\int\nolimits_{-\infty}^{+\infty
}d_{\bar{R}}^{n}p\int\nolimits_{-\infty}^{+\infty}d_{L}^{n}y\,g(y^{m}%
)\overset{y}{\circledast}\exp(y^{l}|\text{i}^{-1}p^{i})_{\bar{R},L}\nonumber\\
&  \qquad\qquad\qquad\qquad\overset{p|x}{\odot}_{\hspace{-0.01in}R}%
\exp(\ominus_{\bar{R}}\,x^{j}|\text{i}^{-1}p^{k})_{\bar{R},L}\nonumber\\
&  \qquad=\,\frac{1}{\text{vol}_{\bar{R}}}\int\nolimits_{-\infty}^{+\infty
}d_{\bar{R}}^{n}p\int\nolimits_{-\infty}^{+\infty}d_{L}^{n}y\,g(y^{m}%
)\overset{y|x}{\odot}_{\hspace{-0.01in}\bar{L}}\exp((\ominus_{\bar{R}}\,\kappa
x^{j})\oplus_{\bar{L}}y^{l}|\text{i}^{-1}p^{i})_{\bar{R},L}\nonumber\\
&  \qquad=\,\frac{1}{\text{vol}_{\bar{R}}}\int\nolimits_{-\infty}^{+\infty
}d_{\bar{R}}^{n}p\int\nolimits_{-\infty}^{+\infty}d_{L}^{n}y\,g((\ominus
_{\bar{L}}(\ominus_{\bar{R}}\,\kappa x^{j}))\oplus_{\bar{L}}y^{m})\nonumber\\
&  \qquad\qquad\qquad\qquad\overset{\tilde{x}}{\circledast}\exp(y^{l}%
|\text{i}^{-1}p^{i})_{\bar{R},L}\nonumber\\
&  \qquad=\,\frac{1}{\text{vol}_{\bar{R}}}\int\nolimits_{-\infty}^{+\infty
}d_{\bar{R}}^{n}p\int\nolimits_{-\infty}^{+\infty}d_{L}^{n}y\,g((\kappa
x^{j})\oplus_{\bar{L}}y^{m})\overset{y}{\circledast}\exp(y^{l}|\text{i}%
^{-1}p^{i})_{\bar{R},L}\nonumber\\
&  \qquad=\,\frac{1}{\text{vol}_{\bar{R}}}\int\nolimits_{-\infty}^{+\infty
}d_{L}^{n}y\,g((\kappa x^{j})\oplus_{\bar{L}}y^{m})\overset{y}{\circledast
}\delta_{\bar{R}}^{n}(y^{l})\nonumber\\
&  \qquad=\,\frac{1}{\text{vol}_{\bar{R}}}\,\text{vol}_{L}\,g((\kappa
x^{j})\oplus_{\bar{L}}0)=g((\kappa x^{j})). \label{RechInv1N}%
\end{align}
For the first and second equality we inserted the expressions for the
q-deformed Fourier\ transforms. For the third equality we applied the addition
law for q-deformed exponentials. The fourth step uses the relation in
(\ref{IntKoorTrans}) and for the sixth step we identified the expression for a
q-deformed delta function. The final result is a consequence of
(\ref{CharPropDel0}), (\ref{ConVol}), and the axioms of q-translations.

In complete analogy to the identities in (\ref{FunProp1})-(\ref{FunProp2}) the
Fourier transformations $\mathcal{F}_{A}^{\ast}$ fulfill%
\begin{align}
\mathcal{F}_{L}^{\ast}(\text{i}\partial^{j}\overset{p}{\triangleright}%
f(\kappa^{-1}p^{l}))(\kappa^{-1}x^{k})  &  =\kappa^{n}x^{j}\overset
{x}{\circledast}\mathcal{F}_{L}^{\ast}(f)(x^{k}),\nonumber\\
\mathcal{F}_{\bar{L}}^{\ast}(\text{i}\hat{\partial}^{j}\,\overset{p}%
{\bar{\triangleright}}\,f(\kappa p^{l}))(\kappa x^{k})  &  =\kappa^{-n}%
x^{j}\overset{x}{\circledast}\mathcal{F}_{\bar{L}}^{\ast}(f)(x^{k}%
),\label{FunPropInv1}\\[0.16in]
\mathcal{F}_{R}^{\ast}(\text{i}f(\kappa p^{l})\overset{p}{\triangleleft}%
\hat{\partial}^{j})(\kappa x^{k})  &  =\kappa^{-n}\mathcal{F}_{R}^{\ast
}(f)(x^{k})\overset{x}{\circledast}x^{j},\nonumber\\
\mathcal{F}_{\bar{R}}^{\ast}(\text{i}f(\kappa^{-1}p^{l})\,\overset{p}%
{\bar{\triangleleft}}\,\partial^{j})(\kappa^{-1}x^{k})  &  =\kappa
^{n}\mathcal{F}_{\bar{R}}^{\ast}(f)(x^{k})\overset{x}{\circledast}x^{j},
\label{FunPropInv2}%
\end{align}
and%
\begin{align}
\mathcal{F}_{L}^{\ast}(\text{i}^{-1}p^{j}\overset{p}{\circledast}f(\kappa
^{-1}p^{l}))(\kappa^{-1}x^{k})  &  =\kappa^{n}\partial^{j}\,\overset{x}%
{\bar{\triangleright}}\,\mathcal{F}_{L}^{\ast}(f)(x^{k}),\nonumber\\
\mathcal{F}_{\bar{L}}^{\ast}(\text{i}^{-1}p^{j}\overset{p}{\circledast
}f(\kappa p^{l}))(\kappa x^{k})  &  =\kappa^{-n}\hat{\partial}^{j}\overset
{x}{\triangleright}\mathcal{F}_{\bar{L}}^{\ast}(f)(x^{k}),
\label{InfTransFour}\\[0.16in]
\mathcal{F}_{R}^{\ast}(f(\kappa p^{l})\overset{p}{\circledast}(\text{i}%
^{-1}p^{j}))(\kappa x^{k})  &  =\kappa^{-n}\mathcal{F}_{R}^{\ast}%
(f)(x^{k})\,\overset{x}{\bar{\triangleleft}}\,\hat{\partial}^{j},\nonumber\\
\mathcal{F}_{\bar{R}}^{\ast}(f(\kappa^{-1}p^{l})\overset{p}{\circledast
}(\text{i}^{-1}p^{j}))(\kappa^{-1}x^{k})  &  =\kappa^{n}\mathcal{F}_{\bar{R}%
}^{\ast}(f)(x^{k})\overset{x}{\triangleleft}\partial^{j}.
\label{InfTransFour2}%
\end{align}
The above identities are a direct consequence of the relations in
(\ref{FunProp1})-(\ref{FunProp2}) if we take into account the relations in
(\ref{InvFourAnf2a})-(\ref{InvFourAnf1b}). This can be seen as follows:%
\begin{align}
&  \mathcal{F}_{L}(f\overset{x}{\triangleleft}\partial^{j})(p^{k}%
)=\mathcal{F}_{L}(f)(p^{k})\overset{p}{\circledast}(\text{i}^{-1}%
p^{j})\nonumber\\
\Rightarrow\  &  f\overset{x}{\triangleleft}\partial^{j}=\mathcal{F}_{\bar{R}%
}^{\ast}(\mathcal{F}_{L}(f)(p^{k})\overset{p}{\circledast}(\text{i}^{-1}%
p^{j}))(\kappa^{-1}x^{l})\nonumber\\
\Rightarrow\  &  \mathcal{F}_{\bar{R}}^{\ast}(f)(x^{l})\overset{x}%
{\triangleleft}\partial^{j}=\kappa^{-n}\mathcal{F}_{\bar{R}}^{\ast}%
(f(\kappa^{-1}p^{m})\overset{p}{\circledast}(\text{i}^{-1}p^{j}))(\kappa
^{-1}x^{l}),
\end{align}
and%
\begin{align}
&  \mathcal{F}_{L}(f\overset{x}{\circledast}x^{j})(p^{k})=\text{i}%
\mathcal{F}_{L}(f)(p^{k})\,\overset{p}{\bar{\triangleleft}}\,\partial
^{j}\nonumber\\
\Rightarrow\  &  f\overset{x}{\circledast}x^{j}=\mathcal{F}_{\bar{R}}^{\ast
}(\text{i}\mathcal{F}_{L}(f)(p^{k})\,\overset{p}{\bar{\triangleleft}%
}\,\partial^{j})(\kappa^{-1}x^{l})\nonumber\\
\Rightarrow\  &  \mathcal{F}_{\bar{R}}^{\ast}(f)(x^{l})\overset{x}%
{\circledast}x^{j}=\kappa^{n}\mathcal{F}_{\bar{R}}^{\ast}(\text{i}%
f(\kappa^{-1}p^{m})\,\overset{p}{\bar{\triangleleft}}\,\partial^{j}%
)(\kappa^{-1}x^{l}).
\end{align}
Similar calculations lead to the other relations in (\ref{FunPropInv1}%
)-(\ref{InfTransFour2}).

The relations in (\ref{FunProp2a}), (\ref{FunProp2}), (\ref{InfTransFour}),
and (\ref{InfTransFour2}) describe in some sense infinitesimal translations of
q-deformed Fourier transforms. We can also write down formulae for global
translations of q-deformed Fourier transforms. Especially, we have%
\begin{align}
&  \mathcal{F}_{L}(f)(p^{k}\oplus_{R}\tilde{p}^{\,l})\nonumber\\
&  \qquad=\int_{-\infty}^{+\infty}d_{L}^{n}x\,f(x^{i})\overset{x}{\circledast
}\exp(x^{j}|\text{i}^{-1}(p^{k}\oplus_{R}\tilde{p}^{\,l}))_{\bar{R}%
,L}\nonumber\\
&  \qquad=\int_{-\infty}^{+\infty}d_{L}^{n}x\,f(x^{i})\overset{x}{\circledast
}\exp(x^{j}|\text{i}^{-1}\tilde{p}^{\,l})_{\bar{R},L}\!\overset{\tilde{p}%
|xp}{\odot}_{\!\!R}\exp(x^{m}|\text{i}^{-1}p^{k})_{\bar{R},L}\nonumber\\
&  \qquad=\mathcal{F}_{L}(f(x^{i})\overset{x}{\circledast}\exp(x^{j}%
|\text{i}^{-1}\tilde{p}^{\,l})_{\bar{R},L})(p^{k}), \label{GlobTrans1}%
\end{align}
and%
\begin{align}
&  \mathcal{F}_{\bar{R}}^{\ast}(f)(x^{i}\oplus_{\bar{R}}y^{j})\nonumber\\
&  \qquad=\frac{1}{\text{vol}_{\bar{R}}}\int_{-\infty}^{+\infty}d_{\bar{R}%
}^{n}\,f(p^{l})\overset{\tilde{p}|xy}{\odot}_{\!\!R}\exp(\ominus_{\bar{R}%
}(x^{i}\oplus_{\bar{R}}y^{j})|\text{i}^{-1}p^{k})_{\bar{R},L}\nonumber\\
&  \qquad=\frac{1}{\text{vol}_{\bar{R}}}\int_{-\infty}^{+\infty}d_{\bar{R}%
}^{n}p\,f(p^{l})\overset{p|xy}{\odot}_{\!\!R}\exp((\ominus_{\bar{R}}%
\,x^{i})\oplus_{\bar{L}}(\ominus_{\bar{R}}\,y^{j})|\text{i}^{-1}p^{k}%
)_{\bar{R},L}\nonumber\\
&  \qquad=\frac{1}{\text{vol}_{\bar{R}}}\int_{-\infty}^{+\infty}d_{\bar{R}%
}^{n}p\,f(p^{l})\overset{p|y}{\odot}_{\hspace{-0.01in}R}\exp(\ominus_{\bar{R}%
}\,y^{j}|\text{i}^{-1}p^{k})_{\bar{R},L}\nonumber\\
&  \qquad\qquad\qquad\qquad\overset{yp|x}{\odot}_{\!\!R}\exp(\ominus_{\bar{R}%
}\,x^{i}|\text{i}^{-1}p^{m})_{\bar{R},L}\nonumber\\
&  \qquad=\mathcal{F}_{\bar{R}}^{\ast}(f(p^{l})\overset{p|y}{\odot}%
_{\hspace{-0.01in}R}\exp(\ominus_{\bar{R}}\,y^{j}|\text{i}^{-1}p^{k})_{\bar
{R},L}),
\end{align}
where again we made use of the addition law for q-exponentials and the axioms
for q-translations. In a similar fashion, we get for the right versions of
q-deformed Fourier transforms:%
\begin{align}
\mathcal{F}_{R}(f)(\tilde{p}^{\,l}\oplus_{L}p^{k})  &  =\int_{-\infty
}^{+\infty}d_{R}^{n}x\,\exp(\text{i}^{-1}p^{k}|x^{m})_{R,\bar{L}}\nonumber\\
&  \qquad\quad\overset{px|\tilde{p}}{\odot}_{\!\!L}\exp(\text{i}^{-1}\tilde
{p}^{\,l}|x^{j})_{R,\bar{L}}\overset{x}{\circledast}f(x^{i})\nonumber\\
&  =\mathcal{F}_{R}(\exp(\text{i}^{-1}\tilde{p}^{\,l}|x^{j})_{R,\bar{L}%
}\overset{x}{\circledast}f(x^{i}))(p^{k}),\\[0.16in]
\mathcal{F}_{\bar{L}}^{\ast}(f)(y^{j}\oplus_{\bar{L}}x^{i})  &  =\frac
{1}{\text{vol}_{\bar{L}}}\int_{-\infty}^{+\infty}d_{\bar{L}}^{n}%
p\,\exp(\text{i}^{-1}p^{m}|\!\ominus_{\bar{L}}x^{i})_{\bar{R},L}\nonumber\\
&  \qquad\quad\overset{x|py}{\odot}_{\!\!L}\exp(\text{i}^{-1}p^{k}%
|\!\ominus_{\bar{L}}y^{j})_{R,\bar{L}}\overset{y|p}{\odot}_{\hspace{-0.01in}%
L}f(p^{l})\nonumber\\
&  =\mathcal{F}_{\bar{L}}^{\ast}(\exp(\text{i}^{-1}p^{k}|\!\ominus_{\bar{L}%
}y^{j})_{R,\bar{L}}\overset{y|p}{\odot}_{\hspace{-0.01in}L}f(p^{l}))(x^{i}).
\label{GlobTrans2}%
\end{align}
The corresponding relations for the other versions of q-deformed Fourier
transforms are obtained most easily from the formulae in (\ref{GlobTrans1}%
)-(\ref{GlobTrans2}) by changing the labels of q-deformed objects according to%
\begin{equation}
L\rightarrow\bar{L},\quad\bar{L}\rightarrow L,\quad R\rightarrow\bar{R}%
,\quad\bar{R}\rightarrow R. \label{SubSym}%
\end{equation}

\subsection{Fourier transforms of q-exponentials and q-deformed delta
functions\label{FTexpDelt}}

It is rather instructive to calculate q-deformed Fourier transforms of
q-exponentials and q-deformed delta functions. We start our considerations
with q-exponentials. For their q-deformed Fourier transforms, we find%
\begin{align}
&  \mathcal{F}_{L}(\exp(\text{i}^{-1}p^{k}|\text{\negthinspace}\ominus
_{L}y^{j})_{\bar{R},L})(x^{i})\nonumber\\
&  \qquad=\,\int_{-\infty}^{+\infty}d_{L}^{n}p\,\exp(\text{i}^{-1}%
p^{k}|\text{\negthinspace}\ominus_{L}y^{j})_{\bar{R},L}\overset{y|p}{\odot
}_{\hspace{-0.01in}\bar{L}}\exp(\text{i}^{-1}p^{l}|x^{i})_{\bar{R}%
,L}\nonumber\\
&  \qquad=\,\delta_{L}^{n}((\ominus_{L}\,y^{j})\oplus_{L}x^{i}),
\label{FTExp1}\\[0.16in]
&  \mathcal{F}_{R}(\exp(\ominus_{R}y^{j}|\text{i}^{-1}p^{k})_{R,\bar{L}%
})(x^{i})\nonumber\\
&  \qquad=\,\int_{-\infty}^{+\infty}d_{R}^{n}p\,\exp(x^{i}|\text{i}^{-1}%
p^{l})_{R,\bar{L}}\overset{p|y}{\odot}_{\hspace{-0.01in}R}\exp(\ominus
_{R}y^{j}|\text{i}^{-1}p^{k})_{R,\bar{L}}\nonumber\\
&  \qquad=\,\delta_{R}^{n}(x^{i}\oplus_{R}(\ominus_{R}\,y^{j})).
\end{align}
Again, the above identities are a direct consequence of the addition law for
q-exponentials.\textbf{ }In the same way we get%
\begin{align}
&  \mathcal{F}_{\bar{R}}^{\ast}(\exp(\text{i\thinspace}y^{j}|p^{k})_{\bar
{R},L})(x^{l})\nonumber\\
&  \qquad=\,\frac{1}{\text{vol}_{\bar{R}}}\int_{-\infty}^{+\infty}d_{\bar{R}%
}^{n}p\,\exp(y^{j}|\text{i}^{-1}p^{k})_{\bar{R},L}\overset{p|x}{\odot
}_{\hspace{-0.01in}R}\exp(\ominus_{\bar{R}}x^{l}|\text{i}^{-1}p^{l})_{\bar
{R},L}\nonumber\\
&  \qquad=\,\frac{1}{\text{vol}_{\bar{R}}}\,\delta_{\bar{R}}^{n}(y^{j}%
\oplus_{\bar{R}}(\ominus_{\bar{R}}\,x^{l})),\\[0.16in]
&  \mathcal{F}_{L}^{\ast}(\exp(p^{k}|\text{i\thinspace}y^{j})_{R,\bar{L}%
})(x^{l})\nonumber\\
&  \qquad=\,\frac{1}{\text{vol}_{\bar{L}}}\int_{-\infty}^{+\infty}d_{\bar{L}%
}^{n}p\,\exp(\text{i}^{-1}p^{m}|\!\ominus_{\bar{L}}x^{l})_{R,\bar{L}}%
\overset{x|p}{\odot}_{\hspace{-0.01in}L}\exp(\text{i}^{-1}p^{k}|y^{j}%
)_{R,\bar{L}}\nonumber\\
&  \qquad=\,\frac{1}{\text{vol}_{\bar{L}}}\,\delta_{\bar{L}}^{n}%
((\ominus_{\bar{L}}\,x^{l})\oplus_{\bar{L}}y^{j}). \label{FTexp2}%
\end{align}
We see that Fourier transformations of q-exponentials lead to q-deformed delta functions.

With the identities in (\ref{VeralDelt1}), (\ref{VeralDelt2}),
(\ref{DeltProAlg0}), and (\ref{DeltProAlg}) we can show that the Fourier
transform of a q-deformed delta function is given by a q-exponential.
Concretely, we have%
\begin{align}
&  \mathcal{F}_{L}(\delta_{\bar{R}}^{n}(y^{j}\oplus_{\bar{R}}(\ominus_{\bar
{R}}\,x^{i})))(p^{k})\nonumber\\
&  \qquad=\,\int_{-\infty}^{+\infty}d_{L}^{n}x\,\delta_{\bar{R}}^{n}%
(y^{j}\oplus_{\bar{R}}(\ominus_{\bar{R}}\,x^{i}))\overset{x}{\circledast}%
\exp(x^{l}|\text{i}^{-1}p^{k})_{\bar{R},L}\nonumber\\
&  \qquad=\int_{-\infty}^{+\infty}d_{L}^{n}x\,\delta_{\bar{R}}^{n}%
((\ominus_{\bar{R}}\,\kappa^{-1}y^{j})\oplus_{\bar{R}}(\kappa x^{i}%
))\overset{x}{\circledast}\exp(x^{l}|\text{i}^{-1}p^{k})_{\bar{R}%
,L}\nonumber\\
&  \qquad=\kappa^{-n}\int_{-\infty}^{+\infty}d_{L}^{n}x\,\delta_{\bar{R}}%
^{n}((\ominus_{\bar{R}}\,\kappa^{-1}y^{j})\oplus_{\bar{R}}x^{i})\overset
{x}{\circledast}\exp(\kappa^{-1}x^{l}|\text{i}^{-1}p^{k})_{\bar{R}%
,L}\nonumber\\
&  \qquad=\kappa^{-n}\text{vol}_{\bar{R}}\exp(\kappa^{-1}y^{j}|\text{i}%
^{-1}p^{k})_{\bar{R},L},\label{FouDelt1}\\[0.16in]
&  \mathcal{F}_{R}(\delta_{\bar{L}}^{n}((\ominus_{\bar{L}}\,x^{i})\oplus
_{\bar{L}}y^{j}))(p^{k})\nonumber\\
&  \qquad=\,\int_{-\infty}^{+\infty}d_{R}^{n}x\,\exp(\text{i}^{-1}p^{k}%
|x^{l})_{R,\bar{L}}\overset{x}{\circledast}\delta_{\bar{L}}^{n}((\ominus
_{\bar{L}}\,x^{i})\oplus_{\bar{L}}y^{j})\nonumber\\
&  \qquad=\,\int_{-\infty}^{+\infty}d_{R}^{n}x\,\exp(\text{i}^{-1}p^{k}%
|x^{l})_{R,\bar{L}}\overset{x}{\circledast}\delta_{\bar{L}}^{n}((\kappa
^{-1}x^{i})\oplus_{\bar{L}}(\ominus_{\bar{L}}\,\kappa y^{j}))\nonumber\\
&  \qquad=\,\kappa^{n}\int_{-\infty}^{+\infty}d_{R}^{n}x\,\exp(\text{i}%
^{-1}p^{k}|\kappa x^{l})_{R,\bar{L}}\overset{x}{\circledast}\delta_{\bar{L}%
}^{n}(x^{i}\oplus_{\bar{L}}(\ominus_{\bar{L}}\,\kappa y^{j}))\nonumber\\
&  \qquad=\,\kappa^{n}\text{vol}_{\bar{L}}\exp(\text{i}^{-1}p^{k}|\kappa
y^{j})_{R,\bar{L}}, \label{FouDelt2}%
\end{align}
and%
\begin{align}
&  \mathcal{F}_{\bar{R}}^{\ast}(\delta_{L}^{n}((\ominus_{L}\,y^{j})\oplus
_{L}x^{i}))(p^{k})\nonumber\\
&  \qquad=\,\frac{1}{\text{vol}_{\bar{R}}}\int_{-\infty}^{+\infty}d_{\bar{R}%
}^{n}x\,\delta_{L}^{n}((\ominus_{L}\,y^{j})\oplus_{L}x^{i})\overset
{yx|p}{\odot}_{\hspace{-0.03in}R}\exp(\text{i}^{-1}p^{k}|\!\ominus_{L}%
x^{l})_{\bar{R},L}\nonumber\\
&  \qquad=\,\exp(\text{i}^{-1}\kappa p^{k}|\!\ominus_{L}y^{j})_{\bar{R}%
,L},\\[0.16in]
&  \mathcal{F}_{\bar{L}}^{\ast}(\delta_{R}^{n}(x^{i}\oplus_{R}(\ominus
_{R}\,y^{j})))(p^{k})\nonumber\\
&  \qquad=\,\frac{1}{\text{vol}_{\bar{L}}}\int_{-\infty}^{+\infty}d_{\bar{L}%
}^{n}x\,\exp(\ominus_{R}\,x^{l}|\text{i}^{-1}p^{k})_{R,\bar{L}}\overset
{p|xy}{\odot}_{\hspace{-0.03in}L}\delta_{R}^{n}(x^{i}\oplus_{R}(\ominus
_{R}\,y^{j}))\nonumber\\
&  \qquad=\,\exp(\ominus_{R}y^{j}|\text{i}^{-1}\kappa^{-1}p^{k})_{R,\bar{L}}.
\end{align}
The above relations can be derived from the results in (\ref{FTExp1}%
)-(\ref{FTexp2}) with the help of the identities in (\ref{InvFourAnf2a}%
)-(\ref{InvFourAnf1b}). Notice that the calculations in (\ref{FouDelt1}) and
(\ref{FouDelt2}) make use of the identities%
\begin{align}
&  \frac{1}{\text{vol}_{L}}\int_{-\infty}^{+\infty}d_{L}^{n}x\,\delta_{\bar
{R}}^{n}((\ominus_{\bar{R}}\,y^{i})\oplus_{\bar{R}}x^{j})\overset
{x}{\circledast}\delta_{\bar{R}}^{n}(x^{k}\oplus_{\bar{R}}(\ominus_{\bar{R}%
}\,\tilde{y}^{l}))\nonumber\\
&  \qquad=\,\delta_{\bar{R}}^{n}((\ominus_{\bar{R}}\,y^{i})\oplus_{\bar{R}%
}(\kappa\tilde{y}^{l}))=\delta_{\bar{R}}^{n}((\kappa y^{i})\oplus_{\bar{R}%
}(\ominus_{\bar{R}}\,\tilde{y}^{l})),\label{DeltIdA}\\[0.1in]
&  \frac{1}{\text{vol}_{R}}\int_{-\infty}^{+\infty}d_{R}^{n}x\,\delta_{\bar
{L}}^{n}((\ominus_{\bar{L}}\,y^{i})\oplus_{\bar{L}}x^{j})\overset
{x}{\circledast}\delta_{\bar{L}}^{n}(x^{k}\oplus_{\bar{L}}(\ominus_{\bar{L}%
}\,\tilde{y}^{l}))\nonumber\\
&  \qquad=\,\delta_{\bar{L}}^{n}((\ominus_{\bar{L}}\,y^{i})\oplus_{\bar{L}%
}(\kappa^{-1}\tilde{y}^{l}))=\delta_{\bar{L}}^{n}((\kappa^{-1}y^{i}%
)\oplus_{\bar{L}}(\ominus_{\bar{L}}\,\tilde{y}^{l})). \label{DeltIdB}%
\end{align}
Lastly, it should be mentioned that we obtain further relations from the above
ones by applying the substitutions in (\ref{SubSym}) together with the
replacement $\kappa\rightarrow\kappa^{-1}$.

\subsection{Convolution products}

In this subsection we turn our attention to q-analogs of the
\textit{convolution product}. They are defined by%
\begin{align}
(f\ast_{A,B}g)(y^{i})  &  \equiv\int_{-\infty}^{+\infty}d_{A}^{n}%
x\,f(x^{j})\overset{x}{\circledast}g((\ominus_{B}\,x^{k})\oplus_{B}%
y^{i}),\nonumber\\
(f\,\tilde{\ast}_{A,B}\,g)(y^{i})  &  \equiv\int_{-\infty}^{+\infty}d_{A}%
^{n}x\,g(y^{i}\oplus_{B}(\ominus_{B}\,x^{k}))\overset{x}{\circledast}f(x^{j}),
\label{DefConPro}%
\end{align}
with $A,B\in\{L,\bar{L},R,\bar{R}\}.$

In Ref. \cite{KM94} it was shown that q-deformed convolution products are
associative. In our formalism this feature of convolution products is modified
as follows:%
\begin{align}
f\ast_{A,B}(g\ast_{A,B}h)  &  =(\kappa_{B})^{-n}(f((\kappa_{B})^{-1}x^{i}%
)\ast_{A,B}g)\ast_{A,B}h,\nonumber\\
(f\,\tilde{\ast}_{A,B}\,g\,)\tilde{\ast}_{A,B}\,h  &  =(\kappa_{B}%
)^{-n}f\,\tilde{\ast}_{A,B}\,(g\,\tilde{\ast}_{A,B}\,h((\kappa_{B})^{-1}%
x^{i})). \label{AssConv}%
\end{align}
The following calculation shall illustrate how to prove this property:%
\begin{align}
&  f\ast_{L,\bar{L}}(g\ast_{L,\bar{L}}h)\nonumber\\
&  \qquad=\,\int_{-\infty}^{+\infty}d_{L}^{n}x\,f(x^{i})\overset
{x}{\circledast}\int_{-\infty}^{+\infty}d_{L}^{n}y\,g(y^{j})\overset
{y}{\circledast}h((\ominus_{\bar{L}}\,y^{k})\oplus_{\bar{L}}((\ominus_{\bar
{L}}\,x^{l})\oplus_{\bar{L}}z^{m}))\nonumber\\
&  \qquad=\,\int_{-\infty}^{+\infty}d_{L}^{n}x\,f(x^{i})\overset
{x}{\circledast}\int_{-\infty}^{+\infty}d_{L}^{n}y\,g(y^{j})\overset
{y}{\circledast}h(((\ominus_{\bar{L}}\,y^{k})\oplus_{\bar{L}}(\ominus_{\bar
{L}}\,x^{l}))\oplus_{\bar{L}}z^{m})\nonumber\\
&  \qquad=\,\int_{-\infty}^{+\infty}d_{L}^{n}x\,f(x^{i})\overset
{x}{\circledast}\int_{-\infty}^{+\infty}d_{L}^{n}y\,g(y^{j})\overset
{y}{\circledast}h((\ominus_{\bar{L}}(y^{k}\oplus_{\bar{R}}x^{l}))\oplus
_{\bar{L}}z^{m})\nonumber\\
&  \qquad=\,\int_{-\infty}^{+\infty}d_{L}^{n}x\,f(x^{i})\overset
{x}{\circledast}\int_{-\infty}^{+\infty}d_{L}^{n}y\,g(y^{j})\overset
{y|x}{\odot}_{\hspace{-0.01in}\bar{L}}h((\ominus_{\bar{L}}((\kappa
x^{l})\oplus_{\bar{L}}y^{k}))\oplus_{\bar{L}}z^{m})\nonumber\\
&  \qquad=\,\int_{-\infty}^{+\infty}d_{L}^{n}x\,f(x^{i})\overset
{x}{\circledast}\int_{-\infty}^{+\infty}d_{L}^{n}y\,g((\ominus_{\bar{L}%
}\,\kappa x^{l})\oplus_{\bar{L}}y^{j})\overset{y}{\circledast}h((\ominus
_{\bar{L}}\,y^{k})\oplus_{\bar{L}}z^{m})\nonumber\\
&  \qquad=\,\int_{-\infty}^{+\infty}d_{L}^{n}y\,\int_{-\infty}^{+\infty}%
d_{L}^{n}x\,f(x^{i})\overset{x}{\circledast}g((\ominus_{\bar{L}}\,\kappa
x^{l})\oplus_{\bar{L}}y^{j})\overset{y}{\circledast}h((\ominus_{\bar{L}%
}\,y^{k})\oplus_{\bar{L}}z^{m})\nonumber\\
&  \qquad=\,\kappa^{-n}(f(\kappa^{-1}x^{i})\ast_{L,\bar{L}}g)\ast_{L,\bar{L}%
}h.
\end{align}
The first equality is clear, since it is a direct consequence of the
definition of q-deformed convolution products. The second equality is the law
of associativity for q-translations, and for the third step we applied a kind
of distributivity law for q-translations. For the fourth step we made use of
the fact that the q-translations $\oplus_{\bar{L}}$ and $\oplus_{\bar{R}}$ are
related to each other by a twist. The sixth step is an application of
(\ref{IntKoorTrans}) and leads to a result that can be recognized as the
right-hand side of the first relation in (\ref{AssConv}).

Next, we consider \textit{opposite convolution products}, which are given by%
\begin{align}
(f\ast_{A,B}^{\bar{L}}g)(y^{i})  &  \equiv((\mathcal{R}_{[2]}\triangleright
g)\ast_{A,B}(\mathcal{R}_{[1]}\triangleright f))(y^{i})\nonumber\\
&  =\int_{-\infty}^{+\infty}d_{A}^{n}x\,\big [f(z^{l})\overset{z|x}{\odot
}_{\hspace{-0.01in}\bar{L}}g(x^{j})\big ]_{z^{l}\rightarrow(\ominus_{B}%
x^{k})\,\oplus_{B}\,y^{i}},\nonumber\\
(f\ast_{A,B}^{L}g)(y^{i})  &  \equiv((\mathcal{R}_{[1]}^{-1}\triangleright
g)\ast_{A,B}(\mathcal{R}_{[2]}^{-1}\triangleright f))(y^{j})\nonumber\\
&  =\int_{-\infty}^{+\infty}d_{A}^{n}x\,\big [f(z^{l})\overset{z|x}%
{\odot\hspace{-0.01in}}_{L}g(x^{j})\big ]_{z^{l}\rightarrow(\ominus_{B}%
x^{k})\,\oplus_{B}\,y^{i}},\\[0.16in]
(f\,\ast_{A,B}^{\bar{R}}\,g)(y^{i})  &  \equiv((g\triangleleft\mathcal{R}%
_{[1]}^{-1})\,\ast_{A,B}\,(f\triangleleft\mathcal{R}_{[2]}^{-1}))(y^{i}%
)\nonumber\\
&  =\int_{-\infty}^{+\infty}d_{A}^{n}x\,\big [f(z^{l})\overset{z|x}{\odot
}_{\hspace{-0.01in}\bar{R}}g(x^{j})\big ]_{z^{l}\rightarrow(\ominus_{B}%
x^{k})\,\oplus_{B}\,y^{i}},\nonumber\\
(f\,\ast_{A,B}^{R}\,g)(y^{i})  &  \equiv((g\triangleleft\mathcal{R}%
_{[2]})\,\ast_{A,B}\,(f\triangleleft\mathcal{R}_{[1]}))(y^{i})\nonumber\\
&  =\int_{-\infty}^{+\infty}d_{A}^{n}x\,\big [f(z^{l})\overset{z|x}{\odot
}_{\hspace{-0.01in}R}g(x^{j})\big ]_{z^{l}\rightarrow(\ominus_{B}%
x^{k})\,\oplus_{B}\,y^{i}},
\end{align}
and%
\begin{align}
(f\,\tilde{\ast}_{A,B}^{\bar{L}}\,g)(y^{i})  &  \equiv((\mathcal{R}%
_{[2]}\triangleright g)\,\tilde{\ast}_{A,B}\,(\mathcal{R}_{[1]}\triangleright
f))(y^{i})\nonumber\\
&  =\int_{-\infty}^{+\infty}d_{A}^{n}x\,\big [f(x^{j})\overset{x|z}%
{\odot\hspace{-0.01in}}_{\bar{L}}g(z^{l})\big ]_{z^{l}\rightarrow\,y^{i}%
\oplus_{B}(\ominus_{B}x^{k})},\nonumber\\
(f\,\tilde{\ast}_{A,B}^{L}\,g)(y^{i})  &  \equiv((\mathcal{R}_{[1]}%
^{-1}\triangleright g)\,\tilde{\ast}_{A,B}\,(\mathcal{R}_{[2]}^{-1}%
\triangleright f))(y^{i})\nonumber\\
&  =\int_{-\infty}^{+\infty}d_{A}^{n}x\,\big [f(x^{j})\overset{x|z}{\odot
}_{\hspace{-0.01in}L}g(z^{l})\big ]_{z^{l}\rightarrow\,y^{i}\oplus_{B}%
(\ominus_{B}x^{k})},\\[0.16in]
(f\tilde{\ast}_{A,B}^{\bar{R}}g)(y^{i})  &  \equiv((g\triangleleft
\mathcal{R}_{[1]}^{-1})\tilde{\ast}_{A,B}(f\triangleleft\mathcal{R}_{[2]}%
^{-1}))(y^{j})\nonumber\\
&  =\int_{-\infty}^{+\infty}d_{A}^{n}x\,\big [f(x^{j})\overset{x|z}{\odot
}_{\hspace{-0.01in}\bar{R}}g(z^{l})\big ]_{z^{l}\rightarrow\,y^{i}\oplus
_{B}(\ominus_{B}x^{k})},\nonumber\\
(f\tilde{\ast}_{A,B}^{R}g)(y^{i})  &  \equiv((g\triangleleft\mathcal{R}%
_{[2]})\tilde{\ast}_{A,B}(f\triangleleft\mathcal{R}_{[1]}))(y^{j})\nonumber\\
&  =\int_{-\infty}^{+\infty}d_{A}^{n}x\,\big [f(x^{j})\overset{x|z}{\odot
}_{\hspace{-0.01in}R}g(z^{l})\big ]_{z^{l}\rightarrow\,y^{i}\oplus_{B}%
(\ominus_{B}x^{k})}.
\end{align}

One can show that for opposite convolution products we have the relations%
\begin{align}
(f\ast_{A,B}^{C}g)\ast_{A,B}^{C}h &  =(\kappa_{B})^{n}f((\kappa_{C})^{-1}%
x^{i})\ast_{A,B}^{C}(g\ast_{A,B}^{C}h(\kappa_{B}\kappa_{C}\tilde{x}%
^{j})),\nonumber\\
f\,\tilde{\ast}_{A,B}^{C}\,(g\,\tilde{\ast}_{A,B}^{C}\,h) &  =(\kappa_{B}%
)^{n}(f(\kappa_{B}\kappa_{C}x^{i})\,\tilde{\ast}_{A,B}^{C}\,g)\,\tilde{\ast
}_{A,B}^{C}\,h((\kappa_{C})^{-1}\tilde{x}^{j}).
\end{align}
The first relation can be verified in the following manner:%
\begin{align}
&  (f\ast_{A,B}^{\bar{L}}g)\ast_{A,B}^{\bar{L}}h\nonumber\\
&  \qquad=\,(\mathcal{R}_{[2]}\triangleright h)\ast_{A,B}(\mathcal{R}%
_{[1]}\triangleright\lbrack(\mathcal{R}_{[2]}^{\prime}\triangleright
g)\ast_{A,B}(\mathcal{R}_{[1]}^{\prime}\triangleright f)])\nonumber\\
&  \qquad=\,(\mathcal{R}_{[2]}\triangleright h)\ast_{A,B}((\mathcal{R}%
_{[1](1)}\mathcal{R}_{[2]}^{\prime}\triangleright g)\ast_{A,B}(\mathcal{R}%
_{[1](2)}\mathcal{R}_{[1]}^{\prime}\triangleright f))\nonumber\\
&  \qquad=\,(\mathcal{R}_{[2]}\mathcal{R}_{[2]}^{\prime\prime}\triangleright
h(\kappa_{\bar{L}}\tilde{x}^{j}))\ast_{A,B}((\mathcal{R}_{[1]}\mathcal{R}%
_{[2]}^{\prime}\triangleright g)\ast_{A,B}(\mathcal{R}_{[1]}^{\prime\prime
}\mathcal{R}_{[1]}^{\prime}\triangleright f))\nonumber\\
&  \qquad=\,((\mathcal{R}_{[2]}\mathcal{R}_{[2]}^{\prime\prime}\triangleright
h(\kappa_{\bar{L}}\kappa_{B}\tilde{x}^{j}))\ast_{A,B}(\mathcal{R}%
_{[1]}\mathcal{R}_{[2]}^{\prime}\triangleright g))\ast_{A,B}(\mathcal{R}%
_{[1]}^{\prime\prime}\mathcal{R}_{[1]}^{\prime}\triangleright f)\nonumber\\
&  \qquad=\,((\mathcal{R}_{[2]}^{\prime}\mathcal{R}_{[2]}\triangleright
h(\kappa_{\bar{L}}\kappa_{B}\tilde{x}^{j}))\ast_{A,B}(\mathcal{R}%
_{[2]}^{\prime\prime}\mathcal{R}_{[1]}\triangleright g))\ast_{A,B}%
(\mathcal{R}_{[1]}^{\prime\prime}\mathcal{R}_{[1]}^{\prime}\triangleright
f)\nonumber\\
&  \qquad=\,((\mathcal{R}_{[2](1)}^{\prime}\mathcal{R}_{[2]}\triangleright
h(\kappa_{\bar{L}}\kappa_{B}\tilde{x}^{j}))\ast_{A,B}(\mathcal{R}%
_{[2](2)}^{\prime}\mathcal{R}_{[1]}\triangleright g))\ast_{A,B}(\mathcal{R}%
_{[1]}^{\prime}\triangleright f)\nonumber\\
&  \qquad=\,((\mathcal{R}_{[2]}^{\prime}\triangleright\lbrack(\mathcal{R}%
_{[2]}\triangleright h(\kappa_{\bar{L}}\kappa_{B}\tilde{x}^{j}))\ast
_{A,B}(\mathcal{R}_{[1]}\triangleright g)])\nonumber\\
&  \qquad\,\hspace{0.16in}\ast_{A,B}(\mathcal{R}_{[1]}^{\prime}\triangleright
f((\kappa_{\bar{L}})^{-1}x^{i}))\nonumber\\
&  \qquad=\,f((\kappa_{\bar{L}})^{-1}x^{i})\ast_{A,B}^{\bar{L}}(g\ast
_{A,B}^{\bar{L}}h(\kappa_{\bar{L}}\kappa_{B}\tilde{x}^{j})).
\end{align}
The first and the last step uses the definition of the opposite convolution
product, while the second and seventh step result from covariance of the
convolution products, i.e. convolution products can be passed through a
braid-crossing up to additional scalings. (Notice that the scaling of $h$ with
$\kappa_{\bar{L}}$ is a consequence of the fact that the definition of a
convolution product contains an integral over the whole space.) For the third
and sixth equality we applied the axioms of quasi-triangularity \cite{Maj95,
KS97, ChDe96}, i.e.
\begin{equation}
(\text{id}\otimes\Delta)\mathcal{R}=\mathcal{R}_{13}\mathcal{R}_{12}%
,\quad(\Delta\otimes\text{id})\mathcal{R}=\mathcal{R}_{13}\mathcal{R}_{23}.
\end{equation}
The fourth step is associativity of the convolution product and for the fifth
equality we made use of the Yang-Baxter equation%
\begin{equation}
\mathcal{R}_{12}\mathcal{R}_{13}\mathcal{R}_{23}=\mathcal{R}_{23}%
\mathcal{R}_{13}\mathcal{R}_{12}.
\end{equation}

There is a property of opposite convolution products worth recording here. It
concerns the observation that q-deformed Fourier transformations map star
products to opposite convolution products and vice versa. To be more specific,
we have%
\begin{align}
\mathcal{F}_{L}(f)(p^{i})\overset{p}{\circledast}\mathcal{F}_{L}(g)(p^{k}) &
=\mathcal{F}_{L}(f(\kappa x^{i})\ast_{L,\bar{L}}^{\bar{L}}g)(p^{k}%
),\nonumber\\
\mathcal{F}_{\bar{L}}(f)(p^{i})\overset{p}{\circledast}\mathcal{F}_{\bar{L}%
}(g)(p^{k}) &  =\mathcal{F}_{\bar{L}}(f(\kappa^{-1}x^{i})\ast_{\bar{L},L}%
^{L}g)(p^{k}),\label{FouCon1}\\[0.16in]
\mathcal{F}_{R}(f)(p^{i})\overset{p}{\circledast}\mathcal{F}_{R}(g)(p^{k}) &
=\mathcal{F}_{R}(f\,\tilde{\ast}_{R,\bar{R}}^{\bar{R}}\,g(\kappa^{-1}%
x^{i}))(p^{k}),\nonumber\\
\mathcal{F}_{\bar{R}}(f)(p^{i})\overset{p}{\circledast}\mathcal{F}_{\bar{R}%
}(g)(p^{k}) &  =\mathcal{F}_{\bar{R}}(f\,\tilde{\ast}_{\bar{R},R}%
^{R}\,g(\kappa x^{i}))(p^{k}).\label{FouCon2}%
\end{align}
The above identities can be checked in the following manner:%
\begin{align}
&  \mathcal{F}_{L}(f)(p^{i})\overset{p}{\circledast}\mathcal{F}_{L}%
(f)(p^{k})\nonumber\\
&  \qquad=\,\int\nolimits_{-\infty}^{+\infty}d_{L}^{n}x\,f(x^{j})\overset
{x}{\circledast}\exp(x^{l}|\text{i}^{-1}p^{i})_{\bar{R},L}\nonumber\\
&  \qquad\qquad\qquad\overset{p}{\circledast}\int\nolimits_{-\infty}^{+\infty
}d_{L}^{n}y\,g(y^{m})\overset{y}{\circledast}\exp(y^{r}|\text{i}^{-1}%
p^{k})_{\bar{R},L}\nonumber\\
&  \qquad=\,\int\nolimits_{-\infty}^{+\infty}d_{L}^{n}x\,f(x^{j})\overset
{x}{\circledast}\int\nolimits_{-\infty}^{+\infty}d_{L}^{n}y\,g(y^{m}%
)\overset{y}{\circledast}\exp(y^{r}\oplus_{\bar{L}}x^{l}|\text{i}^{-1}%
p^{k})_{\bar{R},L}\nonumber\\
&  \qquad=\,\int\nolimits_{-\infty}^{+\infty}d_{L}^{n}x\int\nolimits_{-\infty
}^{+\infty}d_{L}^{n}y\,f(\kappa x^{j})\overset{x|y}{\odot}_{\hspace
{-0.01in}\bar{L}}\big (g(y^{m})\overset{y}{\circledast}\exp(y^{r}\oplus
_{\bar{L}}x^{l}|\text{i}^{-1}p^{k})_{\bar{R},L}\big )\nonumber\\
&  \qquad=\int\nolimits_{-\infty}^{+\infty}d_{L}^{n}y\,(\mathcal{R}%
_{[2]}\triangleright g(y^{m}))\overset{y}{\circledast}\,\int\nolimits_{-\infty
}^{+\infty}d_{L}^{n}x(\mathcal{R}_{[1]}\triangleright f(\kappa x^{j}%
))\nonumber\\
&  \qquad\qquad\qquad\overset{x|y}{\odot}_{\hspace{-0.01in}\bar{L}}\exp
(y^{r}\oplus_{\bar{L}}x^{l}|\text{i}^{-1}p^{k})_{\bar{R},L}\nonumber\\
&  \qquad=\int\nolimits_{-\infty}^{+\infty}d_{L}^{n}y\,(\mathcal{R}%
_{[2]}\triangleright g(y^{m})\,)\overset{y}{\circledast}\int\nolimits_{-\infty
}^{+\infty}d_{L}^{n}x(\mathcal{R}_{[1]}\triangleright f)((\ominus_{\bar{L}%
}\,\kappa y^{r})\oplus_{\bar{L}}(\kappa x^{j}))\nonumber\\
&  \qquad\qquad\qquad\overset{x}{\circledast}\exp(x^{l}|\text{i}^{-1}%
p^{k})_{\bar{R},L}\nonumber\\
&  \qquad=\,\int\nolimits_{-\infty}^{+\infty}d_{L}^{n}x\int\nolimits_{-\infty
}^{+\infty}d_{L}^{n}y\,\big [f(\kappa z^{i})\overset{z|x}{\odot}%
_{\hspace{-0.01in}\bar{L}}g(y^{m})\big ]_{z^{j}\rightarrow(\ominus_{\bar{L}%
}\,y^{r})\oplus_{\bar{L}}\,x^{j}}\nonumber\\
&  \qquad\qquad\qquad\overset{x}{\circledast}\exp(x^{l}|\text{i}^{-1}%
p^{k})_{\bar{R},L}\nonumber\\
&  \qquad=\,\mathcal{F}_{L}(f(\kappa x^{i})\ast_{L,\bar{L}}^{\bar{L}}%
g)(p^{k}).
\end{align}
The first step uses the definition of q-deformed Fourier transformations and
for the second step we apply the addition law for q-deformed exponentials. For
the third and fourth step we rearrange tensor factors, and the fifth equality
holds due to (\ref{IntKoorTrans}).

Last but not least, we would like to show that q-deformed delta functions play
the role of an identity element in the convolution product algebra.
Concretely, it holds%
\begin{align}
(\delta_{B}^{n}\ast_{A,C}f)(y^{k})  &  =(f(\kappa_{D}x^{j})\ast_{A,C}%
^{D}\delta_{B}^{n})(y^{k})\nonumber\\
&  =\int\nolimits_{-\infty}^{+\infty}d_{A}^{n}x\,\delta_{B}^{n}(x^{i}%
)\overset{x}{\circledast}f((\ominus_{C}\,x^{j})\oplus_{C}y^{k})\nonumber\\
&  =\text{vol}_{A,B}\,f(y^{k}),\label{EinConP}\\[0.1in]
(f\,\tilde{\ast}_{A,C}\,\delta_{B}^{n})(y^{k})  &  =(\delta_{B}^{n}%
\,\tilde{\ast}_{A,C}^{D}\,f(\kappa_{D}x^{j}))(y^{k})\nonumber\\
&  =\int\nolimits_{-\infty}^{+\infty}d_{A}^{n}x\,f(y^{k}\oplus_{C}(\ominus
_{C}x^{j}))\overset{x}{\circledast}\delta_{B}^{n}(x^{i})\nonumber\\
&  =\text{vol}_{A,B}\,f(y^{k}), \label{EinConP2}%
\end{align}
and
\begin{align}
(f\ast_{A,C}\delta_{B}^{n})(y^{k})  &  =(\delta_{B}^{n}\ast_{A,C}^{D}%
f(\kappa_{D}x^{i}))(y^{k})\nonumber\\
&  =\int\nolimits_{-\infty}^{+\infty}d_{A}^{n}x\,f(x^{i})\overset
{x}{\circledast}\delta_{B}^{n}((\ominus_{C}\,x^{j})\oplus_{C}y^{k}%
),\nonumber\\
&  =(\kappa_{C})^{n}\,\text{vol}_{A,B}\,f(\kappa_{C}y^{k}), \label{EinConP3}%
\\[0.1in]
(\delta_{B}^{n}\,\tilde{\ast}_{A,C}\,f)(y^{k})  &  =(f(\kappa_{D}%
x^{i})\,\tilde{\ast}_{A,C}^{D}\,\delta_{B}^{n})(y^{k})\nonumber\\
&  =\int\nolimits_{-\infty}^{+\infty}d_{A}^{n}x\,\delta_{B}^{n}(y^{k}%
\oplus_{C}(\ominus_{C}\,x^{j}))\overset{x}{\circledast}f(x^{i})\nonumber\\
&  =(\kappa_{C})^{n}\,\text{vol}_{A,B}\,f(\kappa_{C}y^{k}), \label{EinConP4}%
\end{align}
where $A,B,C,D\in\{L,\bar{L},R,\bar{R}\}.$ Notice that the equalities
concerning opposite convolution products arise from the braiding properties of
q-deformed delta functions [cf. the identities in (\ref{BraPropDelt})]. The
last equality in (\ref{EinConP}) as well as in (\ref{EinConP2}) is\ clear from
the relations in (\ref{VeralDelt0}). The derivations of (\ref{EinConP3}) and
(\ref{EinConP4}) need the identities (\ref{DeltIdA}) and (\ref{DeltIdB}). The
following calculation referring to the case\ $C=L$ shows the line of
reasonings:%
\begin{align}
(f\ast_{A,L}\delta_{B}^{n})(y^{k})  &  =\int\nolimits_{-\infty}^{+\infty}%
d_{A}^{n}x\,f(x^{i})\overset{x}{\circledast}\delta_{B}^{n}((\ominus_{L}%
\,x^{j})\oplus_{L}y^{k})\nonumber\\
&  =\int\nolimits_{-\infty}^{+\infty}d_{A}^{n}x\,f(x^{i})\overset
{x}{\circledast}\delta_{B}^{n}((\kappa x^{j})\oplus_{L}(\ominus_{L}%
\,\kappa^{-1}y^{k}))\nonumber\\
&  =\kappa^{-n}\int\nolimits_{-\infty}^{+\infty}d_{A}^{n}x\,f(\kappa^{-1}%
x^{i})\overset{x}{\circledast}\delta_{B}^{n}(x^{j}\oplus_{L}(\ominus
_{L}\,\kappa^{-1}y^{k}))\nonumber\\
&  =\kappa^{-n}\,\text{vol}_{A,B}\,f(\kappa^{-1}x^{i}).
\end{align}
Notice that for the second step we have to apply a variant of the relation in
(\ref{DeltIdB}) while the last equality holds due to (\ref{DeltProAlg0}).

\subsection{Conjugation properties of Fourier transformations}

We conclude our considerations about q-deformed Fourier transformations by
discussing their conjugation properties. Recalling the conjugation properties
of q-integrals, q-exponentials, and star products (see Ref. \cite{qAn}) we get%
\begin{align}
\overline{\mathcal{F}_{L}(f)(p^{k})}  &  =\overline{\int\nolimits_{-\infty
}^{+\infty}d_{L}^{n}x\,f(x^{i})\overset{x}{\circledast}\exp(x^{j}%
|\text{i}^{-1}p^{k})_{\bar{R},L}}\nonumber\\
&  =(-1)^{n}\int\nolimits_{-\infty}^{+\infty}d_{L}^{n}x\,\exp(\text{i}%
^{-1}p^{k}|x^{j})_{\bar{R},L}\overset{x}{\circledast}\overline{f(x^{i}%
)}\nonumber\\
&  =(-1)^{n}\mathcal{F}_{\bar{R}}(\bar{f})(p^{k}).
\end{align}
For q-deformed exponentials and q-deformed\ volume elements the above
relations respectively imply%
\begin{equation}
\overline{\delta_{L}^{n}(p^{k})}=\overline{\mathcal{F}_{L}(1)(p^{k})}%
=(-1)^{n}\mathcal{F}_{\bar{R}}(1)(p^{k})=(-1)^{n}\delta_{\bar{R}}^{n}(p^{k}),
\end{equation}
and%
\begin{align}
\overline{\text{vol}_{L}}  &  =\overline{\int\nolimits_{-\infty}^{+\infty
}d_{\bar{R}}^{n}p\,\delta_{L}^{n}(p^{k})}=(-1)^{n}\int\nolimits_{-\infty
}^{+\infty}d_{L}^{n}p\,(-1)^{n}\delta_{\bar{R}}^{n}(p^{k})\nonumber\\
&  =\int\nolimits_{-\infty}^{+\infty}d_{\bar{R}}^{n}p\,\delta_{\bar{R}}%
^{n}(p^{k})=\text{vol}_{\bar{R}}.
\end{align}
With the same reasonings we find the conjugation properties of the Fourier
transformations in (\ref{FTtype1}) and (\ref{FTtype2}):%
\begin{align}
\overline{\mathcal{F}_{\bar{R}}^{\ast}(f)(x^{j})}  &  =\overline{\frac
{1}{\text{vol}_{\bar{R}}}\int\nolimits_{-\infty}^{+\infty}d_{\bar{R}}%
^{n}p\,f(p^{l})\overset{p|x}{\odot}_{\hspace{-0.01in}R}\exp(\ominus_{\bar{R}%
}\,x^{j}|\text{i}^{-1}p^{k})_{\bar{R},L}}\nonumber\\
&  =\frac{(-1)^{n}}{\text{vol}_{L}}\int\nolimits_{-\infty}^{+\infty}d_{L}%
^{n}p\,\exp(\text{i}^{-1}p^{k}|\!\ominus_{L}x^{j})_{\bar{R},L}\overset
{x|p}{\odot}_{\bar{L}}\overline{f(p^{l})}\nonumber\\
&  =(-1)^{n}\mathcal{F}_{L}^{\ast}(\bar{f})(x^{k}).
\end{align}
Changing the labels in the above formulae according to the substitutions of
(\ref{SubSym}) yields the relations for the other q-geometries.

A short glance at our results makes it clear that up to an additional minus
sign q-deformed Fourier transforms are transformed into each other by the
operation of conjugation. To avoid the additional minus signs in the above
formulae, we modify the definitions of Fourier transformations, delta
functions, and volume elements by carrying out the following substitutions in
the defining expressions:%
\begin{align}
\int d_{L}^{n}x,\int d_{\bar{R}}^{n}x  &  \longrightarrow\int d_{1}^{n}%
x\equiv\frac{\text{i}^{n}}{2}\Big (\int d_{L}^{n}x+\int d_{\bar{R}}%
^{n}x\Big ),\nonumber\\
\int d_{\bar{L}}^{n}x,\int d_{R}^{n}x  &  \longrightarrow\int d_{2}^{n}%
x\equiv\frac{\text{i}^{n}}{2}\Big (\int d_{\bar{L}}^{n}x+\int d_{R}%
^{n}x\Big ). \label{SubInt}%
\end{align}
In other words, we deal with real integrals, only, and the new objects
obtained this way are distinguished from the original ones by a tilde:%
\begin{gather}
\mathcal{F}_{A}\longrightarrow\mathcal{\tilde{F}}_{A},\quad\mathcal{F}%
_{A}^{\ast}\longrightarrow\mathcal{\tilde{F}}_{A}^{\ast},\nonumber\\
\delta_{A}^{n}\longrightarrow\tilde{\delta}_{A}^{n},\quad\text{vol}%
_{A}\longrightarrow\widetilde{\text{vol}}_{A}. \label{SubTild}%
\end{gather}
It should be mentioned that all considerations and identities so far\ remain
valid under these substitutions.

\section{Sesquilinear forms on quantum spaces\label{SecPart}}

In this section we first introduce sesquilinear forms on q-deformed quantum
spaces. In Paper II these sesquilinear forms will lead us to q-deformed
analogs of the Hilbert space of square integrable functions. We discuss such
properties of q-deformed sesquilinear forms that will prove useful in
formulating a q-deformed version of quantum kinematics. (This task is also be
done in Part II.) This way we focus attention on adjoint operators of symmetry
generators and invariance properties of q-deformed sesquilinear forms. Using
the results of the previous section we finally show that our sesquilinear
forms fulfill q-analogs of Fourier-Plancherel identities.

\subsection{Definition of sesquilinear forms\label{DefSes}}

Let us briefly recall what is meant by a \textit{sesquilinear form} on a
vector space $V.$ It is a mapping%
\begin{equation}
\left\langle .\,,.\right\rangle :V\otimes V\rightarrow\mathbb{C},
\end{equation}
being subject to

\begin{enumerate}
\item $\left\langle x,y_{1}+y_{2}\right\rangle =\left\langle x,y_{1}%
\right\rangle +\left\langle x,y_{2}\right\rangle ,$

\item $\left\langle x_{1}+x_{2},y\right\rangle =\left\langle x_{1}%
,y\right\rangle +\left\langle x_{2},y\right\rangle ,$

\item $\left\langle x,\alpha y\right\rangle =\alpha\left\langle
x,y\right\rangle ,$

\item $\left\langle \alpha x,y\right\rangle =\overline{\alpha}\left\langle
x,y\right\rangle ,$
\end{enumerate}

\noindent for all $x,x_{i},y,y_{i}\in V,$ $\alpha\in\mathbb{C}.$ Likewise, we
can consider the mapping%
\[
\left\langle .\,,.\right\rangle ^{\prime}:V\otimes V\rightarrow\mathbb{C},
\]
with

\begin{enumerate}
\item $\left\langle x,y_{1}+y_{2}\right\rangle ^{\prime}=\left\langle
x,y_{1}\right\rangle ^{\prime}+\left\langle x,y_{2}\right\rangle ^{\prime},$

\item $\left\langle x_{1}+x_{2},y\right\rangle ^{\prime}=\left\langle
x_{1},y\right\rangle ^{\prime}+\left\langle x_{2},y\right\rangle ^{\prime},$

\item $\left\langle \alpha x,y\right\rangle ^{\prime}=\alpha\left\langle
x,y\right\rangle ^{\prime},$

\item $\left\langle x,\alpha y\right\rangle ^{\prime}=\bar{\alpha}\left\langle
x,y\right\rangle ^{\prime},$
\end{enumerate}

\noindent for all $x,x_{i},y,y_{i}\in V,$ $\alpha\in\mathbb{C}.$ A
sesquilinear form is called \textit{symmetrical} when it additionally fulfills%
\begin{equation}
\overline{\left\langle x,y\right\rangle }=\left\langle y,x\right\rangle
\quad\text{for all }x,y\in V.
\end{equation}

It is not very difficult to convince yourself that the following mappings give
sesquilinear forms:%
\begin{align}
\big \langle f,g\big \rangle_{A}  &  \equiv\int_{-\infty}^{+\infty}d_{A}%
^{n}x\,\overline{f(x^{i})}\overset{x}{\circledast}g(x^{j}),\nonumber\\
\big \langle f,g\big \rangle_{A}^{\prime}  &  \equiv\int_{-\infty}^{+\infty
}d_{A}^{n}x\,f(x^{i})\overset{x}{\circledast}\overline{g(x^{j})},
\label{PraSes}%
\end{align}
where $A\in\{L,\bar{L},R,\bar{R}\}.$

Due to the conjugation properties of q-deformed integrals the mappings in
(\ref{PraSes}) behave under conjugation as follows:%
\begin{align}
\overline{\big \langle f,g\big \rangle_{L}}  &  =(-1)^{n}%
\,\big \langle g,f\big \rangle_{\bar{R}},\nonumber\\
\overline{\big \langle f,g\big \rangle_{\bar{L}}}  &  =(-1)^{n}%
\,\big \langle g,f\big \rangle_{R},\\[0.16in]
\overline{\big \langle f,g\big \rangle_{L}^{\prime}}  &  =(-1)^{n}%
\,\big \langle g,f\big \rangle_{\bar{R}}^{\prime},\nonumber\\
\overline{\big \langle f,g\big \rangle_{\bar{L}}^{\prime}}  &  =(-1)^{n}%
\,\big \langle g,f\big \rangle_{R}^{\prime}.
\end{align}
Thus, we conclude that symmetrical sesquilinear forms are given by the
expressions%
\begin{align}
\big \langle f,g\big \rangle_{1}  &  \equiv\frac{\text{i}^{n}}{2}%
\big (\big \langle f,g\big \rangle_{L}+\big \langle f,g\big \rangle_{\bar{R}%
}\big ),\nonumber\\
\big \langle f,g\big \rangle_{2}  &  \equiv\frac{\text{i}^{n}}{2}%
\big (\big \langle f,g\big \rangle_{\bar{L}}+\big \langle f,g\big \rangle_{R}%
\big ),\label{SymSes1}\\[0.16in]
\big \langle f,g\big \rangle_{1}^{\prime}  &  \equiv\frac{\text{i}^{n}}%
{2}\big (\big \langle f,g\big \rangle_{L}^{\prime}%
+\big \langle f,g\big \rangle_{\bar{R}}^{\prime}\big ),\nonumber\\
\big \langle f,g\big \rangle_{2}^{\prime}  &  \equiv\frac{\text{i}^{n}}%
{2}\big (\big \langle f,g\big \rangle_{\bar{L}}^{\prime}%
+\big \langle f,g\big \rangle_{R}^{\prime}\big ). \label{SymSes2}%
\end{align}

\subsection{Adjoint operators and invariance properties of sesquilinear forms}

The sesquilinear forms introduced in the last subsection enable us to assign
linear functionals to a function. This can be achieved by the mappings
$(A\in\{L,\bar{L},R,\bar{R}\})$%
\begin{align}
f_{A}^{\ast} &  :\mathcal{A}_{f}\rightarrow\mathbb{C},\quad g\rightarrow
\big \langle g,f\big \rangle_{A},\nonumber\\
f_{A}^{\prime\ast} &  :\mathcal{A}_{f}^{\prime}\rightarrow\mathbb{C},\quad
g\rightarrow\big \langle f,g\big \rangle_{A}^{\prime},
\end{align}
and $(i\in\{1,2\})$%
\begin{align}
f_{i}^{\ast} &  :\mathcal{A}_{f}\rightarrow\mathbb{C},\quad g\rightarrow
\big \langle g,f\big \rangle_{i},\nonumber\\
f_{i}^{\prime\ast} &  :\mathcal{A}_{f}^{\prime}\rightarrow\mathbb{C},\quad
g\rightarrow\big \langle f,g\big \rangle_{i}^{\prime},
\end{align}
where $\mathcal{A}_{f}$ and $\mathcal{A}_{f}^{\prime}$ denote the subspaces of
the space of formal power series on which the corresponding functionals take
on finite values. This way, our sesquilinear forms should provide a natural
link between suitable subspaces of the space of formal power series and their
dual spaces.

It is well-known that this natural correspondence extends to the operators
defined on these spaces. Each operator on the space of formal power series
induces operators on the above mentioned dual spaces. They are called
\textit{adjoint operators }and their defining equations become
\begin{align}
\big \langle O^{\dag_{A}}\triangleright f,g\big \rangle_{A}  &
=\big \langle f,O\triangleright f\big \rangle_{A},\nonumber\\
\big \langle O^{\dag_{A}^{\prime}}\triangleright f,g\big \rangle_{A}^{\prime}
&  =\big \langle f,O\triangleright f\big \rangle_{A}^{\prime},
\end{align}
and%
\begin{align}
\big \langle O^{\dag_{i}}\triangleright f,g\big \rangle_{i}  &
=\big \langle f,O\triangleright f\big \rangle_{i},\nonumber\\
\big \langle O^{\dag_{i}^{\prime}}\triangleright f,g\big \rangle_{i}^{\prime}
&  =\big \langle f,O\triangleright f\big \rangle_{i}^{\prime}.
\end{align}
In the following it is our aim to find the adjoints of the operators
generating symmetry transformations on quantum spaces.

Let $h$ be an element of a Hopf algebra $\mathcal{H}$ that acts upon our
quantum spaces. Furthermore, we assume that $h$ generates infinitesimal
symmetry transformations. From the definitions in (\ref{PraSes}) it follows
that%
\begin{align}
\big \langle f,h\triangleright g\big \rangle_{A}=  &  \,\int_{-\infty
}^{+\infty}d_{A}^{n}x\,\overline{f(x^{i})}\circledast(h\triangleright
g(x^{j}))\nonumber\\
=  &  \,\int_{-\infty}^{+\infty}d_{A}^{n}x\,h_{(2)}\triangleright
\Big [(\overline{f(x^{i})}\triangleleft h_{(1)})\overset{x}{\circledast
}g(x^{j})\Big ]\nonumber\\
=  &  \,\int_{-\infty}^{+\infty}d_{A}^{n}x\,\epsilon(h_{(2)})\Big [(\overline
{f(x^{i})}\triangleleft h_{(1)})\overset{x}{\circledast}g(x^{j}%
)\Big ]\nonumber\\
=  &  \,\int_{-\infty}^{+\infty}d_{A}^{n}x\,\Big [(\overline{f(x^{i}%
)}\triangleleft h)\overset{x}{\circledast}g(x^{j})\Big ]\nonumber\\
=  &  \,\int_{-\infty}^{+\infty}d_{A}^{n}x\,\Big [(\overline{\overline
{h}\triangleright f(x^{i})})\overset{x}{\circledast}g(x^{j})\Big ]\nonumber\\
=  &  \,\big \langle\overline{h}\triangleright f,g\big \rangle_{A}.
\label{CalAdh1}%
\end{align}
Notice that for the third equality we used the fact that q-integrals over the
whole space behave like scalars (see the discussion in\ Ref. \cite{qAn}). In
very much the same way we have%
\begin{align}
\big \langle f\triangleleft h,g\big \rangle_{A}^{\prime}=  &  \,\int_{-\infty
}^{+\infty}d_{A}^{n}x\,(f(x^{i})\triangleleft h)\overset{x}{\circledast
}\overline{g(x^{j})},\nonumber\\
=  &  \,\int_{-\infty}^{+\infty}d_{A}^{n}x\,\Big [f(x^{i})\overset
{x}{\circledast}(h_{(1)}\triangleright\overline{g(x^{j})})\Big ]\triangleleft
h_{(2)}\nonumber\\
=  &  \,\int_{-\infty}^{+\infty}d_{A}^{n}x\,S^{-1}(h_{(2)})\Big [f(x^{i}%
)\circledast(h_{(1)}\triangleright\overline{g(x^{j})})\Big ]\nonumber\\
=  &  \,\int_{-\infty}^{+\infty}d_{A}^{n}x\,\epsilon(S^{-1}(h_{(2)}%
))\Big [f(x^{i})\overset{x}{\circledast}(h_{(1)}\triangleright\overline
{g(x^{j})})\Big ]\nonumber\\
=  &  \,\int_{-\infty}^{+\infty}d_{A}^{n}x\,\Big [f(x^{i})\overset
{x}{\circledast}(h\triangleright\overline{g(x^{j})})\Big ]\nonumber\\
=  &  \,\int_{-\infty}^{+\infty}d_{A}^{n}x\,\Big [f(x^{i})\overset
{x}{\circledast}(\overline{g(x^{j})\triangleleft\overline{h}}%
)\Big ]\nonumber\\
=  &  \,\big \langle f,g\triangleleft\overline{h}\big \rangle_{A}^{\prime}.
\label{CalAdh2}%
\end{align}

Recalling that integrals over the whole space are invariant under q-deformed
translations, we are able to modify the above considerations in a way that
they carry over to the algebra of partial derivatives. In doing so, we have to
take attention of the fact that for actions of partial derivatives it holds
\cite{qAn}
\begin{align}
\overline{\partial^{i}\triangleright f(x^{j})}  &  =\overline{f(x^{j})}%
\,\bar{\triangleleft}\,\overline{\partial^{i}}, & \overline{f(x^{j}%
)\,\bar{\triangleleft}\,\partial^{i}}  &  =\overline{\partial^{i}%
}\triangleright\overline{f(x^{j})},\nonumber\\
\overline{\hat{\partial}^{i}\,\bar{\triangleright}\,f(x^{j})}  &
=\overline{f(x^{j})}\triangleleft\overline{\hat{\partial}^{i}}, &
\overline{f(x^{j})\triangleleft\hat{\partial}^{i}}  &  =\overline
{\hat{\partial}^{i}}\,\bar{\triangleright}\,\overline{f(x^{j})}.
\end{align}
Applying similar arguments as in (\ref{CalAdh1}) and (\ref{CalAdh2}) we should
now end up with the equations%
\begin{align}
\big \langle f,\partial^{i}\triangleright g\big \rangle_{A}  &
=\big \langle\overline{\partial^{i}}\,\bar{\triangleright}%
\,f,g\big \rangle_{A},\nonumber\\
\big \langle f,\hat{\partial}^{i}\,\bar{\triangleright}\,g\big \rangle_{A}  &
=\big \langle\overline{\hat{\partial}^{i}}\triangleright f,g\big \rangle_{A}%
,\\[0.16in]
\big \langle f\triangleleft\hat{\partial}^{i},g\big \rangle_{A}^{\prime}  &
=\big \langle f,g\,\bar{\triangleleft}\,\overline{\hat{\partial}^{i}%
}\big \rangle_{A}^{\prime},\nonumber\\
\big \langle f\,\bar{\triangleleft}\,\partial^{i},g\big \rangle_{A}^{\prime}
&  =\big \langle f,g\triangleleft\overline{\partial^{i}}\big \rangle_{A}%
^{\prime}.
\end{align}

Without any difficulties our reasonings also apply for the symmetrical
sesquilinear forms defined in (\ref{SymSes1}) and (\ref{SymSes2}). In this
manner, we obtain%
\begin{align}
\big \langle f,h\triangleright g\big \rangle_{i}  &  =\big \langle\overline
{h}\triangleright f,g\big \rangle_{i},\nonumber\\
\big \langle f\triangleleft h,g\big \rangle_{i}^{\prime}  &
=\big \langle f,g\triangleleft\overline{h}\big \rangle_{i}^{\prime},
\end{align}
and%
\begin{align}
\big \langle f,\partial^{j}\triangleright g\big \rangle_{i}  &
=\big \langle\overline{\partial^{j}}\,\bar{\triangleright}%
\,f,g\big \rangle_{i},\nonumber\\
\big \langle f,\hat{\partial}^{j}\,\bar{\triangleright}\,g\big \rangle_{i}  &
=\big \langle\overline{\hat{\partial}^{j}}\triangleright f,g\big \rangle_{i}%
,\\[0.16in]
\big \langle f\triangleleft\hat{\partial}^{j},g\big \rangle_{i}^{\prime}  &
=\big \langle f,g\,\bar{\triangleleft}\,\overline{\hat{\partial}^{j}%
}\big \rangle_{i}^{\prime},\nonumber\\
\big \langle f\,\bar{\triangleleft}\,\partial^{j},g\big \rangle_{i}^{\prime}
&  =\big \langle f,g\triangleleft\overline{\partial^{j}}\big \rangle_{i}%
^{\prime}.
\end{align}

From what we have done so far we can read off the adjoints of symmetry
generators acting on a quantum space algebra. Our results tell us that they
are given by the\ hermitian conjugates of the corresponding operators:
\begin{align}
(h\,\triangleright)^{\dag_{A}}  &  =\overline{h}\,\triangleright, &
(h\,\triangleright)^{\dag_{i}}  &  =\overline{h}\,\triangleright, &
(\triangleleft\,h)^{\dag_{A}^{\prime}}  &  =\triangleleft\,\overline{h}, &
(\triangleleft\,h)^{\dag_{i}^{\prime}}  &  =\triangleleft\,\overline
{h},\nonumber\\
(\partial^{k}\,\triangleright)^{\dag_{A}}  &  =\overline{\partial^{k}}%
\,\bar{\triangleright}, & (\partial^{k}\,\triangleright)^{\dag_{i}}  &
=\overline{\partial^{k}}\,\bar{\triangleright}, & (\triangleleft
\,\hat{\partial}^{k})^{\dag_{A}^{\prime}}  &  =\bar{\triangleleft}%
\,\overline{\hat{\partial}^{k}}, & (\triangleleft\,\hat{\partial}^{k}%
)^{\dag_{i}^{\prime}}  &  =\bar{\triangleleft}\,\overline{\hat{\partial}^{k}%
},\nonumber\\
(\hat{\partial}^{k}\,\bar{\triangleright})^{\dag_{A}}  &  =\overline
{\hat{\partial}^{k}}\,\triangleright, & (\hat{\partial}^{k}\,\bar
{\triangleright})^{\dag_{i}}  &  =\overline{\hat{\partial}^{k}}%
\,\triangleright, & (\triangleleft\,\hat{\partial}^{k})^{\dag_{A}^{\prime}}
&  =\triangleleft\,\overline{\partial^{k}}, & (\triangleleft\,\hat{\partial
}^{k})^{\dag_{i}^{\prime}}  &  =\triangleleft\,\overline{\partial^{k}}.
\label{AdjOp}%
\end{align}

In the case of partial derivatives we have to be aware of the fact that their
adjoints come along with representations being different from those of the
original operator. This observation plays an important role if we are looking
for \textit{self-adjoint operators} in the algebra of partial derivatives. In
this respect, let us recall that in our approach the operation of conjugation
maps the algebra of partial derivatives onto itself. (For a discussion of this
subject and alternative approaches we refer the reader to Refs. \cite{qAn,
Maj94star, OZ92, Fio04, Maj95star}.) However, if we consider actions of
partial derivatives things become more involved, since each partial derivative
can act in different ways upon a quantum space algebra and these actions are
intertwined by the operation of conjugation. On these grounds we have to
combine different\ types of actions if we want to deal with momentum operators
being self-adjoint in the usual sense.

For these ideas to become more clear, we introduce as actions of momentum
operators%
\begin{align}
P^{j}\triangleright f &  \equiv\frac{\text{i}}{2}(\partial^{j}\triangleright
f+\partial^{j}\,\bar{\triangleright}\,f),\nonumber\\
f\triangleleft P^{j} &  \equiv\frac{\text{i}}{2}(f\triangleleft\partial
^{j}+f\,\bar{\triangleleft}\,\partial^{j}).
\end{align}
It is not very difficult to show that we have%
\begin{align}
\overline{P^{j}\triangleright f} &  =\overline{f}\triangleleft\overline{P^{j}%
}=\overline{f}\triangleleft P_{j},\nonumber\\
\overline{f\triangleleft P^{j}} &  =\overline{P^{j}}\triangleright
f=\overline{f}\triangleleft P_{j}.
\end{align}
These identities, in turn, imply
\begin{align}
\overline{\big \langle f,P^{j}\triangleright f\big \rangle}_{i} &
=\big \langle P^{j}\triangleright f,f\big \rangle_{i}\nonumber\\
&  =\big \langle f,\overline{P^{j}}\triangleright f\big \rangle_{i}\nonumber\\
&  =\big \langle f,P_{j}\triangleright f\big \rangle_{i},
\end{align}
and, likewise,%
\begin{equation}
\overline{\big \langle f\triangleleft P^{j},f\big \rangle_{i}^{\prime}%
}=\big \langle f,f\triangleleft P_{j}\big \rangle_{i}^{\prime}.
\end{equation}
As a next step we are looking for a change of basis%
\begin{equation}
\tilde{P}^{j}=\sum_{k}a_{k}P^{k},
\end{equation}
leading to the real components of momentum, i.e.
\begin{equation}
\overline{\tilde{P}^{j}}=\tilde{P}^{j}.
\end{equation}
For the quantum spaces we are dealing with the explicit form of this change of
bases can be found in Appendix \ref{AppQuan}. Finally, a little thought shows
us that the expectation values of the new components of momentum become real
(for a systematic treatment of expectation values see Part II of this paper):%
\begin{align}
\overline{\big \langle f,\tilde{P}^{j}\triangleright f\big \rangle}_{i} &
=\big \langle f,\tilde{P}^{j}\triangleright f\big \rangle_{i},\nonumber\\
\overline{\big \langle f\triangleleft\tilde{P}^{j},f\big \rangle}_{i}^{\prime}
&  =\big \langle f\triangleleft\tilde{P}^{j},f\big \rangle_{i}^{\prime}.
\end{align}

Up to now we considered operators generating infinitesimal symmetry
transformations on quantum spaces and derived the corresponding adjoint
operators. Now, we would like to do the same for operators generating finite
symmetry transformations. To reach this goal let us first recall that finite
symmetry transformations like q-deformed rotations or q-deformed Lorentz
transformations are described by the so-called \textit{canonical elements
}\cite{CSW93a}, which are given by%
\begin{align}
C  &  =\exp(\alpha|h)\equiv\sum\nolimits_{a}e_{\alpha}^{a}\otimes e_{h}^{a}%
\in\mathcal{H}^{\ast}\otimes\mathcal{H},\nonumber\\
C^{^{\prime}}  &  =\exp(h|\alpha)\equiv\sum\nolimits_{a}e_{h}^{a}\otimes
e_{\alpha}^{a}\in\mathcal{H}\otimes\mathcal{H}^{\ast}.
\end{align}
Notice that $\mathcal{H}^{\ast}$ denotes the dual Hopf algebra of
$\mathcal{H}$. Moreover, $\{e_{h}^{a}\}$ is a basis of $\mathcal{H}$ and
$\{e_{\alpha}^{a}\}$ is the corresponding dual basis. With the canonical
elements at hand finite symmetry transformations of functions on position
space can be written in the form \cite{CSW93a}%
\begin{align}
\exp(\alpha|h)\triangleright f(x^{i})  &  =\sum\nolimits_{a}e_{\alpha}%
^{a}\otimes(e_{h}^{a}\triangleright f(x^{i})),\nonumber\\
f(x^{i})\triangleleft\exp(h|\alpha)  &  =\sum\nolimits_{a}(f(x^{i}%
)\triangleleft e_{h}^{a})\otimes e_{\alpha}^{a}.
\end{align}

Now, we have everything together to derive the adjoints of the operators
describing finite symmetry transformations. Towards this end, we proceed as
follows:%
\begin{align}
&  \big \langle f(x^{i}),\exp(\alpha|h)\triangleright g(x^{j})\big \rangle_{A}%
=\,\int_{-\infty}^{+\infty}d_{A}^{n}x\,e_{\alpha}^{a}\otimes\overline
{f(x^{i})}\overset{x}{\circledast}(e_{h}^{a}\triangleright g(x^{j}%
))\nonumber\\
&  \qquad=\,\int_{-\infty}^{+\infty}d_{A}^{n}x\,e_{\alpha}^{a}\otimes
(\,\overline{f(x^{i})}\triangleleft e_{h}^{a})\overset{x}{\circledast}%
g(x^{j})\nonumber\\
&  \qquad=\,\int_{-\infty}^{+\infty}d_{A}^{n}x\,(\,\overline{\overline
{e_{\alpha}^{a}}\otimes\overline{e_{h}^{a}}\triangleright f(x^{i})}%
\,)\overset{x}{\circledast}g(x^{j})\nonumber\\
&  \qquad=\,\int_{-\infty}^{+\infty}d_{A}^{n}x\,(\,\overline{e_{\alpha}%
^{a}\otimes S(e_{h}^{a})\triangleright f(x^{i})}\,)\overset{x}{\circledast
}g(x^{j})\nonumber\\
&  \qquad=\big \langle\exp(\alpha|\!\ominus\!h)\triangleright f(x^{i}%
),g(x^{j})\big \rangle_{A}. \label{AdSym1}%
\end{align}
It should be mentioned that in the above calculation we used the property of
the canonical element to be unitary:%
\begin{align}
\overline{C}  &  =\overline{\exp(\alpha|h)}=\sum\nolimits_{a}\overline
{e_{\alpha}^{a}}\otimes\overline{e_{h}^{a}}\nonumber\\
&  =\sum\nolimits_{a}e_{\alpha}^{a}\otimes S(e_{h}^{a})=\exp(\alpha
|\!\ominus\!h)=C^{-1}.
\end{align}
Notice that this property is equivalent to the requirement that%
\begin{align}
\overline{C\triangleright f(x^{i})}  &  =\overline{\exp(\alpha
|h)\triangleright f(x^{i})}=\sum\nolimits_{a}\overline{e_{\alpha}^{a}}%
\otimes\overline{e_{h}^{a}\triangleright f(x^{i})}\nonumber\\
&  =\sum\nolimits_{a}\overline{e_{\alpha}^{a}}\otimes S^{-1}(\overline
{e_{h}^{a}})\triangleright f(x^{i})=\sum\nolimits_{a}e_{\alpha}^{a}%
\otimes\overline{f(x^{i})}\triangleleft e_{h}^{a}\nonumber\\
&  \overset{!}{=}\exp(\alpha|h)\triangleright\overline{f(x^{i})}%
=C\triangleright\overline{f(x^{i})}.
\end{align}
The explicit form of $C^{-1}$ follows from%
\begin{align}
C\cdot C^{-1}  &  =\exp(\alpha|h)\cdot\exp(\alpha|\!\ominus\!h)\nonumber\\
&  =\sum_{a,b}e_{\alpha}^{a}\cdot e_{\alpha}^{b}\otimes e_{h}^{a}\cdot
S(e_{h}^{b})\nonumber\\
&  =\sum_{a}e_{\alpha}^{a}\otimes(e_{h}^{a})_{(1)}\cdot S((e_{h}^{a}%
)_{(2)})\nonumber\\
&  =\sum_{a}e_{\alpha}^{a}\otimes\epsilon(e_{h}^{a})=1\otimes1,
\end{align}%
\begin{align}
C^{-1}\cdot C  &  =\exp(\alpha|\!\ominus\!h)\cdot\exp(\alpha|h)\nonumber\\
&  =\sum_{a,b}e_{\alpha}^{a}\cdot e_{\alpha}^{b}\otimes S(e_{h}^{a})\cdot
e_{h}^{b}\nonumber\\
&  =\sum_{a}e_{\alpha}^{a}\otimes S((e_{h}^{a})_{(1)})\cdot(e_{h}^{a}%
)_{(2)}\nonumber\\
&  =\sum_{a}e_{\alpha}^{a}\otimes\epsilon(e_{h}^{a})=1\otimes1,
\end{align}
where we applied as characteristic properties of the canonical element
\cite{CSW93a}%
\begin{equation}
(\text{id}_{\mathcal{H}^{\ast}}\otimes\epsilon_{\mathcal{H}})\circ
C=\text{id}_{\mathcal{H}^{\ast}}\otimes\text{id}_{\mathcal{H}},\qquad
C_{12}C_{13}=(\text{id}_{\mathcal{H}^{\ast}}\otimes\Delta_{\mathcal{H}})\circ
C.
\end{equation}

Repeating the same steps as in (\ref{AdSym1}) for the sesquilinear forms with
an apostrophe we get%
\begin{align}
&  \big \langle f(x^{i})\triangleleft\exp(h|\alpha),g(x^{j})\big \rangle_{A}%
^{\prime}=\int_{-\infty}^{+\infty}d_{A}^{n}x\,(f(x^{i})\triangleleft e_{h}%
^{a})\overset{x}{\circledast}\overline{g(x^{j})}\otimes e_{\alpha}%
^{a}\nonumber\\
&  \qquad=\int_{-\infty}^{+\infty}d_{A}^{n}x\,f(x^{i})\overset{x}{\circledast
}(e_{h}^{a}\triangleright\overline{g(x^{j})}\,)\otimes e_{\alpha}%
^{a}\nonumber\\
&  \qquad=\int_{-\infty}^{+\infty}d_{A}^{n}x\,f(x^{i})\overset{x}{\circledast
}\overline{g(x^{j})\triangleleft\overline{e_{h}^{a}}\otimes\overline
{e_{\alpha}^{a}}}\nonumber\\
&  \qquad=\int_{-\infty}^{+\infty}d_{A}^{n}x\,f(x^{i})\overset{x}{\circledast
}\overline{g(x^{j})\triangleleft S(e_{h}^{a})\otimes e_{\alpha}^{a}%
}\nonumber\\
&  \qquad=\big \langle f(x^{i}),g(x^{j})\triangleleft\exp(h|\!\ominus
\!\alpha)\big \rangle_{A}^{\prime}.\label{AdSym2}%
\end{align}
The fourth equality in\ the above calculation is a consequence of the
identities%
\begin{align}
\overline{C^{\prime}} &  =\overline{\exp(h|\alpha)}=\sum\nolimits_{a}%
\overline{e_{h}^{a}}\otimes\overline{e_{\alpha}^{a}}\nonumber\\
&  =\sum\nolimits_{a}S(e_{h}^{a})\otimes e_{\alpha}^{a}\nonumber\\
&  =\exp(\ominus h|\alpha)=(C^{\prime})^{-1}.
\end{align}

Finally, it should be mentioned that the results in (\ref{AdSym1}) and
(\ref{AdSym2}) remain unchanged if we use symmetrical sesquilinear forms
instead . In this manner we conclude that
\begin{align}
(\exp(\alpha|h)\triangleright)^{\dag_{A}/\dag_{i}} &  =\overline{\exp
(\alpha|h)}\,\triangleright=\exp(\alpha|\,\ominus\,h),\nonumber\\
(\triangleleft\exp(h|\alpha))^{\dag_{A}^{\prime}/\dag_{i}^{\prime}} &
=\triangleleft\,\overline{\exp(\alpha|h)}=\exp(\ominus h|\alpha).
\end{align}

From Ref. \cite{qAn} we know that q-deformed exponentials generate finite
translations on q-deformed quantum spaces. It is now our aim to find the
corresponding adjoint operators. To achieve this we apply a method similar to
that for quantum group transformations:%
\begin{align}
&  \big \langle f(x^{i}),\exp(y^{k}|\partial^{l})_{\bar{R},L}\overset
{\partial|x}{\triangleright}g(x^{j})\big \rangle_{A}\nonumber\\
&  \qquad=\,\int_{-\infty}^{+\infty}d_{A}^{n}x\,\overline{f(x^{i})}%
\overset{x|y}{\odot}_{\hspace{-0.01in}R}\big (\exp(y^{k}|\partial^{l}%
)_{\bar{R},L}\overset{\partial|x}{\triangleright}g(x^{j})\big )\nonumber\\
&  \qquad=\,\int_{-\infty}^{+\infty}d_{A}^{n}x\,\big (\exp(y^{k}|\!\ominus
_{R}\hspace{-0.01in}\partial^{l})_{\bar{R},L}\overset{\partial|x}%
{\triangleright}\overline{f(x^{i})}\,\big )\overset{x}{\circledast}%
g(x^{j})\nonumber\\
&  \qquad=\,\int_{-\infty}^{+\infty}d_{A}^{n}x\,\big (\overline{f(x^{i}%
)\,\overset{x|\partial}{\bar{\triangleleft}}\,\exp(\ominus_{\bar{L}}%
\,\partial^{l}|y^{k})_{\bar{R},L}}\big )\overset{x}{\circledast}%
g(x^{j})\nonumber\\
&  \qquad=\,\big \langle f(x^{i})\,\overset{x|\partial}{\bar{\triangleleft}%
}\,\exp(\ominus_{\bar{L}}\,\partial^{l}|y^{k})_{\bar{R},L},g(x^{j}%
)\big \rangle_{A}. \label{AdExp1}%
\end{align}
Notice that the second equality follows from the rule for integration by parts
and trivial braiding of q-exponentials. The third equality makes use of the
conjugation properties of q-exponentials and those of partial derivatives. In
very much the same way we obtain%
\begin{align}
&  \big \langle f(x^{i})\,\overset{x|\partial}{\bar{\triangleleft}}%
\,\exp(\partial^{l}|y^{k})_{\bar{R},L},g(x^{j})\big \rangle_{A}^{\prime
}\nonumber\\
&  \qquad=\,\int_{-\infty}^{+\infty}d_{A}^{n}x\,\big (f(x^{i})\,\overset
{x|\partial}{\bar{\triangleleft}}\,\exp(\partial^{l}|y^{k})_{\bar{R}%
,L}\big )\overset{y|x}{\odot}_{\hspace{-0.01in}\bar{L}}\overline{g(x^{j}%
)}\nonumber\\
&  \qquad=\,\int d_{A}^{n}x\,f(x^{i})\overset{x}{\circledast}\big (\overline
{\exp(y^{k}|\!\ominus_{R}\partial^{l})_{\bar{R},L}\overset{\partial
|x}{\triangleright}g(x^{j})}\big )\nonumber\\
&  \qquad=\,\big \langle f(x^{i}),\exp(y^{k}|\!\ominus_{R}\partial^{l}%
)_{\bar{R},L}\overset{\partial|x}{\triangleright}g(x^{j})\big \rangle_{A}%
^{\prime}. \label{AdExp2}%
\end{align}
Applying the replacements
\begin{equation}
L\leftrightarrow\bar{L},\quad R\leftrightarrow\bar{R},\quad\triangleright
\leftrightarrow\bar{\triangleright},\quad\triangleleft\leftrightarrow
\bar{\triangleleft},\quad\partial\leftrightarrow\hat{\partial} \label{SubKon}%
\end{equation}
to the expressions in (\ref{AdExp1}) and (\ref{AdExp2}) yields further
relations. Again, our results carry over to symmetrical sesquilinear forms.
Finally, we can read off from what we have done so far that%
\begin{align}
(\exp(x^{k}|\partial^{l})_{\bar{R},L}\,\overset{\partial|.}{\triangleright
}\,)^{\dag_{A}/\dag_{i}}  &  =\overset{.|\partial}{\bar{\triangleleft}}%
\,\exp(\partial^{l}|\!\ominus_{R}\,x^{k})_{\bar{R},L},\nonumber\\
(\,\overset{.|\partial}{\bar{\triangleleft}}\,\exp(\partial^{l}|x^{k}%
)_{\bar{R},L})^{\dag_{A}^{\prime}/\dagger_{i}^{\prime}}  &  =\exp
(\ominus_{\bar{L}}\,x^{k}|\partial^{l})_{\bar{R},L}\,\overset{\partial
|.}{\triangleright},\\[0.16in]
(\exp(x^{k}|\hat{\partial}^{l})_{R,\bar{L}}\,\overset{\partial|.}%
{\bar{\triangleright}}\,)^{\dag_{A}/\dag_{i}}  &  =\overset{.|\partial
}{\triangleleft}\exp(\hat{\partial}^{l}|\!\ominus_{R}\,x^{k})_{R,\bar{L}%
},\nonumber\\
(\,\overset{.|\partial}{\triangleleft}\exp(\hat{\partial}^{l}|x^{k}%
)_{R,\bar{L}})^{\dag_{A}^{\prime}/\dagger_{i}^{\prime}}  &  =\exp(\ominus
_{L}\,x^{k}|\hat{\partial}^{l})_{R,\bar{L}}\,\overset{\partial|.}%
{\bar{\triangleright}}.
\end{align}

For the sake of completeness let us mention that we can also assign each
position operator an adjoint operator. To see this, we perform the following
calculation:%
\begin{align}
&  \big \langle f,X^{k}\triangleright g\big \rangle_{A}=\int_{-\infty
}^{+\infty}d_{A}^{n}x\,\overline{f(x^{i})}\overset{x}{\circledast}%
(X^{k}\triangleright g(x^{j}))\nonumber\\
&  \qquad=\,\int_{-\infty}^{+\infty}d_{A}^{n}x\,\overline{f(x^{i})}\overset
{x}{\circledast}(X^{k}\overset{x}{\circledast}g(x^{j}))=\,\int_{-\infty
}^{+\infty}d_{A}^{n}x\,(\,\overline{f(x^{i})}\overset{x}{\circledast}%
X^{k})\overset{x}{\circledast}g(x^{j})\nonumber\\
&  \qquad=\,\int_{-\infty}^{+\infty}d_{A}^{n}x\,\overline{X_{k}\overset
{x}{\circledast}f(x^{i})}\overset{x}{\circledast}g(x^{j})=\int_{-\infty
}^{+\infty}d_{A}^{n}x\,\overline{X_{k}\triangleright f(x^{i})}\overset
{x}{\circledast}g(x^{j})\nonumber\\
&  \qquad=\,\big \langle X_{k}\triangleright f,g\big \rangle_{A},
\end{align}
where we used $\overline{X^{k}}=X_{k}.$ Similar considerations show us that%
\begin{equation}
\big \langle f\triangleleft X^{k},g\big \rangle_{A}^{\prime}%
=\big \langle f,g\triangleleft X_{k}\big \rangle_{A}^{\prime}.
\end{equation}
In this manner we find%
\begin{equation}
(X^{k}\,\triangleright)^{\dag_{A}}=(X^{k}\,\triangleright)^{\dag_{i}}%
=X_{k}\,\triangleright,\quad(\triangleleft\,X^{k})^{\dag_{A}^{\prime}%
}=(\triangleleft\,X^{k})^{\dag_{i}^{\prime}}=\triangleleft\,X_{k}.
\end{equation}

Next, we would like to discuss how our sesquilinear forms behave under
symmetry transformations. In Ref. \cite{qAn} it was shown that q-deformed
integrals over the whole quantum space are invariant under quantum group
symmetries. For this reason, we have%
\begin{align}
h\triangleright\big \langle f,g\big \rangle_{A}  &  =\epsilon
(h)\big \langle f,g\big \rangle_{A}\nonumber\\
&  =\int_{-\infty}^{+\infty}d_{A}^{n}x\,h\triangleright\big (\,\overline
{f(x^{i})}\overset{x}{\circledast}g(x^{j})\big )\nonumber\\
&  =\int_{-\infty}^{+\infty}d_{A}^{n}x\,\big (h_{(1)}\triangleright
\overline{f(x^{i})}\,\big )\overset{x}{\circledast}\big (h_{(2)}\triangleright
g(x^{j})\big )\nonumber\\
&  =\int_{-\infty}^{+\infty}d_{A}^{n}x\,\big (\,\overline{f(x^{i}%
)\triangleleft\overline{h_{(1)}}}\,\big )\overset{x}{\circledast}%
\big (h_{(2)}\triangleright g(x^{j})\big )\nonumber\\
&  =\int_{-\infty}^{+\infty}d_{A}^{n}x\,\big (\,\overline{S^{-1}%
(\,\overline{h_{(1)}}\,)\triangleright f(x^{i})}\,\big )\overset
{x}{\circledast}\big (h_{(2)}\triangleright g(x^{j})\big )\nonumber\\
&  =\big \langle S^{-1}(\,\overline{h_{(1)}}\,)\triangleright f,h_{(2)}%
\triangleright g\big \rangle_{A},
\end{align}
where $h$ again denotes an element of a Hopf algebra $\mathcal{H}$ describing
the symmetry of the quantum space under consideration. Similar arguments give
us%
\begin{equation}
\big \langle f,g\big \rangle_{A}\triangleleft h=\epsilon
(h)\big \langle f,g\big \rangle_{A}=\big \langle f\triangleleft S(\,\overline
{h_{(2)}}\,),g\triangleleft h_{(1)}\big \rangle_{A},
\end{equation}
and%
\begin{align}
h\triangleright\big \langle f,g\big \rangle_{A}^{\prime}  &  =\epsilon
(h)\big \langle f,g\big \rangle_{A}^{\prime}=\big \langle h_{(1)}%
\triangleright f,S^{-1}(\,\overline{h_{(2)}}\,)\triangleright
g\big \rangle_{A}^{\prime},\nonumber\\
\big \langle f,g\big \rangle_{A}^{\prime}\triangleleft h  &  =\epsilon
(h)\big \langle f,g\big \rangle_{A}^{\prime}=\big \langle f\triangleleft
h_{(2)},g\triangleleft S(\,\overline{h_{(1)}}\,)\big \rangle_{A}^{\prime}.
\end{align}
In the same way translation invariance of integrals over the whole space leads
to
\begin{align}
\partial^{k}\triangleright\big \langle f,g\big \rangle_{A}  &  =\epsilon
_{L}(\partial^{k})\big \langle f,g\big \rangle_{A}=\big \langle f\,\bar
{\triangleleft}\,\overline{(\partial^{k})_{(1)}},(\partial^{k})_{(2)}%
\triangleright g\big \rangle_{A},\nonumber\\
\hat{\partial}^{k}\,\bar{\triangleright}\,\big \langle f,g\big \rangle_{A}  &
=\epsilon_{\bar{L}}(\hat{\partial}^{k})\big \langle f,g\big \rangle_{A}%
=\big \langle f\triangleleft\overline{(\hat{\partial}^{k})_{(\bar{1})}}%
,(\hat{\partial}^{k})_{(\bar{2})}\,\bar{\triangleright}\,g\big \rangle_{A}%
,\label{InvAbl1}\\[0.16in]
\big \langle f,g\big \rangle_{A}\triangleleft\hat{\partial}^{k}  &
=\epsilon_{R}(\hat{\partial}^{k})\big \langle f,g\big \rangle_{A}%
=\big \langle\overline{(\hat{\partial}^{k})_{(2)}}\,\bar{\triangleright
}\,f,g\triangleleft(\hat{\partial}^{k})_{(1)}\big \rangle_{A},\nonumber\\
\big \langle f,g\big \rangle_{A}\,\bar{\triangleleft}\,\partial^{k}  &
=\epsilon_{\bar{R}}(\partial^{k})\big \langle f,g\big \rangle_{A}%
=\big \langle\,\overline{(\partial^{k})_{(\bar{2})}}\triangleright
f,g\,\bar{\triangleleft}\,(\partial^{k})_{(\bar{1})}\big \rangle_{A},
\end{align}
and
\begin{align}
\big \langle f,g\big \rangle_{A}^{\prime}\triangleleft\hat{\partial}^{k}  &
=\epsilon(\hat{\partial}^{k})\big \langle f,g\big \rangle_{A}^{\prime
}=\big \langle f\triangleleft(\hat{\partial}^{k})_{(2)},\overline
{(\hat{\partial}^{k})_{(1)}}\,\bar{\triangleright}\,g\big \rangle_{A}^{\prime
},\nonumber\\
\big \langle f,g\big \rangle_{A}^{\prime}\,\bar{\triangleleft}\,\partial^{k}
&  =\bar{\epsilon}(\partial^{k})\big \langle f,g\big \rangle_{A}^{\prime
}=\big \langle f\,\bar{\triangleleft}\,(\partial^{k})_{(\bar{2})}%
,\overline{(\partial^{k})_{(\bar{1})}}\triangleright g\big \rangle_{A}%
^{\prime},\\[0.16in]
\partial^{k}\triangleright\big \langle f,g\big \rangle_{A}^{\prime}  &
=\epsilon(\partial^{k})\big \langle f,g\big \rangle_{A}^{\prime}%
=\big \langle(\partial^{k})_{(1)}\triangleright f,g\,\bar{\triangleleft
}\,\overline{(\partial^{k})_{(2)}}\,\big \rangle_{A}^{\prime},\nonumber\\
\hat{\partial}^{k}\,\bar{\triangleright}\,\big \langle f,g\big \rangle_{A}%
^{\prime}  &  =\bar{\epsilon}(\hat{\partial}^{k}%
)\big \langle f,g\big \rangle_{A}^{\prime}=\big \langle(\hat{\partial}%
^{k})_{(\bar{1})}\,\bar{\triangleright}\,f,g\triangleleft\overline
{(\hat{\partial}^{k})_{(\bar{2})}}\,\big \rangle_{A}^{\prime}, \label{InvAbl2}%
\end{align}
where
\begin{equation}
\Delta_{L}(\partial)=\partial_{(1)}\otimes\partial_{(2)},\quad\Delta_{\bar{L}%
}(\partial)=\partial_{(\bar{1})}\otimes\partial_{(\bar{2})}.
\end{equation}
Notice that all of the above expressions in (\ref{InvAbl1})-(\ref{InvAbl2})
vanish, since $\epsilon_{A}(\partial)=0$ [cf. Eq. (\ref{Coeins})].

Next we try to extend these considerations to finite symmetry transformations.
For the action of the canonical element on a sesquilinear form we have%
\begin{align}
\exp(\alpha|h)\triangleright\big \langle f,g\big \rangle_{A}  &
=\sum\nolimits_{a}e_{\alpha}^{a}\otimes e_{h}^{a}\triangleright
\big \langle f,g\big \rangle_{A}\nonumber\\
&  =\sum\nolimits_{a}e_{\alpha}^{a}\otimes\epsilon(e_{h}^{a}%
)\big \langle f,g\big \rangle_{A}\nonumber\\
&  =\big \langle f,g\big \rangle_{A}, \label{InvSes1}%
\end{align}
and%
\begin{align}
\big \langle f,g\big \rangle_{A}\triangleleft\exp(h|\alpha)  &  =\sum
\nolimits_{a}\big \langle f,g\big \rangle_{A}\triangleleft e_{h}^{a}\otimes
e_{\alpha}^{a}\nonumber\\
&  =\sum\nolimits_{a}\big \langle f,g\big \rangle_{A}\epsilon(e_{h}%
^{a})\otimes e_{\alpha}^{a}\nonumber\\
&  =\big \langle f,g\big \rangle_{A}. \label{InvSes2}%
\end{align}
These equalities also hold for all other types of sesquilinear forms
introduced in the previous subsection. They tell us once more that our
sesquilinear forms behave like scalars.

The results in (\ref{InvSes1}) and (\ref{InvSes2}) can be used to show once
more that finite quantum group transformations described by canonical elements
are unitary:%
\begin{align}
\big \langle f,g\big \rangle_{A}  &  =\exp(\alpha|h)\triangleright
\big \langle f,g\big \rangle_{A}=\sum\nolimits_{a}e_{\alpha}^{a}\otimes
e_{h}^{a}\triangleright\big \langle f,g\big \rangle_{A}\nonumber\\
&  =\int_{-\infty}^{+\infty}d_{A}^{n}x\,\sum\nolimits_{a}e_{\alpha}^{a}%
\otimes\overline{f(x^{i})\triangleleft\overline{(e_{h}^{a})_{(1)}}}\overset
{x}{\circledast}((e_{h}^{a})_{(2)}\triangleright g(x^{j}))\nonumber\\
&  =\int_{-\infty}^{+\infty}d_{A}^{n}x\,\sum\nolimits_{a,b}e_{\alpha}%
^{a}e_{\alpha}^{b}\otimes\overline{f(x^{i})\triangleleft\overline{e_{h}^{a}}%
}\overset{x}{\circledast}(e_{h}^{b}\triangleright g(x^{j}))\nonumber\\
&  =\int_{-\infty}^{+\infty}d_{A}^{n}x\,\overline{\sum\nolimits_{a}%
\overline{e_{\alpha}^{a}}\otimes S^{-1}(\,\overline{e_{h}^{a}}%
\,)\triangleright f(x^{i})}\cdot\sum\nolimits_{b}e_{\alpha}^{b}\otimes
e_{h}^{b}\triangleright g(x^{j})\nonumber\\
&  =\int_{-\infty}^{+\infty}d_{A}^{n}x\,\overline{\sum\nolimits_{a}e_{\alpha
}^{a}\otimes e_{h}^{a}\triangleright f(x^{i})}\cdot\exp(\alpha
|h)\triangleright g(x^{j})\nonumber\\
&  =\int_{-\infty}^{+\infty}d_{A}^{n}x\,\overline{\exp(\alpha|h)\triangleright
f(x^{i})}\cdot\exp(\alpha|h)\triangleright g(x^{j})\nonumber\\
&  =\big \langle\exp(\alpha|h)\triangleright f(x^{i}),\exp(\alpha
|h)\triangleright g(x^{j})\big \rangle_{A}. \label{UnQuanG1}%
\end{align}
Likewise, we have%
\begin{align}
\big \langle f,g\big \rangle_{A}  &  =\big \langle f,g\big \rangle_{A}%
\triangleleft\exp(h|\!\ominus\!\alpha)\nonumber\\
&  =\int_{-\infty}^{+\infty}d_{A}x\sum\nolimits_{a}\overline{f(x^{i}%
)\triangleleft S(\,\overline{(e_{h}^{a})_{(2)}}\,)}\overset{x}{\circledast
}(g(x^{j})\triangleleft(e_{h}^{a})_{(1)})\otimes S(e_{\alpha}^{a})\nonumber\\
&  =\int_{-\infty}^{+\infty}d_{A}x\sum\nolimits_{a,b}\overline{f(x^{i}%
)\triangleleft S(\,\overline{e_{h}^{a}}\,)}\overset{x}{\circledast}%
(g(x^{j})\triangleleft e_{h}^{b})\otimes S(e_{\alpha}^{a})S(e_{\alpha}%
^{b})\nonumber\\
&  =\int_{-\infty}^{+\infty}d_{A}x\overline{\sum\nolimits_{a}f(x^{i}%
)\triangleleft S(\,\overline{e_{h}^{a}}\,)\otimes S^{-1}(\,\overline
{e_{\alpha}^{a}}\,)}\cdot\sum\nolimits_{b}g(x^{j})\triangleleft e_{h}%
^{b}\otimes S(e_{\alpha}^{b})\nonumber\\
&  =\int_{-\infty}^{+\infty}d_{A}x\overline{\sum\nolimits_{a}f(x^{i}%
)\triangleleft S(e_{h}^{a})\otimes e_{\alpha}^{a}}\cdot(g(x^{j})\triangleleft
\exp(h|\!\ominus\!\alpha))\nonumber\\
&  =\int_{-\infty}^{+\infty}d_{A}x\overline{\sum\nolimits_{a}f(x^{i}%
)\triangleleft\exp(h|\!\ominus\!\alpha)}\cdot(g(x^{j})\triangleleft
\exp(h|\!\ominus\!\alpha))\nonumber\\
&  =\big \langle f(x^{i})\triangleleft\exp(h|\!\ominus\!\alpha),g(x^{j}%
)\triangleleft\exp(h|\!\ominus\!\alpha)\big \rangle_{A}, \label{UnQuanG2}%
\end{align}
where
\begin{align}
\exp(h|\!\ominus\!\alpha)  &  =\sum\nolimits_{a}e_{h}^{a}\otimes S(e_{\alpha
}^{a})\nonumber\\
&  =\sum\nolimits_{a}S(e_{h}^{a})\otimes e_{\alpha}^{a}=\exp(\ominus
h|\alpha).
\end{align}
For the fourth step in (\ref{UnQuanG1}) as well as the third step in
(\ref{UnQuanG2}) we used the characteristic properties of the canonical
element and the antihomomorphism property of the antipode. For the fourth step
in (\ref{UnQuanG2}) one has to realize that $\overline{S(h)}=S^{-1}%
(\,\overline{h}\,).$ The remaining steps in (\ref{UnQuanG2}) make use of the
identities%
\begin{equation}
\sum\nolimits_{a}\overline{e_{h}^{a}}\otimes\overline{e_{\alpha}^{a}}%
=\sum\nolimits_{a}S(e_{h}^{a})\otimes e_{\alpha}^{a}=\sum\nolimits_{a}%
e_{h}^{a}\otimes S(e_{\alpha}^{a}).
\end{equation}
The following relations can be derived along the same line of reasonings:%
\begin{align}
\big \langle f,g\big \rangle_{A}^{\prime}  &  =\exp(\alpha|h)\triangleright
\big \langle f,g\big \rangle_{A}^{\prime}\nonumber\\
&  =\big \langle\exp(\alpha|h)\triangleright f,\exp(\alpha|h)\triangleright
g\big \rangle_{A}^{\prime},\\[0.1in]
\big \langle f,g\big \rangle_{A}^{\prime}  &
=\big \langle f,g\big \rangle_{A}^{\prime}\triangleleft\exp(h|\!\ominus
\!\alpha)\nonumber\\
&  =\big \langle f\triangleleft\exp(h|\!\ominus\!\alpha),g\triangleleft
\exp(h|\!\ominus\!\alpha)\big \rangle_{A}^{\prime}. \label{UnQuanG4}%
\end{align}
Last but not least, it should be mentioned that these reasonings about
unitarity of finite quantum group transformations also apply\ to symmetrical
sesquilinear forms.

It arises the question whether the operators generating finite
translations\ on quantum spaces\ are also subject to unitary conditions like
those in (\ref{UnQuanG1})-(\ref{UnQuanG4}). That this is indeed the case can
be shown by the following calculation:%
\begin{align*}
\big \langle f,g\big \rangle_{A}  &  =\exp(y^{k}|\partial^{l})_{\bar{R}%
,L}\triangleright\big \langle f,g\big \rangle_{A}\\
&  =\int_{-\infty}^{+\infty}d_{A}^{n}x\,\exp(y^{k}|\partial^{l})_{\bar{R}%
,L}\overset{\partial|x}{\triangleright}\big (\,\overline{f(x^{i})}\overset
{x}{\circledast}g(x^{j})\big )\\
&  =\int_{-\infty}^{+\infty}d_{A}^{n}x\,\sum\nolimits_{a}e_{(y,\bar{R})}%
^{a}\otimes((e_{(\partial,L)}^{a})_{(1)}\triangleright\overline{f(x^{i}%
)}\,)\overset{x}{\circledast}((e_{(\partial,L)}^{a})_{(2)}\triangleright
g(x^{j}))\\
&  =\int_{-\infty}^{+\infty}d_{A}^{n}x\,\sum\nolimits_{a,b}e_{(y,\bar{R})}%
^{a}\overset{y}{\circledast}e_{(y,\bar{R})}^{b}\otimes(e_{(\partial,L)}%
^{b}\triangleright\overline{f(x^{i})}\,)\overset{x}{\circledast}%
(e_{(\partial,L)}^{a}\triangleright g(x^{j}))\\
&  =\int_{-\infty}^{+\infty}d_{A}^{n}x\,\sum\nolimits_{a,b}e_{(y,\bar{R})}%
^{a}\overset{y}{\circledast}\overline{f(x^{i})\,\bar{\triangleleft
}\,e_{(\partial,\bar{R})}^{b}\otimes e_{(y,L)}^{b}}\overset{x}{\circledast
}(e_{(\partial,L)}^{a}\triangleright g(x^{j}))\\
&  =\big \langle f(x^{i})\,\overset{x|\partial}{\bar{\triangleleft}}%
\,\exp(\partial^{r}|y^{m})_{\bar{R},L},\exp(y^{k}|\partial^{l})_{\bar{R}%
,L}\triangleright g(x^{j})\big \rangle_{A}\\
&  =\big \langle f(x^{i})\,\overset{x|\partial}{\bar{\triangleleft}%
}\,\overline{\exp(y^{m}|\partial^{r})_{\bar{R},L}},\exp(y^{k}|\partial
^{l})_{\bar{R},L}\triangleright g(x^{j})\big \rangle_{A}.
\end{align*}
The second equality is due to translation invariance of q-deformed integrals
over the whole space. The fourth equality is an application of the addition
law for q-deformed exponentials. The fifth equality results from\ the
conjugation properties of q-deformed exponentials and the last two expressions
can be viewed as a kind of shorthand notation. With the same reasonings we get%
\begin{align}
\big \langle f,g\big \rangle_{A}^{\prime}  &
=\big \langle f,g\big \rangle_{A}^{\prime}\triangleleft\exp(\partial^{l}%
|y^{k})_{\bar{R},L}\nonumber\\
&  =\int_{-\infty}^{+\infty}d_{A}^{n}x\,\sum\nolimits_{a,b}(f(x^{i}%
)\,\bar{\triangleleft}\,e_{(\partial,\bar{R})}^{a})\overset{x}{\circledast
}\overline{e_{(y,\bar{R})}^{b}\otimes e_{(\partial,L)}^{b}\triangleright
g(x^{j})}\overset{y}{\circledast}e_{(y,L)}^{a}\nonumber\\
&  =\big \langle f(x^{i})\,\overset{x|\partial}{\bar{\triangleleft}}%
\,\exp(\partial^{r}|y^{m})_{\bar{R},L},\exp(y^{k}|\partial^{l})_{\bar{R}%
,L}\overset{\partial|x}{\triangleright}g(x^{j})\big \rangle_{A}^{\prime
}\nonumber\\
&  =\big \langle f(x^{i})\,\overset{x|\partial}{\bar{\triangleleft}}%
\,\exp(\partial^{r}|y^{m})_{\bar{R},L},\overline{\exp(\partial^{l}%
|y^{k})_{\bar{R},L}}\overset{\partial|x}{\triangleright}g(x^{j}%
)\big \rangle_{A}^{\prime}.
\end{align}
Further relations are obtained most easily from the above ones most easily via
the substitutions in (\ref{SubKon}). Again, we are allowed to replace the
sesquilinear forms by their symmetrical versions.

\subsection{Fourier-Plancherel identities}

It is natural to ask about the behavior of our sesquilinear forms under
q-deformed Fourier transformations of their arguments. This question leads us
to q-analogs of Fourier-Plancherel identities, for which we concretely have%
\begin{align}
\big \langle f,g\big \rangle_{L,x}^{\prime}  &  =(-1)^{n}%
\big \langle\mathcal{F}_{L}(f),\mathcal{F}_{\bar{R}}^{\ast}(g)(\kappa
^{-1}p^{i})\big \rangle_{\bar{R},p}^{\prime},\nonumber\\
\big \langle f,g\big \rangle_{\bar{L},x}^{\prime}  &  =(-1)^{n}%
\big \langle\mathcal{F}_{\bar{L}}(f),\mathcal{F}_{R}^{\ast}(g)(\kappa
p^{i})\big \rangle_{R,p}^{\prime},\label{FPId1}\\[0.1in]
\big \langle f,g\big \rangle_{R,x}^{\prime}  &  =(-1)^{n}%
\big \langle\mathcal{F}_{R}^{\ast}(f)(\kappa p^{i}),\mathcal{F}_{\bar{L}%
}(g)\big \rangle_{\bar{L},p}^{\prime},\nonumber\\
\big \langle f,g\big \rangle_{\bar{R},x}^{\prime}  &  =(-1)^{n}%
\big \langle\mathcal{F}_{\bar{R}}^{\ast}(f)(\kappa^{-1}p^{i}),\mathcal{F}%
_{L}(g)\big \rangle_{L,p}^{\prime}, \label{FPId2}%
\end{align}
and%
\begin{align}
\big \langle f,g\big \rangle_{L,x}  &  =(-1)^{n}\big \langle\mathcal{F}%
_{\bar{R}}(f),\mathcal{F}_{L}^{\ast}(g)(\kappa^{-1}p^{i})\big \rangle_{\bar
{R},p},\nonumber\\
\big \langle f,g\big \rangle_{\bar{L},x}  &  =(-1)^{n}\big \langle\mathcal{F}%
_{R}(f),\mathcal{F}_{\bar{L}}^{\ast}(g)(\kappa p^{i})\big \rangle_{R,p}%
,\label{FPId3}\\[0.1in]
\big \langle f,g\big \rangle_{R,x}  &  =(-1)^{n}\big \langle\mathcal{F}%
_{\bar{L}}^{\ast}(f)(\kappa p^{i}),\mathcal{F}_{R}(g)\big \rangle_{\bar{L}%
,p},\nonumber\\
\big \langle f,g\big \rangle_{\bar{R},x}  &  =(-1)^{n}\big \langle\mathcal{F}%
_{L}^{\ast}(f)(\kappa^{-1}p^{i}),\mathcal{F}_{\bar{R}}(g)\big \rangle_{L,p}.
\label{FPId4}%
\end{align}

Now, we come to the proofs of the above relations. First of all, we consider
the first identity in (\ref{FPId1}), which can be proven by the following
calculation:%
\begin{align}
&  \text{vol}_{L}\,\big \langle f,g\big \rangle_{L,x}^{\prime}=\text{vol}%
_{L}\int_{-\infty}^{+\infty}d_{L}^{n}x\,f(x^{i})\circledast\overline{g(x^{j}%
)}\nonumber\\
&  \qquad=\,\int_{-\infty}^{+\infty}d_{L}^{n}x\,\Big (\int_{-\infty}^{+\infty
}d_{L}^{n}y\,f(y^{i})\overset{y|x}{\odot}_{\hspace{-0.01in}\bar{L}}%
\delta_{\bar{R}}^{n}((\ominus_{\bar{R}}\,x^{j})\oplus_{\bar{L}}y^{k}%
)\Big )\overset{x}{\circledast}\overline{g(x^{l})}\nonumber\\
&  \qquad=\,\int_{-\infty}^{+\infty}d_{L}^{n}x\,\Big (\int_{-\infty}^{+\infty
}d_{L}^{n}y\,f(y^{i})\nonumber\\
&  \qquad\qquad\qquad\overset{y|x}{\odot}_{\hspace{-0.01in}\bar{L}}%
\int_{-\infty}^{+\infty}d_{\bar{R}}^{n}p\,\exp((\ominus_{\bar{R}}%
\,x^{j})\oplus_{\bar{L}}y^{k}|\text{i}^{-1}p^{m})_{\bar{R},L}\Big )\overset
{x}{\circledast}\overline{g(x^{l})}\nonumber\\
&  \qquad=\,\int_{-\infty}^{+\infty}d_{L}^{n}x\,\int d_{L}^{n}y\,f(y^{i}%
)\overset{y}{\circledast}\Big (\int_{-\infty}^{+\infty}d_{\bar{R}}^{n}%
p\,\exp(y^{k}|\text{i}^{-1}p^{m})_{\bar{R},L}\nonumber\\
&  \qquad\qquad\qquad\overset{p|x}{\odot}_{\hspace{-0.01in}\bar{L}}%
\exp((\ominus_{\bar{R}}\,\kappa^{-1}x^{j})|\text{i}^{-1}p^{r})_{\bar{R}%
,L}\Big )\overset{x}{\circledast}\overline{g(x^{l})}\nonumber\\
&  \qquad=\,\int_{-\infty}^{+\infty}d_{L}^{n}y\,f(y^{i})\overset
{y}{\circledast}e_{(\bar{R},y)}^{a}\otimes\int_{-\infty}^{+\infty}d_{L}%
^{n}x\,\bar{S}^{\,\,-1}\big (\mathcal{R}_{[2]}\triangleright e_{(\bar
{R},\kappa^{-1}x)}^{b}\big )\overset{x}{\circledast}\overline{g(x^{l}%
)}\nonumber\\
&  \qquad\qquad\qquad\otimes\int_{-\infty}^{+\infty}d_{\bar{R}}^{n}%
p\,\big (\mathcal{R}_{[1]}\triangleright e_{(L,p)}^{a}\big )\overset
{p}{\circledast}e_{(L,p)}^{b}\nonumber\\
&  \qquad=\,\int_{-\infty}^{+\infty}d_{L}^{n}y\,f(y^{i})\overset
{y}{\circledast}e_{(\bar{R},y)}^{a}\otimes\int_{-\infty}^{+\infty}d_{\bar{R}%
}^{n}p\,e_{(L,p)}^{a}\overset{p}{\circledast}\big (\mathcal{R}_{[1]}%
^{-1}\triangleright e_{(L,p)}^{b}\big )\nonumber\\
&  \qquad\qquad\qquad\otimes\int_{-\infty}^{+\infty}d_{L}^{n}x\,\bar{S}%
^{\,-1}\big (\mathcal{R}_{[2]}^{-1}\triangleright e_{(\bar{R},x)}%
^{b}\big )\overset{x}{\circledast}\overline{g(x^{l})}\nonumber\\
&  \qquad=\,\int_{-\infty}^{+\infty}d_{L}^{n}y\,f(y^{i})\overset
{y}{\circledast}e_{(\bar{R},y)}^{a}\otimes\int_{-\infty}^{+\infty}d_{\bar{R}%
}^{n}p\,e_{(L,p)}^{a}\nonumber\\
&  \qquad\qquad\qquad\overset{p}{\circledast}\,\overline{(-1)^{n}\int
_{-\infty}^{+\infty}d_{\bar{R}}^{n}x\,g(x^{l})\overset{x}{\circledast
}S\big (\mathcal{R}_{[1]}^{-1}\triangleright e_{(L,x)}^{b}\big )\otimes
\mathcal{R}_{[2]}^{-1}\triangleright e_{(\bar{R},p)}^{b}}\nonumber\\
&  \qquad=\,(-1)^{n}\int_{-\infty}^{+\infty}d_{L}^{n}y\,f(y^{i})\overset
{y}{\circledast}e_{(\bar{R},y)}^{a}\otimes\int_{-\infty}^{+\infty}d_{\bar{R}%
}^{n}p\,e_{(L,p)}^{a}\nonumber\\
&  \qquad\qquad\qquad\overset{p}{\circledast}\,\overline{\mathcal{R}%
_{[2]}\triangleright e_{(\bar{R},\kappa^{-1}p)}^{b}\otimes\int_{-\infty
}^{+\infty}d_{\bar{R}}^{n}x\,\big (\mathcal{R}_{[1]}\triangleright
g(x^{l})\big )\overset{x}{\circledast}S(e_{(L,x)}^{b})}\nonumber\\
&  \qquad=\,(-1)^{n}\int_{-\infty}^{+\infty}d_{\bar{R}}^{n}p\,\int_{-\infty
}^{+\infty}d_{L}^{n}y\,f(y^{i})\overset{y}{\circledast}\exp(y^{j}%
|\text{i}^{-1}p^{k})_{\bar{R},L}\nonumber\\
&  \qquad\qquad\qquad\overset{p}{\circledast}\overline{\int d_{\bar{R}}%
^{n}x\,g(x^{l})\overset{x|p}{\odot}_{\hspace{-0.01in}\bar{L}}\exp
(\text{i}^{-1}(\kappa^{-1}p^{m})|(\ominus_{L}\,x^{r}))_{\bar{R},L}}\nonumber\\
&  \qquad=\,(-1)^{n}\,\text{vol}_{L}\int d_{\bar{R}}^{n}p\,\mathcal{F}%
_{L}(f)(p^{k})\overset{p}{\circledast}\overline{\mathcal{F}_{\bar{R}}^{\ast
}(g)(\kappa^{-1}p^{m})}\nonumber\\
&  \qquad=\,(-1)^{n}\,\text{vol}_{L}\,\big \langle\mathcal{F}_{L}%
(f),\mathcal{F}_{\bar{R}}^{\ast}(g)(\kappa^{-1}p^{m})\big \rangle_{\bar{R}%
,p}^{\prime}. \label{CalcId1N}%
\end{align}
The first equality is the definition of the sesquilinear form. For the second
equality we apply the first relation in (\ref{VeralDelt1}). Then we rewrite
the q-deformed delta function by using its definition in (\ref{DefDelt2}).
After that\ we use the addition law for q-deformed exponentials. For the
fifth\ step we rewrite our expression in a way that displays in which tensor
factors the various objects live. Notice that the braided antipodes $S$ and
$\bar{S}^{-1}$ represent the operations $\ominus_{L}$ and $\ominus_{\bar{R}}$,
respectively. In the sixth step we rearrange these tensor factors by taking
into account their braiding. This step follows most easily from diagrammatic
considerations (see for example Ref. \cite{Maj93-Int}). Then we extend the
operation of conjugation in a way that it affects the second factor of the
star product on momentum space. To reach this goal we apply the conjugation
properties of q-integrals, q-exponentials, antipodes, and braiding mappings
(for their discussion see Ref. \cite{qAn}). The eighth equality is again
rearranging of tensor factors in a way that allows us to identify the defining
expressions for q-deformed Fourier transformations.

Next, we turn our attention to the other versions of Fourier-Plancherel
identities. The second identity in (\ref{FPId1}) follows from the very same
reasonings as the first one. The identities in (\ref{FPId2}) can be checked
most easily by applying the operation of conjugation to the relations in
(\ref{FPId1}):%
\begin{align}
&  \,\big \langle f,g\big \rangle_{L,x}^{\prime}=(-1)^{n}%
\,\big \langle\mathcal{F}_{L}(f),\mathcal{F}_{\bar{R}}^{\ast}(g)(\kappa
^{-1}p^{i})\big \rangle_{\bar{R},p}^{\prime}\nonumber\\
\Rightarrow &  \,\,\overline{\big \langle f,g\big \rangle_{L,x}^{\prime}%
}=(-1)^{n}\,\overline{\big \langle\mathcal{F}_{L}(f),\mathcal{F}_{\bar{R}%
}^{\ast}(g)(\kappa^{-1}p^{i})\big \rangle_{\bar{R},p}^{\prime}}\nonumber\\
\Rightarrow &  \,\,\big \langle g,f\big \rangle_{\bar{R},x}^{\prime}%
=(-1)^{n}\,\big \langle\mathcal{F}_{\bar{R}}^{\ast}(g)(\kappa^{-1}%
p^{i}),\mathcal{F}_{L}(f)\big \rangle_{L,p}^{\prime}.
\end{align}
Through a slight modification of the considerations leading to the identities
in (\ref{FPId1}) and (\ref{FPId2}) we are able to check the Fourier-Plancherel
identities in (\ref{FPId3}) and (\ref{FPId4}). To illustrate this we give in
Appendix \ref{Proofs} the proof of\ the last identity in (\ref{FPId4}).

If we want to adapt the above considerations to symmetrical sesquilinear forms
we need the Fourier transformations $\mathcal{\tilde{F}}_{A}$ and
$\mathcal{\tilde{F}}_{A}^{\ast}$ [\thinspace see the discussion of
(\ref{SubInt}) and (\ref{SubTild})]. Repeating the same steps as in
(\ref{CalcId1N}) we can verify that%

\begin{align}
\big \langle f,g\big \rangle_{1,x}^{\prime}  &  =\big \langle\mathcal{\tilde
{F}}_{L}(f),\mathcal{\tilde{F}}_{\bar{R}}^{\ast}(g)(\kappa^{-1}p^{i}%
)\big \rangle_{1,p}^{\prime}=\big \langle\mathcal{\tilde{F}}_{\bar{R}}^{\ast
}(f)(\kappa^{-1}p^{i}),\mathcal{\tilde{F}}_{L}(g)\big \rangle_{1,p}^{\prime
},\nonumber\\
\big \langle f,g\big \rangle_{2,x}^{\prime}  &  =\big \langle\mathcal{\tilde
{F}}_{\bar{L}}(f),\mathcal{\tilde{F}}_{R}^{\ast}(g)(\kappa p^{i}%
)\big \rangle_{2,p}^{\prime}=\big \langle\mathcal{\tilde{F}}_{R}^{\ast
}(f)(\kappa p^{i}),\mathcal{\tilde{F}}_{\bar{L}}(g)\big \rangle_{2,p}^{\prime
},
\end{align}
and%
\begin{align}
\big \langle f,g\big \rangle_{1,x}  &  =\big \langle\mathcal{\tilde{F}}%
_{\bar{R}}(f),\mathcal{\tilde{F}}_{L}^{\ast}(g)(\kappa^{-1}p^{i}%
)\big \rangle_{1,p}=\big \langle\mathcal{\tilde{F}}_{L}^{\ast}(f)(\kappa
^{-1}p^{i}),\mathcal{\tilde{F}}_{\bar{R}}(g)\big \rangle_{1,p},\nonumber\\
\big \langle f,g\big \rangle_{2,x}  &  =\big \langle\mathcal{\tilde{F}}%
_{R}(f),\mathcal{\tilde{F}}_{\bar{L}}^{\ast}(g)(\kappa p^{i}%
)\big \rangle_{2,p}=\big \langle\mathcal{\tilde{F}}_{\bar{L}}^{\ast}(f)(\kappa
p^{i}),\mathcal{\tilde{F}}_{R}(g)\big \rangle_{2,p}.
\end{align}

We would like to make some comments about the meaning of q-deformed
Fourier-Plancherel identities. Let us recall that Fourier transformations can
be viewed as a change of basis or a change of representation. The
Fourier-Plancherel identities tell us that sesquilinear forms are invariant
under Fourier transformations of their arguments. For this reason we conclude
that Fourier transformations correspond to unitary operators. However, in the
q-deformed case the situation is a little bit more involved, since the two
arguments of sesquilinear forms have to transform differently. For this to
become more concrete we consider two functions $\psi(x^{i})$ and $\phi(x^{i})$
in position space which we assign functions on momentum space via the Fourier
transformations $\mathcal{F}_{A}$ and $\mathcal{F}_{A}^{\ast}$:%
\begin{equation}
f_{A}(p^{i})\equiv\mathcal{F}_{A}(\psi)(p^{i}),\qquad g_{A}^{\ast}%
(p^{i})\equiv\mathcal{F}_{A}^{\ast}(\phi)(p^{i}).
\end{equation}
With the help of the deformed Fourier-Plancherel identities we obtain%
\begin{align}
\big \langle\psi,\phi\big \rangle_{L,x}^{\prime}  &  =(-1)^{n}%
\big \langle\mathcal{F}_{L}(\psi),\mathcal{F}_{\bar{R}}^{\ast}(\phi
)(\kappa^{-1}p^{i})\big \rangle_{\bar{R},p}^{\prime}\nonumber\\
&  =(-1)^{n}\big \langle f_{L},g_{\bar{R}}^{\ast}(\kappa^{-1}p^{i}%
)\big \rangle_{\bar{R},p}^{\prime},
\end{align}
and likewise%
\begin{align}
\big \langle\psi,\phi\big \rangle_{\bar{L},x}^{\prime}  &  =(-1)^{n}%
\big \langle f_{\bar{L}},g_{R}^{\ast}(\kappa p^{i})\big \rangle_{R,p}^{\prime
},\\[0.1in]
\big \langle\psi,\phi\big \rangle_{L,x}  &  =(-1)^{n}\big \langle f_{\bar{R}%
},g_{L}^{\ast}(\kappa^{-1}p^{i})\big \rangle_{\bar{R},p},\nonumber\\
\big \langle\psi,\phi\big \rangle_{\bar{L},x}  &  =(-1)^{n}\big \langle f_{R}%
,g_{\bar{L}}^{\ast}(\kappa p^{i})\big \rangle_{R,p}.
\end{align}
In very much the same way we get%
\begin{align}
\big \langle\phi,\psi\big \rangle_{R,x}^{\prime}  &  =(-1)^{n}%
\big \langle g_{R}^{\ast}(\kappa p^{i}),f_{\bar{L}}\big \rangle_{\bar{L}%
,p}^{\prime},\nonumber\\
\big \langle\phi,\psi\big \rangle_{\bar{R},x}^{\prime}  &  =(-1)^{n}%
\big \langle g_{\bar{R}}^{\ast}(\kappa^{-1}p^{i}),f_{L}\big \rangle_{L,p}%
^{\prime},\\[0.1in]
\big \langle\phi,\psi\big \rangle_{R,x}  &  =(-1)^{n}\big \langle g_{\bar{L}%
}^{\ast}(\kappa p^{i}),f_{R}\big \rangle_{\bar{L},p},\nonumber\\
\big \langle\phi,\psi\big \rangle_{\bar{R},x}  &  =(-1)^{n}\big \langle g_{L}%
^{\ast}(\kappa^{-1}p^{i}),f_{\bar{R}}\big \rangle_{L,p}.
\end{align}

We would like to close our considerations in this subsection by presenting
examples that confirm the validity of our Fourier-Plancherel identities. To
this end, we first consider the sesquilinear forms%
\begin{align}
&  \big \langle\exp(y^{j}|\text{i}^{-1}p^{k})_{\bar{R},L},\exp(\text{i}%
^{-1}p^{m}|\!\ominus_{L}x^{i})_{\bar{R},L}\big \rangle_{\bar{R},p}^{\prime
}\nonumber\\
&  \qquad=\,\big \langle\exp(\text{i}^{-1}p^{k}|y^{j})_{\bar{R},L}%
,\exp(\ominus_{\bar{R}}\,x^{i}|\text{i}^{-1}p^{m})_{\bar{R},L}%
\big \rangle_{\bar{R},p}\nonumber\\
&  \qquad=\,\int_{-\infty}^{+\infty}d_{\bar{R}}^{n}p\,\exp(y^{j}|\text{i}%
^{-1}p^{k})_{\bar{R},L}\overset{p|x}{\odot}_{\hspace{-0.01in}R}\exp
(\ominus_{\bar{R}}\,x^{i}|\text{i}^{-1}p^{m})_{\bar{R},L}\nonumber\\
&  \qquad=\,\delta_{\bar{R}}^{n}(y^{j}\oplus_{\bar{R}}(\ominus_{\bar{R}%
}\,x^{i})),
\end{align}
and%
\begin{align}
&  \big \langle\exp(\ominus_{\bar{R}}\,x^{i}|\text{i}^{-1}p^{m})_{\bar{R}%
,L},\exp(\text{i}^{-1}p^{k}|y^{j})_{\bar{R},L}\big \rangle_{L,p}^{\prime
}\nonumber\\
&  \qquad=\,\big \langle\exp(\text{i}^{-1}p^{m}|\!\ominus_{L}x^{i})_{\bar
{R},L},\exp(y^{j}|\text{i}^{-1}p^{k})_{\bar{R},L}\big \rangle_{L,p}\nonumber\\
&  \qquad=\,\int_{-\infty}^{+\infty}d_{L}^{n}p\,\exp(\text{i}^{-1}%
p^{m}|\!\ominus_{L}x^{i})_{\bar{R},L}\overset{x|p}{\odot}_{\hspace
{-0.01in}\bar{L}}\exp(\text{i}^{-1}p^{k}|y^{j})_{\bar{R},L}\nonumber\\
&  \qquad=\,\delta_{L}^{n}((\ominus_{L}\,x^{i})\oplus_{L}y^{j}),
\end{align}
where we used the addition law for q-exponentials and the defining expressions
of q-deformed delta functions. We compare the above results with
\begin{align}
&  \big \langle\mathcal{F}_{\bar{R}}^{\ast}(\exp(y^{j}|\text{i}^{-1}%
p^{k})_{\bar{R},L})(\kappa^{-1}\tilde{x}^{l}),\mathcal{F}_{L}(\exp
(\text{i}^{-1}p^{m}|\!\ominus_{L}x^{i})_{\bar{R},L})(\tilde{x}^{r}%
)\big \rangle_{L,\tilde{x}}^{\prime}\nonumber\\
&  \qquad=\,\frac{1}{\text{vol}_{\bar{R}}}\big \langle\delta_{\bar{R}}%
^{n}(y^{j}\oplus_{\bar{R}}(\ominus_{\bar{R}}\,\kappa^{-1}\tilde{x}%
^{l})),\delta_{L}^{n}((\ominus_{L}\,x^{i})\oplus_{L}\tilde{x}^{r}%
)\big \rangle_{L,\tilde{x}}^{\prime}\nonumber\\
&  \qquad=\,\big \langle\mathcal{F}_{L}^{\ast}(\exp(\text{i}^{-1}p^{k}%
|y^{j})_{\bar{R},L})(\kappa^{-1}\tilde{x}^{l}),\mathcal{F}_{\bar{R}}%
(\exp(\ominus_{\bar{R}}\,x^{i}|\text{i}^{-1}p^{m})_{\bar{R},L})(\tilde{x}%
^{r})\big \rangle_{L,\tilde{x}}\nonumber\\
&  \qquad=\,\frac{1}{\text{vol}_{\bar{R}}}\big \langle\delta_{L}^{n}%
((\ominus_{L}\,\kappa^{-1}\tilde{x}^{l})\oplus_{L}y^{j}),\delta_{\bar{R}}%
^{n}(\tilde{x}^{r}\oplus_{\bar{R}}(\ominus_{\bar{R}}\,x^{i}%
))\big \rangle_{L,\tilde{x}}\nonumber\\
&  \qquad=\,\frac{(-1)^{n}}{\text{vol}_{\bar{R}}}\int_{-\infty}^{+\infty}%
d_{L}\tilde{x}\,\delta_{\bar{R}}^{n}(y^{j}\oplus_{\bar{R}}(\ominus_{\bar{R}%
}\,\kappa^{-1}\tilde{x}^{l}))\overset{\tilde{x}}{\circledast}\delta_{\bar{R}%
}^{n}(\tilde{x}^{r}\oplus_{\bar{R}}(\ominus_{\bar{R}}\,x^{i}))\nonumber\\
&  \qquad=\,(-1)^{n}\,\delta_{\bar{R}}^{n}(y^{j}\oplus_{\bar{R}}(\ominus
_{\bar{R}}\,x^{i})), \label{CalFPId1}%
\end{align}
and%
\begin{align}
&  \big \langle\mathcal{F}_{L}(\exp(\text{i}^{-1}p^{m}|\!\ominus_{L}%
x^{i})_{\bar{R},L})(\tilde{x}^{r}),\mathcal{F}_{\bar{R}}^{\ast}(\exp
(y^{j}|\text{i}^{-1}p^{k})_{\bar{R},L})(\kappa^{-1}\tilde{x}^{l}%
)\big \rangle_{\bar{R},\tilde{x}}^{\prime}\nonumber\\
&  \qquad=\,\frac{1}{\text{vol}_{L}}\big \langle\delta_{L}^{n}((\ominus
_{L}\,x^{i})\oplus_{L}\tilde{x}^{r}),\delta_{\bar{R}}^{n}(y^{j}\oplus_{\bar
{R}}(\ominus_{\bar{R}}\,\kappa^{-1}\tilde{x}^{l}))\big \rangle_{\bar{R}%
,\tilde{x}}^{\prime}\nonumber\\
&  \qquad=\,\big \langle\mathcal{F}_{\bar{R}}(\exp(\ominus_{\bar{R}}%
\,x^{i}|\text{i}^{-1}p^{m})_{\bar{R},L})(\tilde{x}^{r}),\mathcal{F}_{L}^{\ast
}(\exp(\text{i}^{-1}p^{k}|y^{j})_{\bar{R},L})(\kappa^{-1}\tilde{x}%
^{l})\big \rangle_{\bar{R},\tilde{x}}\nonumber\\
&  \qquad=\,\frac{1}{\text{vol}_{L}}\big \langle\delta_{\bar{R}}^{n}(\tilde
{x}^{r}\oplus_{\bar{R}}(\ominus_{\bar{R}}\,x^{i})),\delta_{L}^{n}((\ominus
_{L}\,\kappa^{-1}\tilde{x}^{l})\oplus_{L}y^{j})\big \rangle_{\bar{R},\tilde
{x}}\nonumber\\
&  \qquad=\,\frac{(-1)^{n}}{\text{vol}_{L}}\int_{-\infty}^{+\infty}d_{\bar{R}%
}\tilde{x}\,\delta_{L}^{n}((\ominus_{L}\,x^{i})\oplus_{L}\tilde{x}%
^{r})\overset{\tilde{x}}{\circledast}\delta_{L}^{n}((\ominus_{L}\,\kappa
^{-1}\tilde{x}^{l})\oplus_{L}y^{j})\nonumber\\
&  \qquad=\,(-1)^{n}\,\delta_{L}^{n}((\ominus_{L}\,x^{i})\oplus_{L}y^{j}).
\label{CalFPId2}%
\end{align}
For the first and third step in (\ref{CalFPId1}) as well as in (\ref{CalFPId2}%
) we inserted the expressions we obtained in subsection \ref{FTexpDelt} for
Fourier transforms of q-exponentials. The last step in both calculations is an
application of the equalities in (\ref{DeltProAlg0}) and (\ref{DeltProAlg}).

It is not very difficult to realize that our results are in complete agreement
with the Fourier-Plancherel identities in (\ref{FPId1})-(\ref{FPId4}). For the
sake of completeness it should be mentioned that we can repeat the same steps
for the other geometries as well. In doing so, we are led to expressions that
are obtained most easily from the above ones by changing the labels according
to the interchanges%
\begin{equation}
L\leftrightarrow\bar{L},\quad R\leftrightarrow\bar{R}.
\end{equation}

\section{Conclusion\label{SecCon}}

Let us end with some comments on what we have done so far. In this article we
dealt with Fourier transformations and sesquilinear forms on q-deformed
quantum spaces. This way, we laid the foundations for a q-deformed version of
quantum kinematics suitable to describe free particles on q-deformed quantum
spaces. In Part II these results will be\ applied to wave functions on
position and momentum space.

At this point let us also mention that our reasonings are rather similar to
the classical situation. However, we have to be aware of one remarkable
difference between the deformed and the undeformed case. It has to do with the
observation that for each of our q-deformed objects we can find different
realizations, which in the undeformed case become identical. The reason for
this lies in the fact that our constructions apply to a tensor category with
braiding $\Psi$ and one with braiding $\Psi^{-1}$ \cite{Maj91Kat, Maj94Kat,
Maj93-Int}. In general, the two braidings are different and we have to deal
with both categories, since they are linked via the operation of conjugation
\cite{Maj94star, Maj95star}. In the undeformed limit, however, the braidings
as well as the corresponding categories become identical and we regain the
classical situation.\vspace{0.16in}

\noindent\textbf{Acknowledgements}

First of all I am very grateful to Eberhard Zeidler for very interesting and
useful discussions, special interest in my work, and financial support.
Furthermore I would like to thank Alexander Schmidt for useful discussions and
his steady support. Finally, I thank Dieter L\"{u}st for kind hospitality.

\appendix

\section{Quantum spaces\label{AppQuan}}

In this appendix we provide some key notation for the quantum spaces we are
interested in for physical reasons, i.e. Manin plane, q-deformed Euclidean
space in three and four dimensions, and q-deformed Minkowski space. For each
case we give the defining relations, the quantum metric, and the conjugation properties.

The coordinates of the two-dimensional q-deformed quantum plane fulfill the
relation \cite{Man88,SS90}
\begin{equation}
X^{1}X^{2}=qX^{2}X^{1}, \label{2dimQuan}%
\end{equation}
whereas the quantum metric is given by a matrix $\varepsilon^{ij}$ with
non-vanishing entries
\begin{equation}
\varepsilon^{12}=q^{-1/2},\quad\varepsilon^{21}=-q^{1/2}.
\end{equation}
Relation (\ref{2dimQuan}) is compatible with the conjugation assignment
\begin{equation}
\overline{X^{i}}=-\varepsilon_{ij}X^{j},
\end{equation}
where $\varepsilon_{ij}$ denotes the inverse of $\varepsilon^{ij}.$

The commutation relations for the q-deformed Euclidean space in three
dimensions read \cite{LWW97}
\begin{align}
X^{3}X^{+}  &  =q^{2}X^{+}X^{3},\nonumber\\
X^{-}X^{3}  &  =q^{2}X^{3}X^{-},\nonumber\\
X^{-}X^{+}  &  =X^{+}X^{-}+\lambda X^{3}X^{3}. \label{Koord3dimN}%
\end{align}
The non-vanishing elements of the quantum metric are
\begin{equation}
g^{+-}=-q,\quad g^{33}=1,\quad g^{-+}=-q^{-1}.
\end{equation}
The conjugation properties of coordinates are given by
\begin{equation}
\overline{X^{A}}=g_{AB}X^{B}, \label{KonRel}%
\end{equation}
with $g_{AB}$ denoting the inverse of $g^{AB}.$ If we are looking for
coordinates\ subject to $\overline{Y^{i}}=Y^{i}$ we can choose%
\begin{align}
Y^{1}  &  =\frac{\text{i}}{q^{1/2}+q^{-1/2}}(q^{-1/2}X^{+}+q^{1/2}%
X^{-}),\nonumber\\
Y^{2}  &  =\frac{1}{q^{1/2}+q^{-1/2}}(q^{-1/2}X^{+}-q^{1/2}X^{-}),\nonumber\\
Y^{3}  &  =X^{3}. \label{RealKoor3dim}%
\end{align}

For the four-dimensional Euclidean space we have the relations \cite{KS97}
\begin{align}
X^{1}X^{2}  &  =qX^{2}X^{1},\nonumber\\
X^{1}X^{3}  &  =qX^{3}X^{1},\nonumber\\
X^{3}X^{4}  &  =qX^{4}X^{3},\nonumber\\
X^{2}X^{4}  &  =qX^{4}X^{2},\nonumber\\
X^{2}X^{3}  &  =X^{3}X^{2},\nonumber\\
X^{4}X^{1}  &  =X^{1}X^{4}+\lambda X^{2}X^{3}. \label{Algebra4N}%
\end{align}
The non-vanishing components of the corresponding quantum metric read%
\begin{equation}
g^{14}=q^{-1},\quad g^{23}=g^{32}=1,\quad g^{41}=q.
\end{equation}
If $g_{ij}$ again denotes the inverse of $g^{ij}$ it holds
\begin{equation}
\overline{X^{i}}=g_{ij}X^{j}. \label{Metrik}%
\end{equation}
Using this relation it is easy to check that the following independent
coordinates are invariant under conjugation \cite{Oca96}:%
\begin{align}
Y^{1}  &  =\frac{1}{q^{1/2}+q^{-1/2}}(q^{1/2}X^{1}+q^{-1/2}X^{4}),\nonumber\\
Y^{2}  &  =\frac{1}{2}(X^{2}+X^{3}),\nonumber\\
Y^{3}  &  =\frac{\text{i}}{2}(X^{2}-X^{3}),\nonumber\\
Y^{4}  &  =\frac{\text{i}}{q^{1/2}+q^{-1/2}}(q^{1/2}X^{1}-q^{-1/2}X^{4}).
\end{align}

The coordinates of q-deformed Minkowski space obey the relations \cite{LWW97}
\begin{align}
X^{\mu}X^{0}  &  =X^{0}X^{\mu},\quad\mu\in\{0,+,-,3\},\nonumber\\
X^{-}X^{3}-q^{2}X^{3}X^{-}  &  =-q\lambda X^{0}X^{-},\nonumber\\
X^{3}X^{+}-q^{2}X^{+}X^{3}  &  =-q\lambda X^{0}X^{+},\nonumber\\
X^{-}X^{+}-X^{+}X^{-}  &  =\lambda(X^{3}X^{3}-X^{0}X^{3}). \label{MinrelN}%
\end{align}
As non-vanishing components of the corresponding metric we have
\begin{equation}
\eta^{00}=-1,\quad\eta^{33}=1,\quad\eta^{+-}=-q,\quad\eta^{-+}=-q^{-1}.
\end{equation}
(For other deformations of Minkowski spacetime we refer to Refs. \cite{Lu92,
Cas93, Dob94, DFR95, ChDe95, ChKu04, Koch04}.) The conjugation on q-deformed
Minkowski space is determined by
\begin{equation}
\overline{X^{0}}=X^{0},\quad\overline{X^{3}}=X^{3},\quad\overline{X^{\pm}%
}=-q^{\mp1}X^{\mp}.
\end{equation}
A set of independent coordinates being invariant under conjugation is now
given by $Y^{0}=X^{0}$ and the coordinates introduced in (\ref{RealKoor3dim}).

\section{\label{Proofs}Proofs}

In this appendix we present some calculations that are rather lengthy or not
so important for the understanding of the subject. These calculations are
partly based on the relations we derived for objects of q-analysis in Ref.
\cite{qAn}. At some places we make use of the fact that we are dealing with
objects that live in braided tensor categories. This observation enables us to
apply diagrammatic considerations for certain rearrangements. For an
introduction into this subject we refer the reader to Ref. \cite{Maj93-Int}.

\subsection{Proof of the relations in (\ref{DeltAnt})}

We concentrate attention on the first relation in (\ref{DeltAnt}), since the
second one follows from similar arguments. We illustrate the reasonings
leading to (\ref{DeltAnt}) by the following calculation:
\begin{align}
&  \int\nolimits_{-\infty}^{+\infty}d_{\bar{L}}^{n}x\,g(x^{i})\overset
{x}{\circledast}\delta_{R}^{n}(\ominus_{R}\,x^{j})\nonumber\\
&  \qquad=\,\int\nolimits_{-\infty}^{+\infty}d_{\bar{L}}^{n}x\,g(x^{i}%
)\overset{x}{\circledast}\int\nolimits_{-\infty}^{+\infty}d_{R}^{n}%
p\,\exp(\ominus_{R}\,x^{j}|\text{i}^{-1}p^{k})_{R,\bar{L}}\nonumber\\
&  \qquad=\,\int\nolimits_{-\infty}^{+\infty}d_{\bar{L}}^{n}x\,g(x^{i}%
)\overset{x}{\circledast}S^{-1}(e_{(x,R)}^{a})\otimes\int\nolimits_{-\infty
}^{+\infty}d_{R}^{n}p\,e_{(p,\bar{L})}^{a}\nonumber\\
&  \qquad=\,\int\nolimits_{-\infty}^{+\infty}d_{\bar{L}}^{n}%
x\,\big \langle e_{(p,\bar{L})}^{b},g(x^{i})\big \rangle_{\bar{L}%
,R}\,e_{(x,R)}^{b}\overset{x}{\circledast}S^{-1}(e_{(x,R)}^{a})\otimes
\int\nolimits_{-\infty}^{+\infty}d_{R}^{n}p\,e_{(p,\bar{L})}^{a}\nonumber\\
&  \qquad=\,\int\nolimits_{-\infty}^{+\infty}d_{\bar{L}}^{n}%
x\,\big \langle S^{-1}(e_{(p,\bar{L})}^{b}),g(x^{i})\big \rangle_{\bar{L}%
,R}\,S^{-1}(e_{(x,R)}^{b})\overset{x}{\circledast}S^{-1}(e_{(x,R)}%
^{a})\nonumber\\
&  \qquad\qquad\qquad\otimes\int\nolimits_{-\infty}^{+\infty}d_{R}%
^{n}p\,e_{(p,\bar{L})}^{a}\nonumber\\
&  \qquad=\,\int\nolimits_{-\infty}^{+\infty}d_{\bar{L}}^{n}x\,S^{-1}%
([\mathcal{R}_{[1]}^{-1}\triangleright(e_{(x,R)}^{a})]\overset{x}{\circledast
}[\mathcal{R}_{[2]}^{-1}\triangleright(e_{(x,R)}^{b})])\nonumber\\
&  \qquad\qquad\qquad\times\big \langle S^{-1}(e_{(p,\bar{L})}^{b}%
),g(x^{i})\big \rangle_{\bar{L},R}\otimes\int\nolimits_{-\infty}^{+\infty
}d_{R}^{n}p\,e_{(p,\bar{L})}^{a}\nonumber\\
&  \qquad=\,\int\nolimits_{-\infty}^{+\infty}d_{\bar{L}}^{n}x\,S^{-1}%
(e_{(x,R)}^{a}\overset{x}{\circledast}e_{(x,R)}^{b})\big \langle S^{-1}%
(e_{(p,\bar{L})}^{b}),g(\kappa x^{i})\big \rangle_{\bar{L},R}\nonumber\\
&  \qquad\qquad\qquad\otimes\int\nolimits_{-\infty}^{+\infty}d_{R}%
^{n}p\,e_{(p,\bar{L})}^{a}\nonumber\\
&  \qquad=\,\int\nolimits_{-\infty}^{+\infty}d_{\bar{L}}^{n}x\,S^{-1}%
(e_{(x,R)}^{a})\big \langle S^{-1}((e_{(p,\bar{L})}^{a})_{(\bar{R}%
,1)}),g(\kappa x^{i})\big \rangle_{\bar{L},R}\nonumber\\
&  \qquad\qquad\qquad\otimes\int\nolimits_{-\infty}^{+\infty}d_{R}%
^{n}p\,(e_{(p,\bar{L})}^{a})_{(\bar{R},2)}\nonumber\\
&  \qquad=\,\int\nolimits_{-\infty}^{+\infty}d_{\bar{L}}^{n}x\,S^{-1}%
(e_{(x,R)}^{a})\big \langle S^{-1}(1),g(\kappa x^{i})\big \rangle_{\bar{L}%
,R}\otimes\int\nolimits_{-\infty}^{+\infty}d_{R}^{n}p\,e_{(p,\bar{L})}%
^{a}\nonumber\\
&  \qquad=\,g(0)\,\int\nolimits_{-\infty}^{+\infty}d_{\bar{L}}^{n}%
x\int\nolimits_{-\infty}^{+\infty}d_{R}^{n}p\,\exp(\ominus_{R}\,x^{j}%
|\text{i}^{-1}p^{k})_{R,\bar{L}}\nonumber\\
&  \qquad=\kappa^{n}\text{vol}_{R}\,g(0).
\end{align}
For the first step we insert the definition of the q-deformed delta function
[cf. the definitions in (\ref{DefDelt2})]. Then we reformulate the expression
in order to display the tensor product it lives in. Notice that $S^{-1}$
corresponds to the operation $\ominus_{R}$ and can be recognized as antipode
on a braided space \cite{qAn, Maj93-Int}. Next, we rewrite the\ function $g$
by means of a variant of the completeness relation in (\ref{CompRel}). This
step enables us to apply the identity%
\begin{equation}
\sum_{a}e_{(p,\bar{L})}^{a}\otimes e_{(x,R)}^{a}=\sum_{a}S^{-1}(e_{(p,\bar
{L})}^{a})\otimes S^{-1}(e_{(x,R)}^{a})
\end{equation}
together with the axiom
\begin{equation}
m\circ(S^{-1}\otimes S^{-1})=S^{-1}\circ m\circ\Psi^{-1},
\end{equation}
where $m$ denotes multiplication on coordinate space and $\psi^{-1}$ the
braiding mapping induced by $\tau\circ\mathcal{R}^{-1}$. For the sixth
equality we have to realize that the braiding involves a delta function.
However, the trivial braiding of delta functions allows us to neglect the
braiding. The seventh equality is the addition law for q-deformed exponentials
and the eighth equality uses translation invariance of the integral over
momentum space [cf. (\ref{TransGlob})]. The ninth step follows from the same
arguments already applied in (\ref{CharPropDel0}). The last equality is a
consequence of the identities in (\ref{VolSR}).

\subsection{Proof of the relations in (\ref{InvFourAnf2a}) and
(\ref{InvFourAnf1a})\label{Proof1}}%

\begin{align}
&  \mathcal{F}_{L}(\mathcal{F}_{\bar{R}}^{\ast}(g)(x^{i}))(p^{k}%
)=\int\nolimits_{-\infty}^{+\infty}d_{L}^{n}x\,\mathcal{F}_{\bar{R}}^{\ast
}(g)(x^{i})\overset{x}{\circledast}\exp(x^{j}|\text{i}^{-1}p^{k})_{\bar{R}%
,L}\nonumber\\
&  \qquad=\frac{1}{\text{vol}_{\bar{R}}}\int\nolimits_{-\infty}^{+\infty}%
d_{L}^{n}x\Big (\int\nolimits_{-\infty}^{+\infty}d_{\bar{R}}^{n}\tilde
{p}\,g(\tilde{p}^{m})\overset{\tilde{p}|x}{\odot}_{\hspace{-0.01in}R}%
\exp(\ominus_{\bar{R}}\,x^{i}|\text{i}^{-1}\tilde{p}^{l})_{\bar{R}%
,L}\Big )\nonumber\\
&  \qquad\qquad\qquad\qquad\overset{x}{\circledast}\exp(x^{j}|\text{i}%
^{-1}p^{k})_{\bar{R},L}\nonumber\\
&  \qquad=\frac{1}{\text{vol}_{\bar{R}}}\int\nolimits_{-\infty}^{+\infty}%
d_{L}^{n}x\Big (\int\nolimits_{-\infty}^{+\infty}d_{\bar{R}}^{n}\tilde
{p}\,g(\tilde{p}^{m})\overset{\tilde{p}|x}{\odot}_{\hspace{-0.01in}R}%
\exp(x^{i}|\!\ominus_{L}\!(\text{i}^{-1}\tilde{p}^{l}))_{\bar{R}%
,L}\Big )\nonumber\\
&  \qquad\qquad\qquad\qquad\overset{x}{\circledast}\exp(x^{j}|\text{i}%
^{-1}p^{k})_{\bar{R},L}\nonumber\\
&  \qquad=\frac{1}{\text{vol}_{\bar{R}}}\int\nolimits_{-\infty}^{+\infty
}d_{\bar{R}}^{n}\tilde{p}\,g(\kappa\tilde{p}^{m})\nonumber\\
&  \qquad\qquad\qquad\qquad\overset{\tilde{p}|p}{\odot}_{\hspace{-0.01in}%
R}\int\nolimits_{-\infty}^{+\infty}d_{L}^{n}x\,\exp(x^{i}|(\text{i}^{-1}%
p^{k})\oplus_{R}(\ominus_{L}(\text{i}^{-1}\tilde{p}^{l})))_{\bar{R},L}.
\label{Proof2}%
\end{align}
Let us make a few comments on what we have done so far. We inserted the
defining expressions for q-deformed Fourier transformations. In the left
exponential we switched the antipode from space coordinates to momentum
coordinates by applying
\begin{equation}
\exp(\ominus_{\bar{R}}\,x^{i}|\text{i}^{-1}p^{k})_{\bar{R},L}=\exp
(x^{i}|\!\ominus_{L}\!(\text{i}^{-1}p^{k}))_{\bar{R},L}, \label{UmtAnt}%
\end{equation}
Finally, we used the addition law for q-exponentials.

In contrast to the calculation in (\ref{RechInv1N}), we are not yet in a
position to apply the identities for q-deformed delta functions in
(\ref{VeralDelt1}) and (\ref{VeralDelt2}). Thus, we have to work a little bit
harder. In what follows, it is helpful\ to rewrite the last expression in
(\ref{Proof2}) in such a way that tensor factors are explicitly displayed:%
\begin{align}
&  \frac{1}{\text{vol}_{\bar{R}}}\int\nolimits_{-\infty}^{+\infty}d_{\bar{R}%
}^{n}\tilde{p}\,g(\kappa\tilde{p}^{m})\overset{\tilde{p}|p}{\odot}%
_{\hspace{-0.01in}R}\int\nolimits_{-\infty}^{+\infty}d_{L}^{n}x\,\exp
(x^{i}|\text{i}^{-1}(p^{k}\oplus_{R}(\ominus_{L}\,\tilde{p}^{l})))_{\bar{R}%
,L}\nonumber\\
&  \qquad=\frac{1}{\text{vol}_{\bar{R}}}\int\nolimits_{-\infty}^{+\infty}%
d_{L}^{n}x\,e_{(x,\bar{R})}^{a}\otimes\mathcal{R}_{[2]}\triangleright
(e_{(p,L)}^{a})_{(R,1)}\nonumber\\
&  \qquad\qquad\qquad\otimes\int\nolimits_{-\infty}^{+\infty}d_{\bar{R}}%
^{n}\tilde{p}\,S\big ((\mathcal{R}_{[2]}^{\prime}\triangleright(e_{(\tilde
{p},L)}^{a})_{(R,2)})\nonumber\\
&  \qquad\qquad\qquad\overset{\tilde{p}}{\circledast}S^{-1}(\mathcal{R}%
_{[1]}^{\prime}\mathcal{R}_{[1]}\triangleright g(\kappa\tilde{p}^{m}))\big )\\
&  \qquad=\frac{1}{\text{vol}_{\bar{R}}}\int\nolimits_{-\infty}^{+\infty}%
d_{L}^{n}x\,e_{(x,\bar{R})}^{a}\otimes(e_{(p,L)}^{a})_{(R,1)}\nonumber\\
&  \qquad\qquad\qquad\otimes\int\nolimits_{-\infty}^{+\infty}d_{\bar{R}}%
^{n}\tilde{p}\,S\big ((e_{(\tilde{p},L)}^{a})_{(R,2)}\overset{\tilde{p}%
}{\circledast}S^{-1}(g(\tilde{p}^{m}))\big )\nonumber\\
&  \qquad=\frac{1}{\text{vol}_{\bar{R}}}\int\nolimits_{-\infty}^{+\infty}%
d_{L}^{n}x\,e_{(x,\bar{R})}^{a}\otimes S\big (\mathcal{R}_{[2]}\triangleright
(S^{-1}g(p^{k}))_{(R,1)}\big )\nonumber\\
&  \qquad\qquad\qquad\otimes\int\nolimits_{-\infty}^{+\infty}d_{\bar{R}}%
^{n}\tilde{p}\,S(\mathcal{R}_{[1]}\triangleright e_{(\tilde{p},L)}%
^{a})\overset{\tilde{p}}{\circledast}(S^{-1}g(\tilde{p}^{m}))_{(R,2)}%
\nonumber\\
&  \qquad=\frac{1}{\text{vol}_{\bar{R}}}\int\nolimits_{-\infty}^{+\infty}%
d_{L}^{n}x\,e_{(x,\bar{R})}^{a}\otimes S(S^{-1}g(\kappa^{-1}p^{k}%
))_{(R,1)}\nonumber\\
&  \qquad\qquad\qquad\otimes\int\nolimits_{-\infty}^{+\infty}d_{\bar{R}}%
^{n}\tilde{p}\,S\big (e_{(\tilde{p},L)}^{a}\overset{\tilde{p}}{\circledast
}(S^{-1}g(\tilde{p}^{m}))_{(R,2)}\big ). \label{HerPro1}%
\end{align}
For the first step we applied the axiom%
\begin{equation}
S^{-1}\circ m=m\circ(S\otimes S)\circ\psi.
\end{equation}
The braiding mappings in the second expression determine how the function $g$
commutes with a delta function. Due to the trivial braiding of delta
functions, we are able to neglect the braiding. To understand the third step
one has to realize that the expressions%
\begin{equation}
\int\nolimits_{-\infty}^{+\infty}d_{A}^{n}p\,f(\ominus_{B}\,p^{i}),
\label{TransInS}%
\end{equation}
fulfill%
\begin{align}
\int\nolimits_{-\infty}^{+\infty}d_{A}^{n}p\,f(\ominus_{B}\,p^{i})  &
=\int\nolimits_{-\infty}^{+\infty}d_{A}^{n}p\,f((\ominus_{B}\,p^{i})\oplus
_{C}\tilde{p}^{j})\nonumber\\
&  =\int\nolimits_{-\infty}^{+\infty}d_{A}^{n}p\,f(\,\tilde{p}^{j}\oplus
_{C}(\ominus_{B}\,p^{i})), \label{InsSTrans}%
\end{align}
where $A,B,C\in\{L,\bar{L},R,\bar{R}\}.$ This can easily be seen by
calculations of the following form:%
\begin{align}
&  \int\nolimits_{-\infty}^{+\infty}d_{\bar{R}}^{n}p\,f((\ominus_{L}%
\,p^{i})\oplus_{R}\tilde{p}^{j})\nonumber\\
&  \qquad=\,\int\nolimits_{-\infty}^{+\infty}d_{\bar{R}}^{n}p\,f(\ominus
_{L}(\,p^{i}\oplus_{L}(\ominus_{R}\,\tilde{p}^{j})))=\int\nolimits_{-\infty
}^{+\infty}d_{\bar{R}}^{n}p\,f(\ominus_{L}\,p^{i}). \label{TranInvIntS}%
\end{align}
The point now is that due to (\ref{InsSTrans}) relation (\ref{IntKoorTrans})
remains valid if we replace the integrals over the whole space by the
expressions in (\ref{TransInS}) and this observation finally leads to the
third equality in (\ref{HerPro1}). In the last step we exploit the trivial
braiding of q-integral and q-exponential to rewrite the expression in
(\ref{HerPro1}) without any braiding.

To proceed any further, we need another identity, which we now derive:%
\begin{align}
&  \int\nolimits_{-\infty}^{+\infty}d_{L}^{n}x\,e_{(x,\bar{R})}^{a}\otimes
\int\nolimits_{-\infty}^{+\infty}d_{\bar{R}}^{n}\tilde{p}\,S\big ((e_{(\tilde
{p},L)}^{a})\overset{\tilde{p}}{\circledast}f(\tilde{p}^{i})\big )\nonumber\\
&  \qquad=\int\nolimits_{-\infty}^{+\infty}d_{L}^{n}x\,e_{(x,\bar{R})}%
^{a}\otimes\int\nolimits_{-\infty}^{+\infty}d_{\bar{R}}^{n}\tilde
{p}\,S\big (e_{(\tilde{p},L)}^{a}\overset{\tilde{p}}{\circledast}e_{(\tilde
{p},L)}^{b}\big )\nonumber\\
&  \qquad\qquad\qquad\times\big \langle f(p^{i}),e_{(x,\bar{R})}%
^{b}\big \rangle_{L,\bar{R}}\nonumber\\
&  \qquad=\int\nolimits_{-\infty}^{+\infty}d_{L}^{n}x\,(e_{(x,\bar{R})}%
^{a})_{(R,1)}\otimes\int\nolimits_{-\infty}^{+\infty}d_{\bar{R}}^{n}\tilde
{p}\,S(e_{(\tilde{p},L)}^{a})\nonumber\\
&  \qquad\qquad\qquad\times\big \langle f(p^{i}),(e_{(x,\bar{R})}^{a}%
)_{(R,2)}\big \rangle_{L,\bar{R}}\nonumber\\
&  \qquad=\int\nolimits_{-\infty}^{+\infty}d_{L}^{n}x\,e_{(x,\bar{R})}%
^{a}\otimes\int\nolimits_{-\infty}^{+\infty}d_{\bar{R}}^{n}\tilde
{p}\,S(e_{(\tilde{p},L)}^{a})\big \langle f(p^{i}),1\big \rangle_{L,\bar{R}}.
\end{align}
The first equality results from the completeness relation (\ref{CompRel}). The
second equality is the addition law for q-exponentials. Finally, we make use
of translation invariance of the integral over position space.

By virtue of the above identity the last expression in (\ref{HerPro1}) becomes%
\begin{align}
&  \,\frac{1}{\text{vol}_{\bar{R}}}\int\nolimits_{-\infty}^{+\infty}d_{L}%
^{n}x\,e_{(x,\bar{R})}^{a}\otimes S(S^{-1}g(\kappa^{-1}p^{k}))_{(R,1)}%
\nonumber\\
&  \qquad\qquad\qquad\otimes\int\nolimits_{-\infty}^{+\infty}d_{\bar{R}}%
^{n}\tilde{p}\,S\big (e_{(\tilde{p},L)}^{a}\overset{\tilde{p}}{\circledast
}(S^{-1}g(\tilde{p}^{m}))_{(R,2)}\big )\nonumber\\
=  &  \,\frac{1}{\text{vol}_{\bar{R}}}\int\nolimits_{-\infty}^{+\infty}%
d_{L}^{n}x\,e_{(x,\bar{R})}^{a}\otimes S(S^{-1}g(\kappa^{-1}p^{k}%
))_{(R,1)}\big \langle(S^{-1}g(p^{k}))_{(R,2)},1\big \rangle_{L,\bar{R}%
}\nonumber\\
&  \qquad\qquad\qquad\otimes\int\nolimits_{-\infty}^{+\infty}d_{\bar{R}}%
^{n}\tilde{p}\,S(e_{(\tilde{p},L)}^{a})\nonumber\\
=  &  \,\frac{1}{\text{vol}_{\bar{R}}}\int\nolimits_{-\infty}^{+\infty}%
d_{L}^{n}x\,e_{(x,\bar{R})}^{a}\otimes S(S^{-1}g(\kappa^{-1}p^{k}))\otimes
\int\nolimits_{-\infty}^{+\infty}d_{\bar{R}}^{n}\tilde{p}\,S(e_{(\tilde{p}%
,L)}^{a})\nonumber\\
=  &  \,\kappa^{-n}g(\kappa^{-1}p^{k}).
\end{align}
Notice that for the second step we applied%
\begin{equation}
f_{(R,1)}\otimes\langle f_{(R,2)},1\rangle_{L,\bar{R}}=f_{(R,1)}%
\otimes\epsilon_{R}(f_{(R,2)})=f,
\end{equation}
while the last step needs the relation%
\begin{equation}
\text{vol}_{R}=\kappa^{n}\int\nolimits_{-\infty}^{+\infty}d_{L}^{n}%
x\int\nolimits_{-\infty}^{+\infty}d_{\bar{R}}^{n}p\,\exp(x^{i}|\!\ominus
_{L}\!(\text{i}^{-1}p^{k}))_{\bar{R},L}. \label{VolS}%
\end{equation}

It remains to prove identity (\ref{VolS}). To this end we show that the
following equalities hold:%
\begin{align}
\kappa^{-n}(\text{vol}_{\bar{R}})^{2} &  =\int\nolimits_{-\infty}^{+\infty
}d_{L}^{n}\tilde{x}\int\nolimits_{-\infty}^{+\infty}d_{L}^{n}x\,\delta
_{\bar{R}}^{n}((\ominus_{\bar{R}}\,x^{i})\oplus_{\bar{R}}\tilde{x}%
^{j}))\overset{\tilde{x}}{\circledast}\delta_{\bar{R}}^{n}(\tilde{x}%
^{l})\nonumber\\
&  =\text{vol}_{\bar{R}}\int\nolimits_{-\infty}^{+\infty}d_{L}^{n}%
x\int\nolimits_{-\infty}^{+\infty}d_{\bar{R}}^{n}p\,\exp(x^{i}|\!\ominus
_{L}\!(\text{i}^{-1}p^{k}))_{\bar{R},L}.\label{Lem1}%
\end{align}
The first equality can be verified in a straightforward manner:%
\begin{align}
&  \int\nolimits_{-\infty}^{+\infty}d_{L}^{n}\tilde{x}\int\nolimits_{-\infty
}^{+\infty}d_{L}^{n}x\,\delta_{\bar{R}}^{n}((\ominus_{\bar{R}}\,x^{i}%
)\oplus_{\bar{R}}\tilde{x}^{j})\overset{\tilde{x}}{\circledast}\delta_{\bar
{R}}^{n}(\tilde{x}^{l})\nonumber\\
&  \qquad=\,\int\nolimits_{-\infty}^{+\infty}d_{L}^{n}\tilde{x}\int
\nolimits_{-\infty}^{+\infty}d_{L}^{n}x\Big (\int\nolimits_{-\infty}^{+\infty
}d_{\bar{R}}^{n}p\,\exp((\ominus_{\bar{R}}\,x^{i})\oplus_{\bar{R}}\tilde
{x}^{j}|\text{i}^{-1}p^{k})_{\bar{R},L}\Big )\nonumber\\
&  \qquad\qquad\qquad\overset{\tilde{x}}{\circledast}\int\nolimits_{-\infty
}^{+\infty}d_{\bar{R}}^{n}\tilde{p}\,\exp(\tilde{x}^{l}|\text{i}^{-1}\tilde
{p}^{m})_{\bar{R},L}\nonumber\\
&  \qquad=\,\int\nolimits_{-\infty}^{+\infty}d_{L}^{n}\tilde{x}\int
\nolimits_{-\infty}^{+\infty}d_{L}^{n}x\Big (\int\nolimits_{-\infty}^{+\infty
}d_{\bar{R}}^{n}p\,\exp(x^{i}|\text{i}^{-1}p^{k})_{\bar{R},L}\Big )\nonumber\\
&  \qquad\qquad\qquad\overset{x|\tilde{x}}{\odot}_{\hspace{-0.02in}\bar{R}%
}\int\nolimits_{-\infty}^{+\infty}d_{\bar{R}}^{n}\tilde{p}\,\exp(\tilde{x}%
^{l}\oplus_{\bar{R}}x^{j}|\text{i}^{-1}\tilde{p}^{m})_{\bar{R},L}\nonumber\\
&  \qquad=\,\int\nolimits_{-\infty}^{+\infty}d_{L}^{n}x\Big (\int
\nolimits_{-\infty}^{+\infty}d_{\bar{R}}^{n}p\,\exp(x^{i}|\text{i}^{-1}%
p^{k})_{\bar{R},L}\Big )\nonumber\\
&  \qquad\qquad\qquad\overset{x}{\circledast}\int\nolimits_{-\infty}^{+\infty
}d_{L}^{n}\tilde{x}\int\nolimits_{-\infty}^{+\infty}d_{\bar{R}}^{n}\tilde
{p}\,\exp((\kappa\tilde{x}^{l})\oplus_{\bar{R}}x^{j}|\text{i}^{-1}\tilde
{p}^{m})_{\bar{R},L}\nonumber\\
&  \qquad=\,\int\nolimits_{-\infty}^{+\infty}d_{L}^{n}x\Big (\int
\nolimits_{-\infty}^{+\infty}d_{\bar{R}}^{n}p\,\exp(x^{i}|\text{i}^{-1}%
p^{k})_{\bar{R},L}\Big )\nonumber\\
&  \qquad\qquad\qquad\times\int\nolimits_{-\infty}^{+\infty}d_{L}^{n}\tilde
{x}\int\nolimits_{-\infty}^{+\infty}d_{\bar{R}}^{n}\tilde{p}\,\exp
(\kappa\tilde{x}^{l}|\text{i}^{-1}\tilde{p}^{m})_{\bar{R},L}\nonumber\\
&  \qquad=\,\kappa^{-n}(\text{vol}_{R})^{2}.
\end{align}
For the first step we insert the defining expressions for q-deformed delta
functions. Then we are ready to apply (\ref{IntKoorTrans}). The third equality
is a consequence of the braiding properties of integrals over the whole space.
For the fourth step we use translation invariance of integrals over the whole space.

Now, we come to the second equality in (\ref{Lem1}), for which we likewise
have%
\begin{align}
&  \int\nolimits_{-\infty}^{+\infty}d_{L}^{n}\tilde{x}\int\nolimits_{-\infty
}^{+\infty}d_{\bar{R}}^{n}x\,\delta_{\bar{R}}^{n}((\ominus_{\bar{R}}%
\,x^{i})\oplus_{\bar{R}}\tilde{x}^{j})\overset{\tilde{x}}{\circledast}%
\,\delta_{\bar{R}}^{n}(\tilde{x}^{l})\nonumber\\
&  \quad=\int\nolimits_{-\infty}^{+\infty}d_{L}^{n}\tilde{x}\int
\nolimits_{-\infty}^{+\infty}d_{L}^{n}x\Big (\int\nolimits_{-\infty}^{+\infty
}d_{\bar{R}}^{n}p\,\exp((\kappa\tilde{x}^{j})\oplus_{\bar{L}}(\ominus_{\bar
{R}}\,x^{i})|\text{i}^{-1}p^{k})_{\bar{R},L}\Big )\nonumber\\
&  \qquad\qquad\qquad\overset{\tilde{x}}{\circledast}\int\nolimits_{-\infty
}^{+\infty}d_{\bar{R}}^{n}\tilde{p}\,\exp(\tilde{x}^{l}|\text{i}^{-1}\tilde
{p}^{m})_{\bar{R},L}\nonumber\\
&  \quad=\int\nolimits_{-\infty}^{+\infty}d_{L}^{n}\tilde{x}\int
\nolimits_{-\infty}^{+\infty}d_{L}^{n}x\Big (\int\nolimits_{-\infty}^{+\infty
}d_{\bar{R}}^{n}p\,\exp(\ominus_{\bar{R}}((\ominus_{\bar{L}}\kappa\tilde
{x}^{j})\oplus_{\bar{L}}x^{i})|\text{i}^{-1}p^{k})_{\bar{R},L}%
\Big )\nonumber\\
&  \qquad\qquad\qquad\overset{\tilde{x}}{\circledast}\int\nolimits_{-\infty
}^{+\infty}d_{\bar{R}}^{n}\tilde{p}\,\exp(\tilde{x}^{l}|\text{i}^{-1}\tilde
{p}^{m})_{\bar{R},L}\nonumber\\
&  \quad=\int\nolimits_{-\infty}^{+\infty}d_{L}^{n}\tilde{x}\int
\nolimits_{-\infty}^{+\infty}d_{L}^{n}x\Big (\int\nolimits_{-\infty}^{+\infty
}d_{\bar{R}}^{n}p\,\exp((\ominus_{\bar{L}}\kappa\tilde{x}^{j})\oplus_{\bar{L}%
}x^{i}|\!\ominus_{L}\!(\text{i}^{-1}p^{k}))_{\bar{R},L}\Big )\nonumber\\
&  \qquad\qquad\qquad\overset{\tilde{x}}{\circledast}\int\nolimits_{-\infty
}^{+\infty}d_{\bar{R}}^{n}\tilde{p}\,\exp(\tilde{x}^{l}|\text{i}^{-1}\tilde
{p}^{m})_{\bar{R},L}\nonumber\\
&  \quad=\int\nolimits_{-\infty}^{+\infty}d_{L}^{n}x\int\nolimits_{-\infty
}^{+\infty}d_{\bar{R}}^{n}p\,\exp(x^{i}|\!\ominus_{L}\!(\text{i}^{-1}%
p^{k}))_{\bar{R},L}\nonumber\\
&  \qquad\qquad\qquad\times\int\nolimits_{-\infty}^{+\infty}d_{L}^{n}\tilde
{x}\int\nolimits_{-\infty}^{+\infty}d_{\bar{R}}^{n}\tilde{p}\,\exp(\tilde
{x}^{l}|\text{i}^{-1}\tilde{p}^{m})_{\bar{R},L}\nonumber\\
&  \quad=\text{vol}_{\bar{R}}\int\nolimits_{-\infty}^{+\infty}d_{L}^{n}%
x\int\nolimits_{-\infty}^{+\infty}d_{\bar{R}}^{n}p\,\exp(x^{i}|\!\ominus
_{L}\!(\text{i}^{-1}p^{k}))_{\bar{R},L}.
\end{align}
Again let us make some comments on the above calculation. First, we plug in
the expressions for q-deformed delta functions. Simultaneously, we change the
order of the tensor factors corresponding to the coordinates $\tilde{x}^{j}$
and $x^{i}$ by making use of
\begin{equation}
f(x^{i}\oplus_{\bar{L}}y^{j})=f(y^{j}\oplus_{\bar{R}}x^{i}).
\end{equation}
In doing so, there appears a scaling, since we integrate over the coordinate
$x^{i}.$ Then we make use of the distributive law for q-deformed translations.%
\begin{equation}
f(\ominus_{\bar{L}}(x^{i}\oplus_{\bar{L}}y^{j}))=f((\ominus_{\bar{L}}%
\,x^{i})\oplus_{\bar{R}}(\ominus_{\bar{L}}\,y^{j})).
\end{equation}
(Notice that $f(\ominus_{\bar{L}}(\ominus_{\bar{R}}\,x^{i}))=f(x^{i}).$) The
third equality is the property (\ref{UmtAnt}) and the fourth equality again
results from translation invariance of integrals over the whole space.

For the sake of completeness it should be mentioned that in addition to
(\ref{Lem1}) we also have%
\begin{align}
\kappa^{n}(\text{vol}_{\bar{R}})^{2}  &  =\int\nolimits_{-\infty}^{+\infty
}d_{L}^{n}\tilde{x}\int\nolimits_{-\infty}^{+\infty}d_{L}^{n}x\,\delta
_{\bar{R}}^{n}((\ominus_{\bar{L}}x^{i})\oplus_{\bar{L}}\tilde{x}^{j}%
))\overset{\tilde{x}}{\circledast}\delta_{\bar{R}}^{n}(\tilde{x}%
^{l})\nonumber\\
&  =\text{vol}_{\bar{R}}\int\nolimits_{-\infty}^{+\infty}d_{L}^{n}%
x\int\nolimits_{-\infty}^{+\infty}d_{\bar{R}}^{n}p\,\exp(x^{i}|\!\ominus
_{R}(\text{i}^{-1}p^{k}))_{\bar{R},L},
\end{align}
which, in turn, leads to%
\begin{equation}
\text{vol}_{\bar{R}}=\kappa^{-n}\int\nolimits_{-\infty}^{+\infty}d_{L}%
^{n}x\int\nolimits_{-\infty}^{+\infty}d_{\bar{R}}^{n}p\,\exp(x^{i}%
|\!\ominus_{R}(\text{i}^{-1}p^{k}))_{\bar{R},L}.
\end{equation}
These relations can be proved in very much the same way as those in
(\ref{Lem1}). To sum up, the volume elements defined in (\ref{DefVol1}) and
(\ref{DefVol2}) can alternatively be represented as
\begin{align}
\text{vol}_{L}=\text{vol}_{\bar{R}}  &  =\kappa^{n}\int\nolimits_{-\infty
}^{+\infty}d_{L}^{n}x\int\nolimits_{-\infty}^{+\infty}d_{\bar{R}}^{n}%
p\,\exp(x^{i}|\!\ominus_{L}\!(\text{i}^{-1}p^{k}))_{\bar{R},L}\nonumber\\
&  =\kappa^{-n}\int\nolimits_{-\infty}^{+\infty}d_{L}^{n}x\int
\nolimits_{-\infty}^{+\infty}d_{\bar{R}}^{n}p\,\exp(x^{i}|\!\ominus
_{R}(\text{i}^{-1}p^{k}))_{\bar{R},L},\\
\text{vol}_{\bar{L}}=\text{vol}_{R}  &  =\kappa^{-n}\int\nolimits_{-\infty
}^{+\infty}d_{\bar{L}}^{n}x\int\nolimits_{-\infty}^{+\infty}d_{R}^{n}%
p\,\exp(x^{i}|\!\ominus_{\bar{L}}\!(i^{-1}p^{k}))_{R,\bar{L}}\nonumber\\
&  =\kappa^{n}\int\nolimits_{-\infty}^{+\infty}d_{\bar{L}}^{n}x\int
\nolimits_{-\infty}^{+\infty}d_{R}^{n}p\,\exp(x^{i}|\!\ominus_{\bar{R}%
}(\text{i}^{-1}p^{k}))_{R,\bar{L}}. \label{VolSR}%
\end{align}
Let us note that there are similar identities for the more general volume
elements vol$_{A,B}.$

\subsection{Proof of the last relation in (\ref{FPId4})}

To prove the last identity in (\ref{FPId4}) we adapt the calculation of
(\ref{CalcId1N}) as follows:%
\begin{align}
&  \text{vol}_{\bar{R}}\big \langle f,g\big \rangle_{\bar{R},x}=\text{vol}%
_{\bar{R}}\int_{-\infty}^{+\infty}d_{\bar{R}}^{n}x\,\overline{f(x^{i}%
)}\circledast g(x^{j})\nonumber\\
&  =\,\int_{-\infty}^{+\infty}d_{\bar{R}}^{n}x\,\overline{f(x^{i})}\overset
{x}{\circledast}\int_{-\infty}^{+\infty}d_{\bar{R}}^{n}y\,\delta_{L}^{n}%
(y^{k}\oplus_{R}(\ominus_{L}\!x^{j}))\overset{x|y}{\odot}_{\hspace{-0.01in}%
R}g(y^{l})\nonumber\\
&  =\,\int_{-\infty}^{+\infty}d_{\bar{R}}^{n}x\,\overline{f(x^{i})}\overset
{x}{\circledast}\int_{-\infty}^{+\infty}d_{\bar{R}}^{n}y\,\int_{-\infty
}^{+\infty}d_{L}^{n}p\,\exp(p^{m}|y^{k}\oplus_{R}(\ominus_{L}\!x^{j}%
))_{\bar{R},L}\nonumber\\
&  \qquad\qquad\qquad\overset{x|y}{\odot}_{\hspace{-0.02in}R}g(y^{l}%
)\nonumber\\
&  =\,\int_{-\infty}^{+\infty}d_{\bar{R}}^{n}x\,\int_{-\infty}^{+\infty
}d_{\bar{R}}^{n}y\,\overline{f(x^{i})}\overset{x}{\circledast}\int_{-\infty
}^{+\infty}d_{L}^{n}p\,\exp(p^{r}|\!\ominus_{L}\!\kappa x^{j})_{\bar{R}%
,L}\nonumber\\
&  \qquad\qquad\qquad\overset{x|p}{\odot}_{\hspace{-0.02in}R}\exp(p^{m}%
|y^{k})_{\bar{R},L}\overset{y}{\circledast}g(y^{l})\nonumber\\
&  =\,\int_{-\infty}^{+\infty}d_{L}^{n}p\,e_{(\bar{R},p)}^{a}\overset
{p}{\circledast}\big (\mathcal{R}_{[2]}\triangleright e_{(\bar{R},p)}%
^{b}\big )\otimes\int_{-\infty}^{+\infty}d_{\bar{R}}^{n}x\,\overline{f(x^{i}%
)}\overset{x}{\circledast}S\big (\mathcal{R}_{[1]}\triangleright e_{(L,\kappa
x)}^{a}\big )\nonumber\\
&  \qquad\qquad\qquad\otimes\int_{-\infty}^{+\infty}d_{\bar{R}}^{n}%
y\,e_{(L,y)}^{b}\overset{y}{\circledast}g(y^{l})\nonumber\\
&  =\,\int_{-\infty}^{+\infty}d_{\bar{R}}^{n}x\,\overline{f(x^{i})}\overset
{x}{\circledast}S\big (\mathcal{R}_{[1]}^{-1}\triangleright e_{(L,x)}%
^{a}\big )\otimes\int_{-\infty}^{+\infty}d_{L}^{n}p\,\big (\mathcal{R}%
_{[2]}^{-1}\triangleright e_{(\bar{R},p)}^{a}\big )\overset{p}{\circledast
}e_{(\bar{R},p)}^{b}\nonumber\\
&  \qquad\qquad\qquad\otimes\int_{-\infty}^{+\infty}d_{\bar{R}}^{n}%
y\,e_{(L,y)}^{b}\overset{y}{\circledast}g(y^{l})\nonumber\\
&  =\,\int_{-\infty}^{+\infty}d_{L}^{n}p\,\overline{\mathcal{R}_{[1]}%
^{-1}\triangleright e_{(L,p)}^{a}\otimes(-1)^{n}\int_{-\infty}^{+\infty}%
d_{L}^{n}x\,\bar{S}^{-1}\big (\mathcal{R}_{[2]}^{-1}\triangleright e_{(\bar
{R},x)}^{a}\big )\overset{x}{\circledast}f(x^{i})}\nonumber\\
&  \qquad\qquad\qquad\overset{p}{\circledast}e_{(\bar{R},p)}^{b}\otimes
\int_{-\infty}^{+\infty}d_{\bar{R}}^{n}y\,e_{(L,y)}^{b}\overset{y}%
{\circledast}g(y^{l})\nonumber\\
&  =(-1)^{n}\int_{-\infty}^{+\infty}d_{L}^{n}p\,\overline{\int_{-\infty
}^{+\infty}d_{L}^{n}x\,\bar{S}^{-1}(e_{(\bar{R},x)}^{a})\overset
{x}{\circledast}\big (\mathcal{R}_{[2]}\triangleright f(x^{i})\big )\otimes
\mathcal{R}_{[1]}\triangleright e_{(L,\kappa^{-1}p)}^{a}}\nonumber\\
&  \qquad\qquad\qquad\overset{p}{\circledast}e_{(\bar{R},p)}^{b}\otimes
\int_{-\infty}^{+\infty}d_{\bar{R}}^{n}y\,e_{(L,y)}^{b}\overset{y}%
{\circledast}g(y^{l})\nonumber\\
&  =\,(-1)^{n}\int_{-\infty}^{+\infty}d_{L}^{n}p\,\overline{\int_{-\infty
}^{+\infty}d_{L}^{n}x\,\exp(\ominus_{\bar{R}}\,x^{r}|\kappa^{-1}p^{m}%
)_{\bar{R},L}\overset{p|x}{\odot}_{\hspace{-0.01in}R}f(x^{i})}\nonumber\\
&  \qquad\qquad\qquad\overset{p}{\circledast}\int_{-\infty}^{+\infty}%
d_{\bar{R}}^{n}y\,\exp(p^{k}|y^{j})_{\bar{R},L}\overset{y}{\circledast}%
g(y^{l})\nonumber\\
&  =\,(-1)^{n}\text{vol}_{\bar{R}}\int_{-\infty}^{+\infty}d_{L}^{n}%
p\,\overline{\mathcal{F}_{L}^{\ast}(f)(\kappa^{-1}p^{m})}\overset
{p}{\circledast}\mathcal{F}_{\bar{R}}(g)(p^{k})\nonumber\\
&  =\,(-1)^{n}\text{vol}_{\bar{R}}\big \langle\mathcal{F}_{L}^{\ast}%
(f)(\kappa^{-1}p^{m}),\mathcal{F}_{\bar{R}}(g)\big \rangle_{L,p}.
\end{align}
The line of reasonings is the same as for the calculation in\ (\ref{CalcId1N}).

\end{document}